\documentclass[aps,twocolumn,groupedaddress,preprintnumbers,nofootinbib,balancelastpage,longbibliography]{aastex6}

\usepackage{graphicx}
\usepackage{epstopdf}
\usepackage{dcolumn}
\usepackage{bm}
\usepackage[utf8]{inputenc}
\usepackage{subfigure}
\usepackage{color}
\usepackage{hhline}

\usepackage{amsmath}
\usepackage{booktabs}
\usepackage{array}
\usepackage{makecell}
\usepackage[framemethod=TikZ]{mdframed}
\usepackage{pifont}
\usepackage{longtable}

\usepackage[english]{babel}

\begin{document}


\def\a{\alpha}
\def\b{\beta}
\def\c{\varepsilon}
\def\d{\delta}
\def\e{\epsilon}
\def\f{\phi}
\def\g{\gamma}
\def\h{\theta}
\def\k{\kappa}
\def\l{\lambda}
\def\m{\mu}
\def\n{\nu}
\def\p{\psi}
\def\q{\partial}
\def\r{\rho}
\def\s{\sigma}
\def\t{\tau}
\def\u{\upsilon}
\def\v{\varphi}
\def\w{\omega}
\def\x{\xi}
\def\y{\eta}
\def\z{\zeta}
\def\D{\Delta}
\def\G{\Gamma}
\def\H{\Theta}
\def\L{\Lambda}
\def\F{\Phi}
\def\P{\Psi}
\def\S{\Sigma}

\def\o{\over}
\def\beq{\begin{align}}
\def\eeq{\end{align}}
\newcommand{\gsim}{ \mathop{}_{\textstyle \sim}^{\textstyle >} }
\newcommand{\lsim}{ \mathop{}_{\textstyle \sim}^{\textstyle <} }
\newcommand{\vev}[1]{ \left\langle {#1} \right\rangle }
\newcommand{\bra}[1]{ \langle {#1} | }
\newcommand{\ket}[1]{ | {#1} \rangle }
\newcommand{\mEV}{ {\rm meV} }
\newcommand{\EV}{ {\rm eV} }
\newcommand{\KEV}{ {\rm keV} }
\newcommand{\MEV}{ {\rm MeV} }
\newcommand{\GEV}{ {\rm GeV} }
\newcommand{\TEV}{ {\rm TeV} }
\newcommand{\1}{\mbox{1}\hspace{-0.25em}\mbox{l}}
\newcommand{\headline}[1]{\noindent{\bf #1}}
\def\diag{\mathop{\rm diag}\nolimits}
\def\Spin{\mathop{\rm Spin}}
\def\SO{\mathop{\rm SO}}
\def\O{\mathop{\rm O}}
\def\SU{\mathop{\rm SU}}
\def\U{\mathop{\rm U}}
\def\Sp{\mathop{\rm Sp}}
\def\SL{\mathop{\rm SL}}
\def\tr{\mathop{\rm tr}}
\def\mpl{M_{\rm Pl}}

\def\IJMP{Int.~J.~Mod.~Phys. }
\def\MPL{Mod.~Phys.~Lett. }
\def\NP{Nucl.~Phys. }
\def\PL{Phys.~Lett. }
\def\PR{Phys.~Rev. }
\def\PRL{Phys.~Rev.~Lett. }
\def\PTP{Prog.~Theor.~Phys. }
\def\ZP{Z.~Phys. }

\def\dd{\mathrm{d}}
\def\ff{\mathrm{f}}
\def\BH{{\rm BH}}
\def\inf{{\rm inf}}
\def\ev{{\rm evap}}
\def\eq{{\rm eq}}
\def\SM{{\rm sm}}
\def\Mpl{M_{\rm Pl}}
\def\GeV{{\rm GeV}}
\def\I{$I_{2-10}$}

\newcommand{\es}[2] {\begin{equation} \label{#1} \begin{split} #2 \end{split} \end{equation}}

\newcommand{\CD}[1]{\textcolor{red}{[CD]: #1}}
\newcommand{\BS}[1]{\textcolor{blue}{[BS]: #1}}
\newcommand{\JF}[1]{\textcolor{orange}{[JF]: #1}}

\def\newpar{\vskip4pt}

\title{
Hard X-ray Excess from the Magnificent Seven Neutron Stars
}
\preprint{LCTP-19-25}

\author{Christopher Dessert, Joshua W. Foster, and Benjamin R. Safdi}
\affiliation{Leinweber Center for Theoretical Physics, Department of Physics, University of Michigan, Ann Arbor, MI 48109, USA}

\begin{abstract}
We report significant hard X-ray excesses in the energy range 2-8 keV for two nearby isolated neutron stars: RX J1856.6-3754 and RX J0420.0-5022.  These neutron stars have previously been observed in soft X-rays to have nearly thermal spectra at temperatures $\sim$100 eV, which are thought to arise from the warm neutron star surfaces.   We find nontrivial hard X-ray spectra well above the thermal surface predictions with archival data from the XMM-Newton and Chandra X-ray telescopes.
 We analyze possible systematic effects that could generate such spurious signals, such as nearby X-ray point sources and pileup of soft X-rays, but we find that the hard X-ray excesses are robust to these systematics to the extent that is possible to test. We also investigate possible sources of hard X-ray emission from the neutron stars and find no satisfactory explanation with known mechanisms, suggesting that a novel source of X-ray emission is at play.  We do not find high-significance hard X-ray excesses from the other five Magnificent Seven isolated neutron stars.
\end{abstract}

\date{\today}

\maketitle

\section{Introduction}

The Magnificent Seven neutron stars (NSs) are a group of seven nearby X-ray dim isolated NSs (XDINSs) that emit near-thermal soft X-ray emission with relatively low luminosities.  They were first discovered in the ROSAT All Sky Survey data: \object{RX J1856.6-3754} by~\cite{1996Natur.379..233W}, \object{RX J0720.4-3125} by~\cite{1997AA...326..662H}, \object{RX J0806.4-4123} by~\cite{1998AN....319...97H}, \object{RX J1308.6+2127} by~\cite{Schwope:1998gp}, \object{RX J1605.3+3249} by~\cite{Motch:1999xy}, \object{RX J0420.0-5022} by~\cite{Haberl:1999ck}, and \object{RX J2143.0+0654} by~\cite{Zampieri:2001ewa}. They were later identified as a distinct class of objects by their spectral and temporal properties; see, e.g.,~\cite{Haberl2007} for a review.  Until now, no hard X-ray flux has been observed from the XDINSs.  In this work, we use archival XMM-Newton (hereafter XMM) and Chandra data to search for hard X-ray excesses in the 2-8 keV energy range from the XDINSs.  We find that such excesses exist for the NSs RX J1856.6-3754 and  RX J0420.0-5022. We characterize the spectral shapes of the hard excesses, search for evidence of variability and time dependence, and discuss possible origins of the flux.     

Each of the XDINSs is radio-quiet (but see~\cite{2007ApSS.308..211M}) and characterized by a near-blackbody continuum in soft X-rays with distortions due to attenuation by the interstellar medium as well as potential absorption lines from the NS atmospheres. The near-thermal emission suggests we are viewing the NS surfaces, with temperatures ranging from approximately 50 eV to 100 eV. The low interstellar attenuation implies that the XDINSs are within hundreds of pc of Earth, confirmed in some cases by parallax measurements~\citep{vanKerkwijk2007}. The origin of the absorption lines is thought to be cyclotron resonance absorption~\citep{Haberl:2003zn}. Each NS also has an optical counterpart with a flux larger than expected from the Rayleigh-Jeans tail of the X-ray blackbody~\citep{Kaplan:2011ay}, although this may be associated with the NS atmosphere. 

Six of the NSs are known to pulsate in X-rays with spin periods on the order of seconds. Assuming the NSs were born with millisecond spin periods, spin-down via magnetodipole radiation suggests large magnetic fields $\sim$$10^{13}$ G and ages of around $10^6$ yr~\citep{Page:2005fq}. Coherent timing solutions have confirmed the field strengths, which roughly agree with the field strengths inferred from the absorption lines assuming they are due to cyclotron resonance by protons. The ages, along with the proper motions, point to a single birth place in the Gould Belt~\citep{Motch:2006vy}.

The hard X-ray excesses identified in this work could have a variety of physical origins.  One exotic origin, which we discuss in a companion paper~\citep{companion}, is that the excesses arise due to the presence of a new ultralight particle of nature called the axion.  The axions may be produced thermally in the cores of the NSs, which are expected to have temperatures of a few keV.  The axions then escape the NSs and convert into X-rays in the strong magnetic fields surrounding the stars.  The resulting spectrum is then nearly thermal at the core temperature, though some deviations away from the thermal spectrum are expected~\citep{Raffelt:1987im}.  Less exotic explanations of the excess flux include nonthermal emission from charged-particle acceleration in the magnetospheres and X-ray emission from accretion of surrounding material.  However, we point out issues with these explanations later in this work.  

The remainder of this paper is organized as follows.  In Section~\ref{Sec:2}, we describe our data reduction and analysis procedure.  In Section~\ref{1856}, we present detailed results for RX J1856.6-3754, the NS in which we find the most significant hard X-ray excess with a statistical significance around 5$\sigma$. In Section~\ref{Sec:4}, we present our main results for the hard X-ray spectra of the remaining XDINSs, showing that while an excess is found robustly for RX J0420.0-5022 we cannot conclusively say whether similar excesses exist for the other five XDINSs. We consider possible origins of the flux in Section~\ref{Sec:5}, and conclude with a discussion in Section~\ref{Sec:6}.

\section{X-ray data reduction and analysis}
\label{Sec:2}

We use archival XMM and Chandra data to investigate the hard X-ray fluxes from the XDINSs. It is valuable to use data from both instruments because each is optimized for a different objective. {\it A priori}, Chandra should be the superior instrument for observation of the XDINSs with its excellent point source sensitivity in the hard X-ray range, which can be attributed to its small point spread function (PSF). However, the instrument is highly susceptible to X-ray pileup, which can artificially raise the event energies reported in an observation.  This is potentially an issue when searching for hard X-ray flux in the presence of a significant soft X-ray spectrum.  Meanwhile, XMM has the superior effective area and collectively the most exposure time of the XDINSs. Furthermore, pileup is likely to be an insignificant contributor to the hard XMM spectra for these relatively dim NSs. However, the large PSF of XMM also allows for contamination due to nearby sources, which could bias either our estimates of the signal or background spectra.

The fact that the hard XDINS spectra are consistent between the two instruments, as we show, is a promising sign that the reported excesses are not due to systematic effects. It is unlikely that a point source in the XMM spectra would also contaminate the Chandra spectra.  Moreover, the consistency between the XMM and Chandra spectra suggests that pileup, which strongly depends on the source count rate, is not responsible for the excesses. Nevertheless, we incorporate systematic tests for these issues into our analysis.

In this section, we outline our data reduction procedures for XMM and Chandra. We further discuss our MARX simulations of the Chandra detector, which diagnose possible pileup effects and which we use to cut data if it appears pileup could be significant. We then discuss our analysis procedures for reconstructing the XDINS 2-8 keV spectra.

\subsection{Data reduction}

Here, we describe the methods we use to process the publicly available data from XMM and Chandra into the spectra and images analyzed in this work.  The observation identification numbers for the observations used in this work are given in Appendix~\ref{app:IDs}, and the reduced data is given are given in Appendix~\ref{sec:data}.
 
\subsubsection{XMM-Newton Observations}

The data products for XMM are downloaded from the XMM-Newton Science Archive\footnote{http://nxsa.esac.esa.int}. To perform the processing, we use \textit{XMM-Newton} Science Analysis System (SAS)~\citep{XMM-SAS} version 17.0.

We first generate summary information for the dataset by generating the Calibration Index File (CIF) using the SAS task \texttt{cifbuild}, which locates the Current Calibration File (CCF). The CCF provides information about the state of the detector at observation time, which is necessary for future processing. We then run the task \texttt{odfingest}, which generates the Observation Data Files (ODF) containing general information on the detector. 

For any individual observation, there may be multiple exposures for each camera, which are individual datasets taken during the observation time. Only a subset, called the ``science exposures,'' are useful for analysis. The relevant science exposures for each observation ID to use for data reduction are determined from the Pipeline Processing Subsystem summary file. We only use science exposures in imaging mode, which we refer to simply as exposures for the remainder of the text.

From this information, we reprocess the ODF for the MOS and PN cameras with the tasks \texttt{emproc} and \texttt{epproc}, respectively. These tasks create calibrated and concatenated but otherwise unfiltered event lists. We then generate the filtered event lists for each science exposure with the task \texttt{espfilt}, which filters the light curves for soft proton (SP) contamination, which can significantly enhance the count rates for short periods of time. An observation affected by SP will have a count rate histogram that is approximately Gaussian with a peak at the unaffected rate but with a long high-count rate tail due to the contamination. The \texttt{espfilt} task establishes thresholds at $\pm 1.5\sigma$ of the count rate distribution and creates a good time interval (GTI) file containing the time intervals where the count rate is contained within the thresholds. This task returns a filtered event list, which contains only the events arriving during the GTIs. We then use only the filtered events in the analysis going forward.

We create images with \texttt{evselect} in the energy bins 2-4, 4-6, and 6-8 keV with the standard pixel sizes, $4\farcs 1$ for PN and $1\farcs 1$ for MOS. For the PN camera, we select only events with \texttt{FLAG==0} and \texttt{PATTERN<=4} (i.e., single and double events), while for the MOS camera, we select events with \texttt{PATTERN<=12}. We run a point-source detection algorithm, \texttt{edetect\_chain}, simultaneously on the images, which returns a list of point-source locations. We use this source detection to determine the location of the NS in each exposure\textemdash the coordinates are subject to variations between exposures due to calibration uncertainties and the NS proper motion. In addition to a list of resolved point sources, this task returns exposure maps.  
 We then run \texttt{rmfgen} and \texttt{arfgen}, which compute the detector redistribution matrix file (RMF) and the ancillary response file (ARF). The former accounts for the energy resolution of the detector while the latter accounts for the energy-dependent effective area. We correct the RMF for pileup in the case of the PN camera with the parameter setting \texttt{correctforpileup=yes}; however, a correction for MOS is not possible at this time. In its place, we run the task \texttt{epatplot} which estimates the amount of pileup in a spectrum; however, it is of limited use in our specific case. The \texttt{Epatplot} task compares the observed spectrum of the single and quadruple events in a single observation to the theoretical spectrum assuming no pileup. For the XDINSs, while there are significant counts below 2 keV, there are very few counts above 2 keV. With so few counts in individual observations, any deviations from the no-pileup spectrum are not statistically significant. However, we do run the pileup test of~\cite{Jethwa}, which estimates several measures of pileup in the XMM cameras.
 
 To fit for the X-ray spectra, we begin with the image files around the NSs.  We use the images created by \texttt{evselect}/\texttt{edetect\_chain}. 
 We create images for each exposure $e$: counts image $c_i^{p,e}$ (units (counts)) and exposure images $w_i^{p,e}$ (units (cm$^2$\ s\ keV)) for each of the energy bands $i$, where $p$ indexes the pixels.  In the high-energy analysis, we stack the images over exposures on a uniform R.A.-decl. grid, while for the low-energy analyses, we use a joint likelihood over the individual exposures because the instrument responses are more important at low energies where the energy resolution plays an important role. To create the stacked images, we separately stack the images in each detector (MOS or PN) over the individual exposures in each energy band in the following way. In each image, we first redefine the coordinate system such that the origin is at the source location $(\textrm{R.A.}_0,\textrm{decl.}_0)$. To correct for astrometric errors, the source location in the image is determined through PSF-based template fitting to the counts data in the low-energy data ($<$2 keV). Because each NS is bright in soft X-rays, this permits the localization of the NS in each image with subpixel accuracy. This corrects for the fact that the NS location may not be identical in each image due to a combination of calibration errors and proper motions. We then downbin the images $I^e = \{c^e,w^e\}$ from the individual exposures into the stacked images $I = \{c,w\}$ on a uniform grid of R.A. and decl. according to
\begin{align}
	c_{i}^{ p^\prime} &= \sum_e \sum_{p \in p^\prime} c_i^{p,e} \\
	w_{i}^{ p^\prime} &= \sum_e \dfrac{\sum_{p \in p^\prime} w_{i}^{p,e}}{\sum_{p \in p^\prime} } \,,
\end{align}
where the pixel sums are over pixels $p$ that have coordinates contained within the downbinned pixel indexed by $p'$.

The above stacking procedure leaves us with images in each energy band for each NS and detector with which we perform our fiducial high-energy analyses.  To extract the spectra in each energy band, we define a signal region $R_S$ and a background region $R_B$. Extraction regions are defined relative to the energy-averaged 90\% encircled energy fraction (EEF) radius. The signal region is a circle centered at the source location with radius 50\% of the energy-averaged 90\% EEF radius, whereas the background region is an annulus centered at the source location extending from the edge of the signal region to an outer radius of 75\% of the energy-averaged 90\% EEF radius.  We keep the background region compact to mitigate possible contamination from point sources.
For the XMM EEF model, we use the `Medium' mode PSF description assuming an on-axis source. 
Using the EEFs, we compute that in $R_S$ there is a fraction of the signal $\chi_{S,i}$, while in $R_B$ there is a signal fraction $\chi_{B,i}$.

Energy containment fractions are calculated by a Monte Carlo procedure in which test counts are placed with displacements from source location are drawn from the angular radius distribution specified by the PSF, then binned at our downbinned pixel resolution. The fraction of test counts placed within signal (background) region pixels defines the signal containment of the signal (background) region. Note that these fractions are energy-dependent because the region sizes are energy-independent, and that 50\% of the energy-averaged 90\% EEF radius is inequivalent to the energy-averaged 50\% EEF radius. For instance, while 100\% of the energy-averaged 90\% EEF radius of $0\farcm 6$ for PN corresponds to a signal containment of 88\% in the 2-4 keV bin, 50\% of that energy-averaged 90\% EEF radius corresponds to a signal containment of 74\% in that same bin.

\subsubsection{Chandra Observations}

For the Chandra analyses we use the Chandra Interactive Analysis of Observations (CIAO) \citep{2006SPIE.6270E..1VF} version 4.11.
We choose all ACIS \texttt{Timed Exposure} observations of the XDINSs for analysis, irrespective of the grating and the spectral (-S) or imaging (-I) component. We will refer to these observations as Chandra observations for the remainder of the text. We use the CIAO task \texttt{download\_chandra\_obsid} to download the observations and reprocess them using \texttt{chandra\_repro}. This task yields a filtered events file. We run \texttt{fluximage} on the events file to create images in the same bands as for XMM, along with exposure maps with pixel sizes of $0\farcs 492$. We then run the source detection algorithm \texttt{celldetect} on the images, yielding the source coordinates. 
We use the \texttt{specextract} task to produce the detector response matrices. 

We create Chandra images using the task  \texttt{fluximage}.  The signal and background regions are defined in a manner analogous to that used for XMM, except that for Chandra, the outer radius of the background region is taken to be 250\% of the 90\% energy-averaged EEF radius. This is done because, for Chandra, nearby point sources are less of a concern, given the superior PSF.  For the Chandra EEF model, we use the CIAO tool \texttt{psf}.

\subsection{MARX simulations}
\label{sec:marx}

In this section, we discuss our Model of AXAF Response to X-rays (MARX)~\citep{2012SPIE.8443E..1AD} simulation framework. We use MARX version 5.4.0 to perform two simulations for each Chandra observation\textemdash each with the best-fit soft thermal spectrum from the data, but one with a hard X-ray tail of constant $10^{-15}$~erg~cm$^{-2}$~s$^{-1}$~keV$^{-1}$ and one without. To create the spectral file, we use Interactive Spectral Interpretation System version 1.6.2~\citep{2000ASPC..216..591H} to generate a parameter file, and then we use the MARX tool \texttt{marxflux} to convert it to the MARX-friendly format. In order to reproduce the observation conditions as closely as possible, to negate systematic errors, we simulate the NS with \texttt{marx} at the same detector coordinates and with the same detector configuration as in the original observation. We create an events file, an aspect solution file, an ARF, and an RMF with \texttt{marx2fits}, \texttt{marxasp}, \texttt{mkarf}, and \texttt{mkrmf}, respectively.

At this point, images can be created with \texttt{fluximage} as in the Chandra processing. For grating observations, the dispersed events must be order-sorted with the CIAO tool \texttt{tg\_create\_mask} and then filtered out with \texttt{tg\_resolve\_events} to create events files compatible with \texttt{fluximage}. The resultant images will not include the effects of pileup. To simulate pileup effects, we run \texttt{marxpileup} and then convert the results to an events file and image with \texttt{marx2fits} and \texttt{fluximage} as before. 

\subsection{Data analysis}

We bin the data in 25 energy bins, with bin widths of 0.05 keV from 0 to 1 keV, one bin from 1 to 2 keV, and four bins of width 2 keV from 2 to 10 keV. Because of the energy range of the calibration on both instruments, we do not analyze data outside the range 0.5 keV-8 keV. We also exclude observations that have a flaring time greater than 50\%. In some detector operating modes ({\it e.g.}, \texttt{Small Window}), due to the placement of the source in the detector, there is a limited extraction region available for background estimation within the vicinity of the source, and we exclude these as well. Furthermore, some XMM observations are excluded due to the presence of spurious source detections in the wings of the PSF.

In this subsection, we discuss our analysis of the soft spectra, including our computation of the 0.5-2 keV spectra and the analysis of the 0.5-1 keV data, in which the NSs are significantly thermally emitting. We then outline our analysis of the 2-8 keV data, in which we search for hard X-ray emission. Both analyses are of a frequentist nature. 

\subsubsection{Soft Spectral Analysis}

We measure the soft X-ray spectra from 0.5 to 1 keV from the XDINSs, both to confirm that we reproduce the previously observed spectra and so that we may fit the soft spectra and extrapolate into the hard X-ray band.  
That is, we fit for the soft thermal flux in order to verify that the exponential tail of the surface blackbody cannot account for the hard X-ray excesses. We find that, although extrapolating the best-fit blackbody suggests the 2-8 keV bins are not contaminated by thermal surface emission, pileup of the soft photons can impact the hard spectra for some NSs and instruments. Furthermore, modeling the soft flux instead with NS atmosphere models indicates that, depending on the NS and the NS surface composition, the 2-4 keV flux may be partially contaminated by thermal emission.  We discuss these points later in this work.

For the soft analysis, we use different extraction regions $R_S$ and $R_B$ relative to the high-energy analysis, since in the low-energy analysis we are more concerned with mismodeling the PSF than with misestimating the background.  As such, we take $R_S$ to be 150\% of the 90\% EEF radius, and for $R_B$ to be 250\% of the 90\% EEF radius, for both XMM and Chandra.  
 We perform the following procedure for each detector (MOS, PN, or Chandra) independently. First, we construct the NS soft spectrum from 0.5 to 2 keV in the following manner. We let the data in $R_S$ in (counts) be denoted $d_S^e = \{c_{S,i}^e\}$, where $i$ runs over the 10 relevant energy bins and $e$ runs over the number of exposures passing the quality cuts for a specific detector.  Similarly, we let the background data in (counts) be denoted $d_B^{e} = \{c_{B,i}^{e}\}$.
 Recall that we do not work with the images stacked over exposures in the low-energy analyses.
 We use this background data to compute the mean expected background counts within the signal region, $\mu_{{\rm B},i}^e = \frac{\Omega_S}{\Omega_B} c_{B,i}^{e}$,  where $\Omega_S$ ($\Omega_B$) is the solid angle of $R_S$ ($R_B$).

Having obtained the source spectrum, we must now put our source spectral model in the same form. Assuming a source flux $S(E|{\bm \theta}_{\rm S})$ in (counts~cm$^{-2}$~s$^{-1}$~keV$^{-1}$) where ${\bm \theta}_{\rm S}$ are generic model parameters describing the soft flux, we obtain the expected source counts using forward modeling
\es{}{
	\mu_{S,i}^e({\bm \theta}_{\rm S}) =t^e \int dE^\prime{\rm RMF}_i^e(E^\prime) {\rm ARF}^e(E^\prime) S(E^\prime|{\bm \theta}_{\rm S}) \,.
}
Above, we have designated $t^e$ the observation time for the exposure in (s). The ARF (a function of true X-ray energy $E^\prime$) represents the effective area of the detector in (cm$^2$) and the RMF (dimensionless) is a probability distribution function for the probability to observe an X-ray photon in (reconstructed) energy bin $i$ given its true energy $E^\prime$\textemdash in short, it accounts for the energy resolution of the detector. 

To fit the data $d$ we use the Poisson likelihood
\es{}{
	\mathcal{L}(d|{\bm \theta}_{\rm S}) = \prod_{e,i} { \left( \mu_{S,i}^e({\bm \theta}_{\rm S}) + \mu_{{\rm B},i}^e \right)^{c_{S,i}^e} e^{- \left( \mu_{S,i}^e({\bm \theta}_{\rm S}) +  \mu_{{\rm B},i}^e \right)}\over c_{S,i}^e  !}
	}
joint over all exposures and energy bins. Note a slight subtlety: we may consider $\mu_{{\rm B},i}^e$ to be known because the background region from which it is measured is much larger than the signal region.  This is not true in the high-energy analysis. In the high-energy analysis, the background counts also play a more important role, so we treat them more carefully. 

Except when discussing more complicated atmosphere models, we limit ourselves to the three signal parameters ${\bm \theta}_S = \{I,T,N_H\}$, which are the intensity and surface temperature of the NS in (erg~cm$^{-2}$~s$^{-1}$~keV$^{-1}$) and (keV), respectively, along with the integrated hydrogen column density $N_{\rm H}$ in (atoms~cm$^{-2}$). That is, we assume a blackbody spectrum $dN/dE \sim E^2/(e^{E/T}-1)$ with the hydrogen absorption model presented in~\cite{Wilms:2000ez}. 
Deviations from pure thermal spectra have been observed in the XDINSs, however, and we study this further in Sec.~\ref{sec:atmo}.

\subsubsection{Hard Spectral Analysis}

In the high-energy analyses, we assume that the background is Poisson distributed with a \textbf{flux $\{B_i\}$} in each energy bin. 
Assuming the source has source fluxes $\{S_i\}$, the expected number of counts in $R_B$ in energy bin $i$ is $\mu_{B,i} = \sum_{p \in R_B} w_i^p B_i + \chi_{B,i} w_i^q S_i$, where $q$ is the pixel at the source location. We compare this to the number of counts in $R_B$, which we denote $c_{B,i} = \sum_{p \in R_B} c_i^p$.  Note that we present the explicit numbers of counts in Appendix~\ref{sec:data}.

We then expect that the count in $R_S$ is similarly $\mu_{S,i} = \sum_{p \in R_S} w_i^p B_i + \chi_{S,i} w_i^q S_i$, where again the former is the background contribution and the latter is the signal contribution. Letting the number of counts in the signal region be $c_{S,i} = \sum_{p \in R_S} c_i^p$ leads to the joint Poisson likelihood over both $R_S$ and $R_B$:
\es{eq:likelihood}{
	\mathcal{L}_{i,{\rm hard}}(\{c_{S,i}, c_{B,i} \}|S_i,B_i) = &\dfrac{(\mu_{S,i})^{c_{S,i}}e^{-\mu_{S,i}}}{c_{S,i}!}  \times \\&\dfrac{(\mu_{B,i})^{c_{B,i}}e^{-\mu_{B,i}}}{c_{B,i}!} \,.
}
In each energy bin, we construct the profile likelihood over the signal flux $S_i$ treating the background flux $B_i$ as a nuisance parameter, leading to 
\begin{equation}
	\mathcal{L}_{i,{\rm hard}}(\{c_{S,i}, c_{B,i} \}|S_i) = \max_{B_i} \mathcal{L}_{i,{\rm hard}}(d|S_i,B_i) \,.
\end{equation}
The best-fit flux in energy bin $i$, which we denote by $\hat S_i$, is then given by the value of $S_i$ that maximizes the profile likelihood.

\subsection{Statistical analysis}
\label{sec:stats}

We determine the confidence intervals on the fluxes in the individual energy bins using the test statistic (TS)
\es{eq:TS-confidence}{
	q_{i,{\rm hard}}(S_i) = 2 \times \left[  \log \mathcal{L}_{i,{\rm hard}}(\{c_{S,i}, c_{B,i} \}|\hat S_i) - \right. \\
	\left. \log \mathcal{L}_{i,{\rm hard}}(\{c_{S,i}, c_{B,i} \}|S_i) \right] \,.
}
The 1$\sigma$ frequentist confidence interval for the flux is asymptotically given by the range of $S_i$ in which the $q$ is within one of its minimum (see,~{\it e.g.},~\citep{Cowan:2010js}).  Note that, for consistency, we must consider negative $S_i$ values.  To more accurately compute the confidence intervals, we must compute the distribution of TSs from Monte Carlo, given that the number of counts may be small for some observations so that we are not in the asymptotic limit.

Away from the asymptotic limit, we wish to determine the $n$$\sigma$ confidence interval for a parameter of interest $S$ with best-fit parameter $\hat S$.  The confidence interval is defined as  the range of values below the $n$$\sigma$ upper limit $S_{+n}$ and above the $n$$\sigma$ lower limit $S_{-n}$. The upper limit $S_{+n}$ is defined by the maximum value of the parameter such that simulated data generated by the model with that parameter would satisfy the condition $P(\hat S' \geq \hat S) = \Phi(n)$, where $\hat S'$ denotes the best-fit values from the simulated data, $\Phi$ is the cumulative distribution function of the normal distribution, and $P(\hat S' \geq \hat S)$ is the probability that $\hat S' \geq \hat S$.  The lower limit $S_{-n}$ is defined similarly.

We apply this frequentist confidence interval procedure to our data in the following way. Given our data, we maximize the likelihood profile, which is profiled over nuisance parameters $\bm \theta$,  to determine $\hat S$.

To determine the upper limit, we first consider a particular value $S' \geq \hat S$ and maximize our likelihood at this fixed $S'$ to find the best-fit nuisance parameters $\hat{\bm \theta}'$.  We then generate many Monte Carlo realizations of the data under the model defined by $S'$ and $\hat{\bm \theta}'$.  From the simulated data, we determine the distribution of the best-fit $\hat{S}'$.
  In this way, we are able to determine the percentile of the observed best-fit $\hat S$ (from the actual data) in the distribution of $\hat{S}'$ generated under $\{S',\hat{\bm \theta}'\}$ and ultimately determine $\hat S_{+n}$. An analogous procedure using $S' \leq \hat S$ enables the determination of $\hat S_{-n}$.  In practice, we find that this procedure reproduces the asymptotic expectation except in a few specific cases, such as those with Chandra data, where the number of counts is low. 
    
  In addition to determining the fluxes in the individual energy bins, we also fit power-law spectral models across energy bins.  As will be further described later, these models have parameters of interest $I$ and $n$, where $I$ denotes an intensity over the full energy range and $n$ is the spectral index.  The parameters $I$ and $n$ may be constrained in the frequentist way by constructing the joint likelihood over the relevant energy bins and datasets.  The confidence intervals on these parameters are determined using the Monte Carlo method described above, which matches the asymptotic expectation in most, though not all, cases.  
  
  When fitting the spectral models, we are also interested in the evidence for the nontrivial spectral model over the null hypothesis of no hard X-ray flux from the source.  To quantify the statistical significance of the model (i.e., the evidence), we need to define a TS for discovery:
  \es{eq:TS-discovery}{
  {\rm TS} = 2 \times \left[ \log \mathcal{L}_{{\rm hard}}(d|\hat I) - \log \mathcal{L}_{i,{\rm hard}}(d| {\bf 0}) \right] \,,
  }
  unless the best-fit intensity $\hat I < 0$ in which case ${\rm TS} = 0$.  Note that ${\bf 0}$ denotes the null hypothesis $I = 0$ and $\mathcal{L}_{{\rm hard}}(d|\hat I) $ is the profile likelihood for the intensity over all energy bins, with the index $n$ also profiled over, evaluated at the best-fit intensity ({\it i.e.}, the maximum log-likelihood for the signal hypothesis).  Here, $d$ denotes the combination of datasets under consideration.  In the asymptotic limit, the TS may be straightforwardly interpreted in terms of significance, considering that our signal model has two model parameters of interest (see~\citep{Cowan:2010js}).  However, as we are often away from the asymptotic limit, we determine the significance directly through Monte Carlo.
To do so, we first determine the best-fit null model, and then we generate Monte Carlo data from the null model parameters.  We calculate the distribution of the TS on that Monte Carlo data using~\eqref{eq:TS-discovery}. The fraction of TSs generated under Monte Carlo that exceed the TS evaluated on the observed data defines a $p$-value with standard interpretation in terms of detection significance.  Again, we find that in most (but not all) fits, the recovered $p$-value matches the asymptotic expectation.

\subsection{Point-source Detection}

As a systematic test, we consider the effect that nearby point sources might have on the recovered spectra for our sources of interest.
Point sources within the signal or background extraction regions of the PN and MOS data could potentially bias our determinations of the source spectra.  While point sources could, in principle, also be an issue for Chandra observations, the superior angular resolution of that detector means that the issue is much more important for XMM.  We search for sources by first constructing a high-density (R.A., decl.) grid within the vicinity of the source and background regions.  At each (R.A., decl.) point, we determine a signal and background region as we do for our source of interest. We identify point sources by calculating a TS at each grid point for excess counts. Because point sources are expected to appear across a range of energies, we sum the counts maps over the 2-8 keV range.  The point-source discovery TS is defined analogously to~\eqref{eq:TS-discovery}. 
 We join the PN and MOS TSs together to form a joint test statistic at each (R.A., decl.) point.
 We identify point sources at those locations where the joint test statistics is greater than or equal to nine and the TS is the maximum TS on a region with an angular extent of the 50\% of the 90\% EEF radius. We then construct a point-source mask by masking out regions with radius 50\% of the 90\% EEF radius centered at any location where a point source was identified. Later, we demonstrate that the impact of masking point sources on the recovered spectra is relatively minor. 

\section{Hard X-ray excess in RX J1856.6-3754}
\label{1856}

\begin{figure}[htb!]
\begin{center}
\includegraphics[width = 1.0\columnwidth]{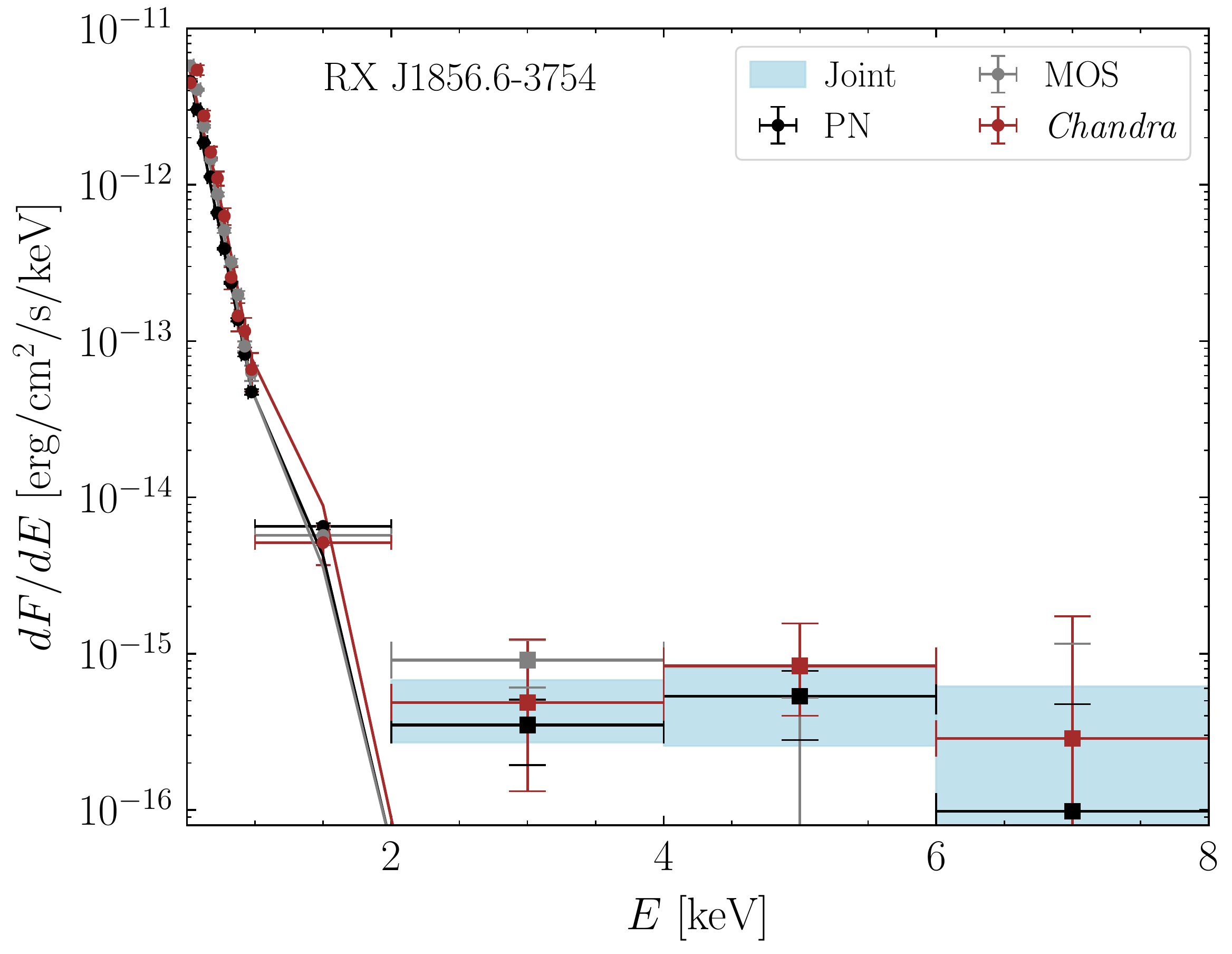}
\end{center}
\caption{Background-subtracted X-ray spectrum of RX J1856.6-3754 for each of the three cameras individually and combined. Data points were constructed by stacking all available exposures from the source, with best-fit spectral points and associated 68\% confidence intervals indicated.  In all three cameras, there is a clear and consistent excess above the background in the hard X-ray range of 2-8 keV, and because of the complementary strengths of the individual cameras, we believe this excess is robust. Solid curves denote the best-fit thermal spectra with hydrogen absorption fit from 0.5 to 1 keV, and as can be seen, the extrapolations of these spectra to the hard energy range does not account for the observed excess.
}
\label{fig:hard1856}
\end{figure}

In this section, we show results for the NS RX J1856.6-3754, for which the hard X-ray excess is detected with the greatest significance. Note that departures from a thermal model have previously been observed around 1 keV~\citep{Yoneyama:2017xth}; here, we focus on the 2-8 keV range.
In Fig.~\ref{fig:hard1856}, we show the background-subtracted X-ray spectrum $dF/dE$ over the energy range from 0.5 to 8 keV for PN, MOS, and Chandra. More precisely, what is shown is the observed number of counts per second in each energy channel divided by the diagonal entry on the forward modeling matrix that gives the effective area at that energy.  This subtlety is important below $\sim$1 keV, because at these energies the observed thermal spectrum is significantly affected by the energy resolution of the detector.  For this reason, it is not correct to interpret Fig.~\ref{fig:hard1856} as a plot of the true flux, since that would require inverting the forward modeling matrix, which is very much not diagonal at the low energies.  On the other hand, we are primarily interested in energies above 2 keV, and the forward modeling matrix is effectively diagonal at these energies with our energy binning, such that Fig.~\ref{fig:hard1856} is effectively a plot of the true flux at these energies. 

We also emphasize that these data points arise from joining all of the exposures from the individual cameras together into one single counts map per camera.  This is important because, as further discussed later in this work, the individual exposures do not have high enough statistics to detect the hard X-ray excess.  To construct this spectrum, we used 40 observations for a total of 1.0 Ms of exposure for PN, 18 for a total of 0.69 Ms for MOS, and 9 for a total of 0.23 Ms for Chandra.
We fit a thermal model, including the effect of hydrogen absorption at low energies, to the spectrum from 0.5 to 1 keV.  We find best-fit temperatures $T = 71.1 \pm 0.2$ eV ($T = 66.2 \pm 0.3$ eV) ($T = 67.8 \pm 0.9$ eV) for PN (MOS) (Chandra).  We note that these uncertainties are statistical only and do not capture possible systematic discrepancies in the true spectrum from thermal, in possible variations of the surface temperature over time, or in systematic uncertainties in the detector response.  For all cameras, we use the forward modeling matrices, constructed for each individual exposure, that account for both the effective area and the distribution of true flux to observed flux between energy channels in the low-energy analyses.  However, only the PN forward modeling matrix includes the effect of pileup.  We do not investigate the surface temperature uncertainties in more detail because it is not the main focus of this work.  Rather, as we illustrate in Fig.~\ref{fig:hard1856}, the thermal distribution, whose best-fit background-subtracted spectra are shown as solid curves, is able to account for the emission seen at and below $\sim$2 keV but is not able to account for the high-energy emission above 2 keV.  We will show later on that this statement remains true even for more complicated NS atmosphere models.

Below, we provide more detail for the spectral characterization of the high-energy flux and systematic tests that investigate the robustness of the signal.

\subsection{Spectral characterization of the RX J1856.6-3754 hard X-ray emission}

We fit a power-law model $dF / dE \propto E^n$ to the data to measure both the intensity of the signal and the hardness of the signal as indicated by the spectral index $n$. We quantify the intensity through $I_{2-8} = \int_{2 \, \, {\rm keV}}^{8 \, \, {\rm keV}} dE \, dF/dE$ in units of~erg~cm$^{-2}$~s$^{-1}$.\footnote{Note that later in this work we will consider, for some NSs, only the energy range 4-8 keV.  In those cases, we will quote $I_{4-8} = \int_{4 \, \, {\rm keV}}^{8 \, \, {\rm keV}} dE \, dF/dE$.}
The statistical procedure that we use for constraining $I_{2-8}$ and $n$ is outlined in Sec.~\ref{sec:stats}.

The results of the spectral fits for the three different cameras are given in Table~\ref{tab:1856}.
\renewcommand{\arraystretch}{1.25}
\begin{table}[htb]
\centering
\begin{tabular}{|c || c | c |c |}
\hline
 camera & $I_{2-8}$ ($10^{-15}$~erg~cm$^{-2}$~s$^{-1}$) & $n$ & $\sigma$ \\ \hline
  PN & $2.1_{-0.9}^{+0.9}$ & $-0.03_{-1.1}^{+0.89}$ & 3.0 \\ 
MOS & $1.4_{-0.7}^{+1.1}$ & $<-0.93$ & 2.9 \\  
 Chandra  & $3.8_{-1.8}^{+2.6}$ & $-0.28_{-1.89}^{+1.71}$ & 3.4 \\ 
 \hline
 Joint  & $1.53_{-.63}^{+.70}$ & $-0.28_{-0.75}^{+0.65}$ & 4.6 \\ 
 \hline
\end{tabular}
\caption{\label{tab:1856} Best-fit results for a power-law spectrum in RX J1856.6-3754. The fluxes and spectral indices are consistent between cameras, although the latter is not well constrained.  We also show the results from the joint-likelihood analysis over all cameras. For the fit to the MOS data, we are only able to place upper limits on the power-law index and report the 84$^\mathrm{th}$ percentile upper limit.}

\end{table}
 Interestingly, all three cameras give consistent flux measurements for $I_{2-8}$; moreover, for PN and Chandra the detection is at high statistical significance.  The computations of statistical significance are summarized in Sec.~\ref{sec:stats}.  The consistency between the three cameras is important because each has its own strengths and weaknesses.  
The fact that the high-energy signal is detected in each camera thus gives confidence that the high-energy signal is real and arises from the NS itself.
 
 The spectral index $n$ is not well-constrained by any of the individual cameras, which is perhaps not too surprising given the modest significance of the detections and the fact that we only have three independent energy bins to constrain the power law.  However, combining the results from all three cameras, we find the relatively hard energy index $n = -0.28_{-0.75}^{+0.65}$.  This index suggests that the emission is not the high-energy tail of the thermal surface emission, which should have a soft spectrum in this energy range.
 
 \subsection{Systematic tests for the RX J1856.6-3754 hard X-ray excess}
 
 The hard X-ray excess, suggesting nonthermal\footnote{More appropriately, if the spectrum is thermal then the temperature would need to be significantly higher than the NS surface temperature.} X-ray emission from RX J1856.6-3754, is detected at high statistical significance with PN and Chandra and at marginal significance with MOS.  However, each of these instruments is subject to systematic uncertainties, which we now examine in more detail. 
 
 \subsubsection{Test statistic maps and nearby point sources}
 
 One of the central differences between the Chandra and XMM detectors is the significantly better angular resolution of Chandra as compared to XMM.  This is important because it is possible that the observed hard X-ray excess does not arise from RX J1856.6-3754; it may instead come from a nearby source that is unrelated to RX J1856.6-3754 but happens to be at a similar angular position on the sky.  Pertinently, it is also possible that the hard X-ray excess is the result of misinterpreting  the background statistics.  That is, if a significant fraction of the background flux arises from relatively bright point sources, then the assumption that we may use the observed number of counts in the background region to infer the mean number of background counts in the signal region, with the probability distribution then being Poisson distributed about this mean, could break down.  It is reassuring that, for these reasons, we observe the excess both with Chandra and XMM.  Still, it is worth investigating the XMM counts maps visually and quantitatively in order to make sure that they do not show significant nearby point source emission or other sources of emission that would violate our assumptions.
 
\begin{figure*}[htb]
\begin{center}
\includegraphics[height = 0.23\textheight]{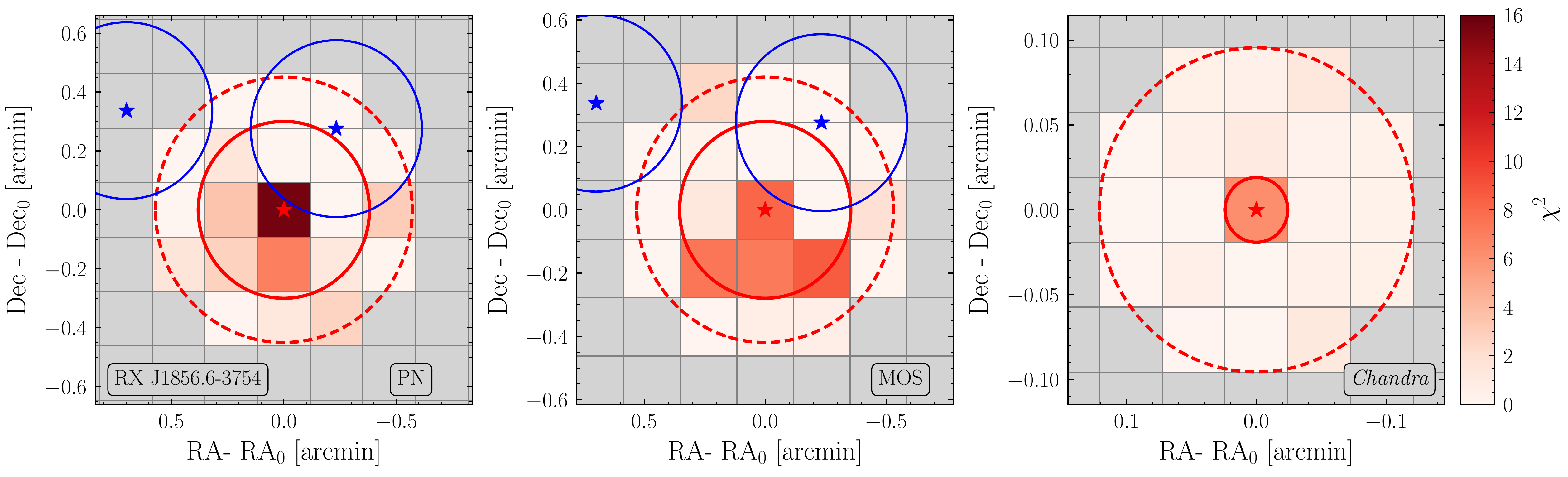}
\end{center}
\caption{Chi-squared maps (summed over 2-8 keV and all exposures) for each camera centered around the location of RX J1856.6-3754. In each case, the inner red ring denotes the radius within which the source data is extracted. Background data are extracted from the annulus between the inner and outer red rings.  Maps are presented downbinned for presentation purposes only. Blue rings, where present, indicate the location of point sources identified in a joint analysis of PN and MOS data with a local TS greater than nine.  Masking the identified point sources has little effect on the spectrum. Pixels that do not reside within signal or background extraction regions are displayed in gray.
(\textit{Left}) PN data show a significant excess in the signal region. Due to the large XMM PSF it is not confined to a single full-resolution pixel and is only apparent after downbinning. 
(\textit{Center}) The MOS data show a less clear excess as compared to the PN data.
\textit{(Right}) The Chandra data show a central pixel excess with no other clear point sources visible in the background region appearing with approximately 3$\sigma$ significance. Note that the axis scale for Chandra is much smaller than for the XMM cameras due to Chandra's improved PSF: in fact, the entire Chandra map would fit within the XMM source regions.  }
\label{fig:hard1856_map}
\end{figure*}

 In Fig.~\ref{fig:hard1856_map} we show pixel-by-pixel $\chi^2$ maps, with downbinned pixels, within the vicinity of the NS, which is located at R.A.$_0$ and decl.$_0$. The $\chi^2$ value in pixel $i$ is defined by $S^{-1}_{\chi}\left(S_p(O_i | E_i)\right)$, where $E_i$ is the expected number of counts, $O_i$ is the observed number of counts, $S_p$ is the Poisson-distribution survival function, and $S^{-1}_\chi$ is the inverse survival function for the chi-squared distribution with one degree of freedom.  We have downbinned the maps for visualization purposes.   This figure uses the sum of the counts from 2 to 8 keV.
 The background flux level is estimated from the background region, which is the region between the outer dashed circle and the inner solid circle.  As a reminder, the actual pixel sizes that we use are significantly smaller than indicated for both XMM and Chandra.  The predicted background flux level elsewhere in the map is calculated by assuming that the background flux is simply proportional to the exposure template (without accounting for vignetting), as would be expected if the background is predominantly from particle background.  Accounting for vignetting, as would be the case if the background was dominantly from astrophysical X-rays, leads to virtually indistinguishable results because all source observations were on-axis.  In a given pixel, we may then compute the expected number of background counts.  The higher the $\chi^2$ value, the more likely that the photon flux within that pixel arose from source emission and not a statistical fluctuation of the background\textemdash explicitly, the $p$-value is given by $S_\chi(\chi^2)$, where $S_\chi$ is the survival function for the chi-squared distribution with one degree of freedom.

In the right panel of Fig.~\ref{fig:hard1856_map}, which shows the results for the Chandra observations, it is clearly seen that there is a significant excess of X-ray counts over the background in the central pixel within the extraction region, which is the inner circle.  In this case, the extraction region is approximately $1\farcs1$ in radius, while the outer circle of the background region is approximately $11\farcs5$ in radius.  The Chandra image strongly suggests that there is indeed excess hard X-ray flux arising from this NS between 2 and 8 keV.  On the other hand, the Chandra images are the most subject to pileup. As we will show shortly, however, we do not believe that pileup is responsible for the Chandra results. It is useful, though, to examine the image for PN, which is less subject to pileup and also shows a significant excess, but has much worse angular resolution.  The corresponding image for the PN data is given in the left panel.  Note that, in this case, the source extraction region (inner circle) has a radius of $18\farcs0$ while the background region has an outer radius of $27\farcs0$. In this case, a visual excess is still observable within the signal region, as compared to the background region, which is the region between the two circles, and a less prominent excess can also be seen in the MOS image in the central panel. 

\subsubsection{Validating PN and MOS background extraction regions}
Due to the comparatively worse angular resolution of the PN and MOS instruments, the signal and background extraction regions used in the analysis of PN and MOS data are necessarily larger in angular extent than in the corresponding Chandra analyses. Our treatment of the background count rate, which assumes a uniform particle background resulting in a pixel-by-pixel count rate that depends only on the total exposure in each pixel, may be violated by the presence of point sources.\footnote{Later, we attempt to mitigate this possibility with point source identification and masking, and show that it has a minimal effect on the spectrum.}
 In order to validate our assumptions for the MOS and PN data, we perform a goodness-of-fit test on the pixelated counts data for both instruments. 
 
  \begin{figure*}[htb]
\begin{center}
\includegraphics[width = 0.45\textwidth]{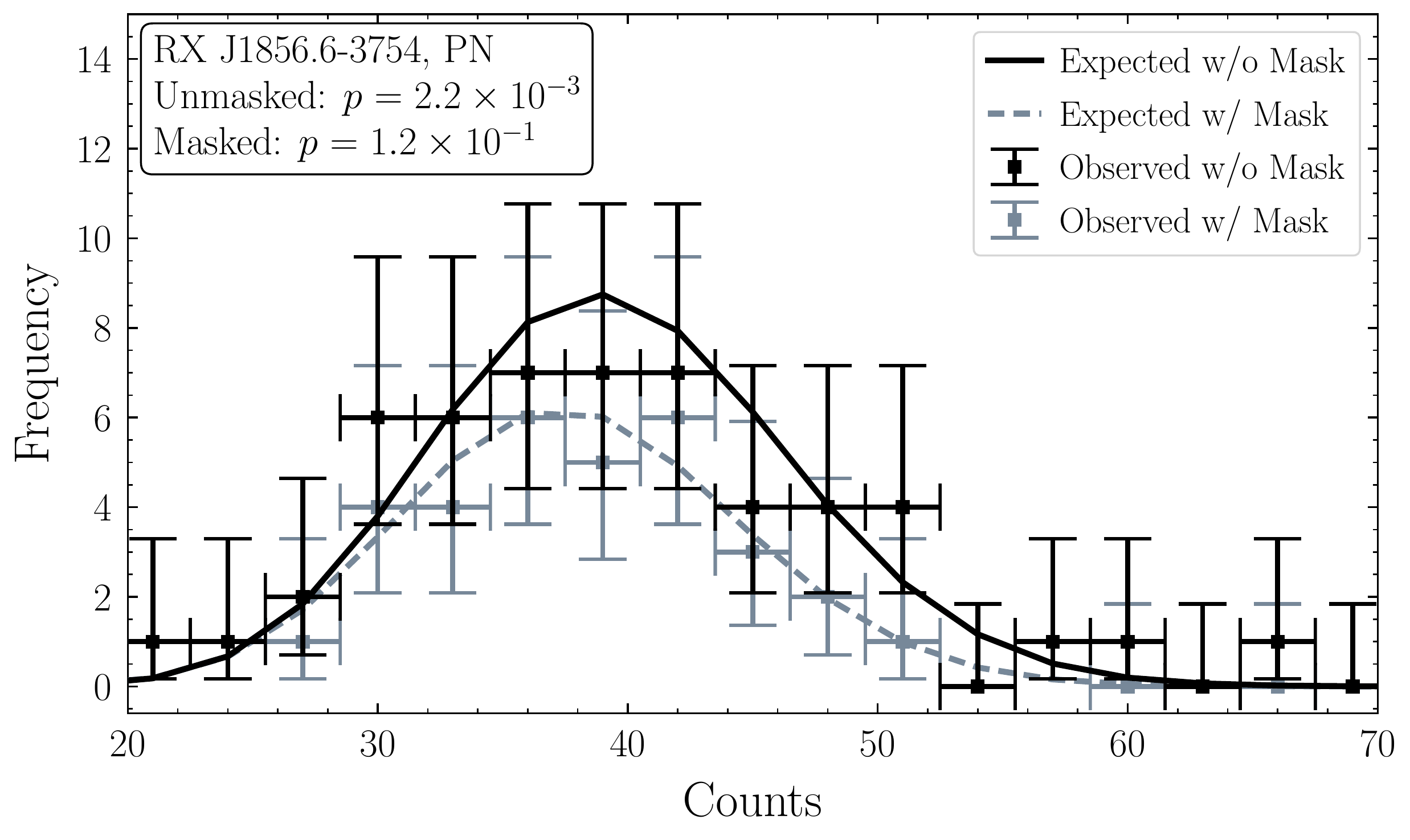} \includegraphics[width = 0.45\textwidth]{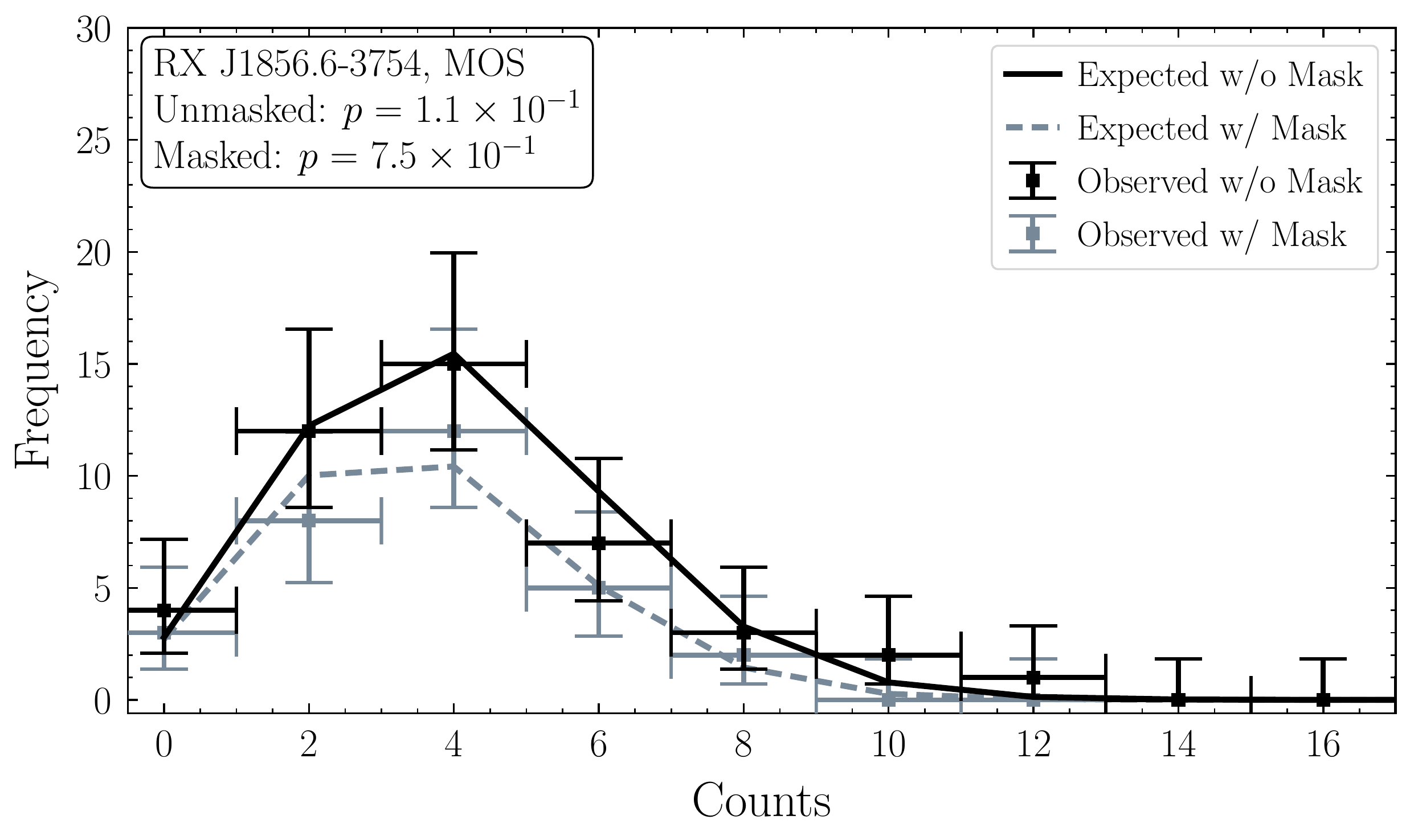}
\end{center}
\caption{Comparison of the observed distributions of counts in the background regions summed over energies from 2 to 8 keV with the expected counts distribution under the best-fit uniform background model for PN (\textit{left}) and MOS (\textit{right}). Fitted distributions and observed counts are shown both with and without the application of a point source mask, and $p$-values are indicated.}
\label{fig:PoissonDistTest}
\end{figure*}

In our goodness-of-fit test, we sum over energies to obtain the set $\{c^p\}$ of total counts over energies 2-8 keV at the $p^\mathrm{th}$ pixel in the background extraction region. Likewise, we obtain the total exposure map summed over energies 2-8 keV at each pixel, denoted $\{w^p\}$. Assuming a uniform Poisson rate for events in the background region, the best-fit expected mean number of counts in the $p^\mathrm{th}$ pixel is $\lambda w^p$, with $\lambda = (\sum_p c^p) / (\sum_p w^p)$.
 We compute a likelihood value for the data, assuming the best-fit parameter $\lambda$ by
\begin{equation}
\mathcal{L}(\lambda | \{c^p\})= \prod_p \frac{(\lambda w^p)^{c^p}e^{\lambda w^p}}{c^p!} \,.
\label{eq:PixelLikelihood}
\end{equation}
We can then determine the $p$-value for the observed data by generating Monte Carlo data under the assumed background rate $\lambda$, then determining the fraction of likelihood values in the Monte Carlo ensemble that are less than the likelihood as evaluated on the observed data. This fraction then represents a $p$-value, where smaller values indicate a worse goodness of fit of the data under the fitted model. We emphasize that this Monte Carlo test is performed on the pixelated data directly, with the joint likelihood over pixels as given in~\eqref{eq:PixelLikelihood}, and not on the higher-level photon-count histogram data shown in, e.g. Fig.~\ref{fig:Systematic}.

 In Fig.~\ref{fig:PoissonDistTest}, we compare the observed and fitted counts distributions for PN and MOS data with and without the application of a point source mask, with associated $p$-values indicated.  Note that, for this test, we use the high-resolution pixels and not the downbinned pixels. Some tension between the data and the fitted background is observed in the PN data without the point source mask\footnote{We formally define a relation between $p$-values and equivalent significance as measured in standard deviations by $Z = \Phi^{-1}(1-p)$ where $\Phi^{-1}$ is the inverse of the standard normal cumulative distribution function. This does not assume statistics following an underlying standard normal distribution; it merely serves the purpose of enabling easy qualitative comparison of significances.}. The tension is at the $\sim$$3.9\sigma$ level, although this falls to the marginal $\sim$$1.9\sigma$ level after the application of the point source mask. The MOS data show good consistency with the fitted background model both before and after the application of the mask. As we will subsequently show, the reconstructed fluxes in PN and MOS are robust with respect to the applied mask. In particular, because the excess appears in the PN data before the application of the mask, the exclusion of high-count pixels from the background extraction region will only serve to increase the reconstructed intensity and associated significance of the fit to the signal model.

\subsubsection{Systematic tests of the XMM X-ray spectrum} 
\begin{figure*}[htb]
\begin{center}
\includegraphics[width = .99\textwidth]{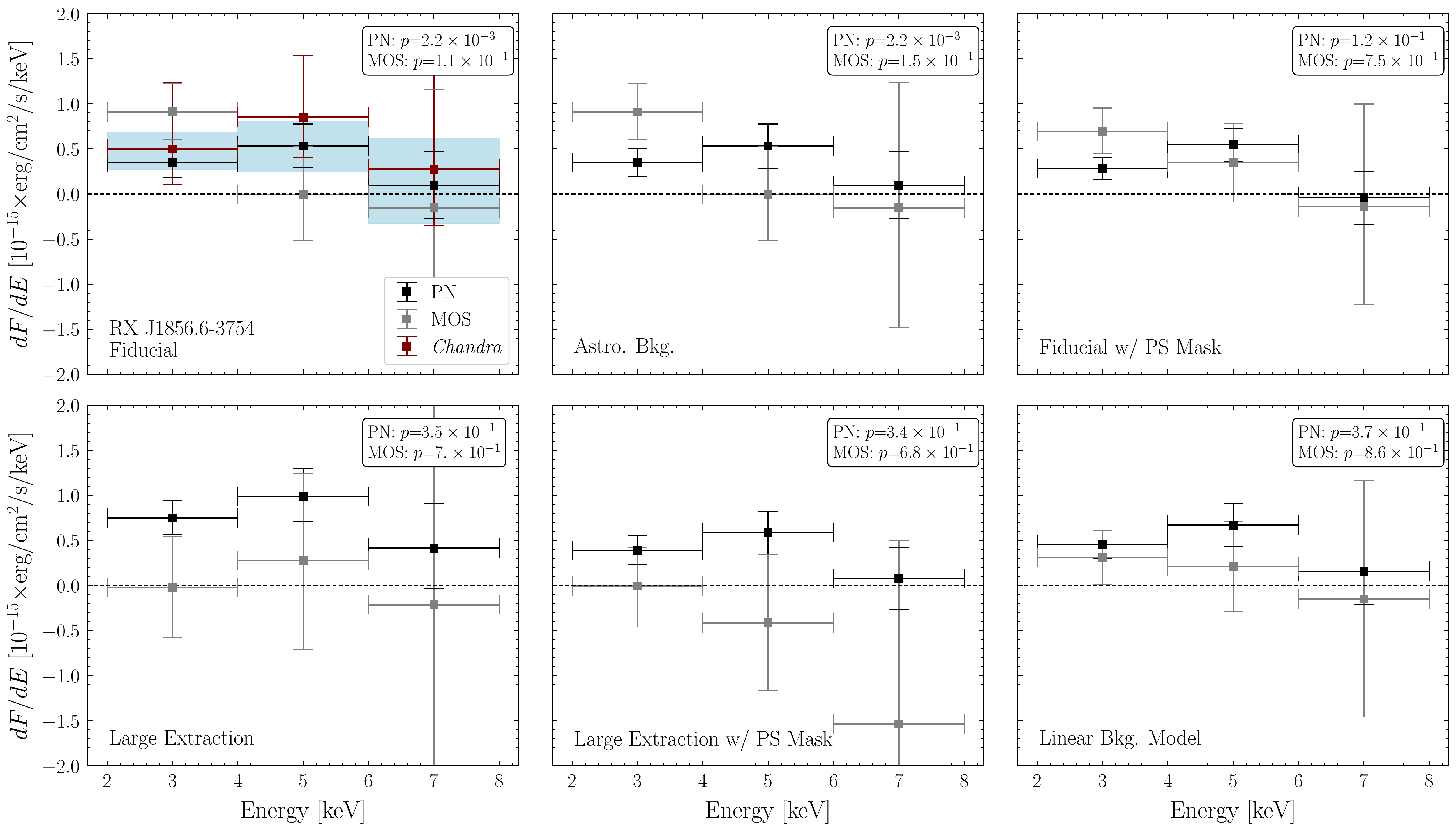}
\end{center}
\caption{(\textit{Upper Left}) Results of the fiducial analysis procedure. Results shown for PN, MOS, and Chandra, with $p$-values corresponding to our goodness-of-fit test of the background model in the background region in the upper right corner. Excess can be seen in all bins for PN and Chandra, while such an excess, if present, is not clearly visible in MOS. (\textit{Upper Center}) Identical signal and background extraction regions as in the fiducial analysis, but fitting the background using the astrophysical exposure, which accounts for vignetting, within signal and background extraction regions. (\textit{Upper Right}) The fiducial analysis with the inclusion of a point source (PS) mask. The $p$-value for the goodness of fit in the PN data has markedly improved while the spectra remain largely unchanged. (\textit{Lower Left}) As in the fiducial analysis, but with the signal extraction region increased to 75\% of the energy-averaged 90\% EEF radius. For PN and MOS, the background extraction radius is doubled but is unchanged for the Chandra extraction. Limits and fits do not change significantly, but the $p$-values for the goodness of fit in PN and MOS show tension, perhaps attributable to nearby point sources. (\textit{Lower Center}) Analysis using the larger extraction regions and the point source mask. The $p$-values increase, demonstrating an improved goodness-of-fit after masking. (\textit{Lower Right}) Analysis using the larger  extraction regions and point source mask, but using a background template linearly varying in R.A. and decl.}
\label{fig:Systematic}
\end{figure*}

We test the robustness of the observed hard X-ray excesses in XMM data for RX J1856.6-3754  by systematically varying our analysis procedure.  The results of the different analyses are shown in Fig.~\ref{fig:Systematic}.
In the top left panel, we show our fiducial recovered spectrum for PN, MOS, and Chandra, along with the joint spectrum from combining all three datasets (68\% confidence intervals indicated).  We also indicate the $p$-values for the background-only fits in the background regions for the PN and MOS datasets.  The other five panels consider various systematic analysis variations, which in principle should all return consistent spectra if large systematic uncertainties are not present.  Indeed, we find that this is the case.  In the middle column of the top row, we change the assumption that the background is dominantly particle background to the assumption that the background is dominantly astrophysical.  The difference between the two is that we include the vignetting correction for the astrophysical background.  This is seen to make a minimal difference, which arises from the fact that the vignetting correction is small over our region of interest.  

The top right panel of Fig.~\ref{fig:Systematic} investigates the spectrum when the point source mask is included.  The spectral points move up slightly, as expected because we are masking high-flux background pixels, but the spectra are broadly consistent with the unmasked versions.  Note that the background $p$-values improve greatly, relative to the unmasked case, as previously noted.  In the bottom left panel, we increase the radius of the background region to 1.5 times the EEF radius.  The background $p$-values decrease, suggesting that our recovered spectra are more susceptible to systematic biases in this case, but again the spectra become slightly larger relative to our fiducial case.
Masking point sources with the large background region, as shown in the middle lower panel, increases the $p$-values but at the same time leads to a similar spectrum as in the unmasked case.  Last, in the bottom right panel, we consider an alternative analytical approach where we allow the background model to vary linearly in the R.A. and decl. directions.  That is, our background model in this case has three nuisance parameters instead of one.  We profile over these nuisance parameters when determining the recovered spectra.  Note that we apply this analysis to the large-background region and with the point source mask.  As expected, given the additional model parameters, the $p$-values improve relative to the case where the background model only has a single nuisance parameter.  In this case, the spectra become even larger relative to our fiducial analysis, though still consistent within uncertainties.

Note that we do not show the results of these tests on Chandra data because, e.g., point sources are less of a concern in this case, though we have still checked that similar systematic analysis variations return consistent results in that case as well.

\subsubsection{Flaring}

\begin{figure*}[htb]
\begin{center}
\includegraphics[width = .90\textwidth]{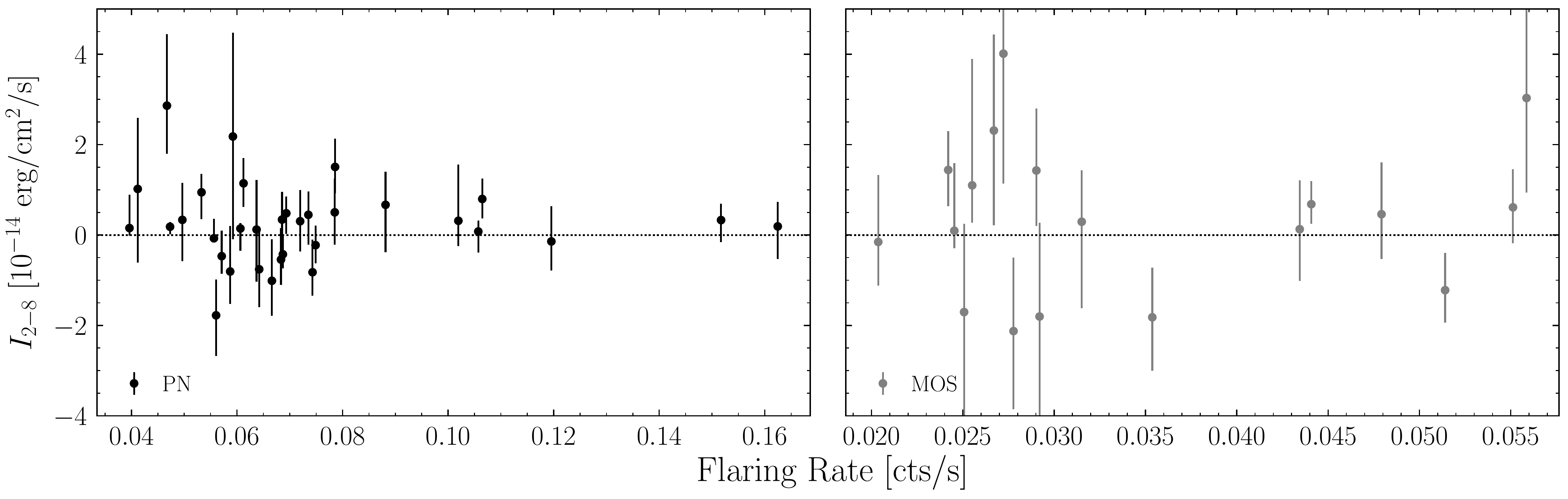}\\
\includegraphics[width = .90\textwidth]{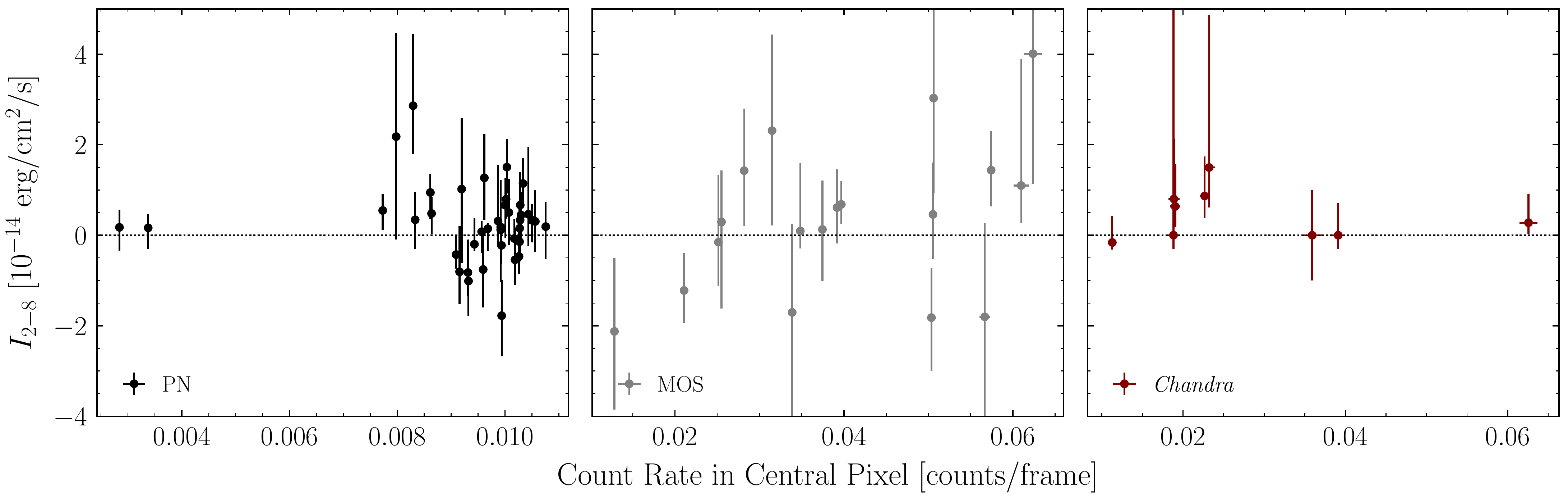}\\
\includegraphics[width = .90\textwidth]{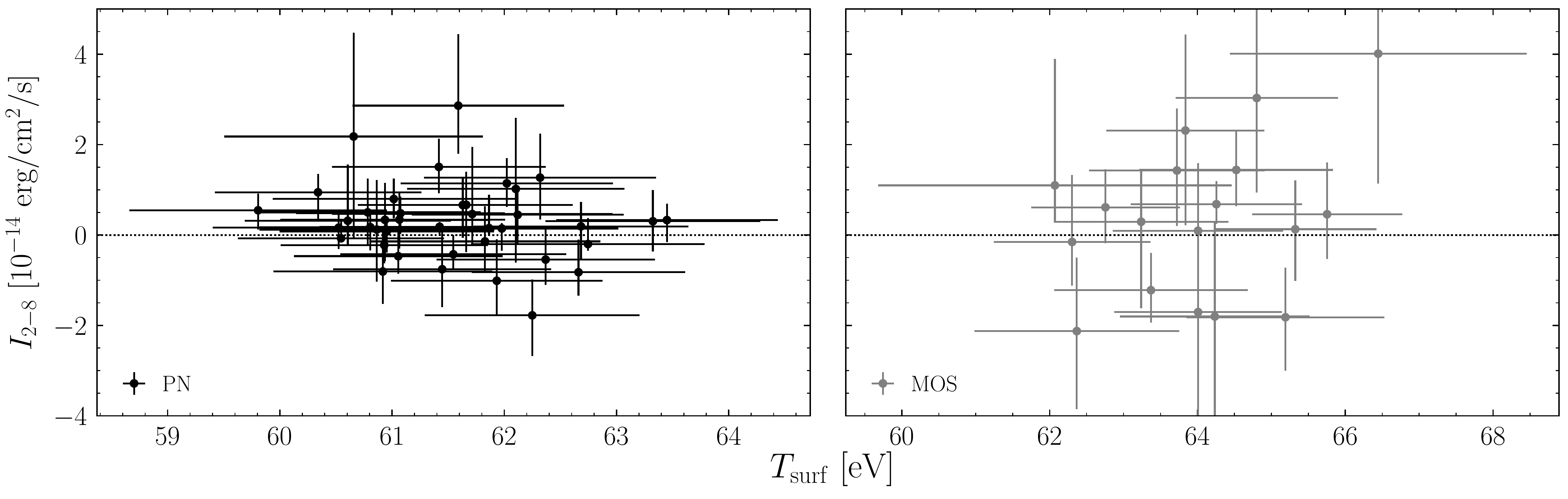}
\end{center}
\caption{(\textit{Top}) Best-fit $I_{2-8}$ in ($10^{-14}$~erg~cm$^{-2}$~s$^{-1}$) for the individual exposures (left\textendash PN; right\textendash MOS) for RX J1856.6-3754 plotted against the flaring rate for that exposure, if flaring was observed. (\textit{Middle}) Best-fit $I_{2-8}$ in ($10^{-14}$~erg~cm$^{-2}$~s$^{-1}$) (left\textendash PN; middle\textendash MOS; right\textendash Chandra) plotted against the count rate in the central pixel for that exposure. (\textit{Bottom}) Best-fit $I_{2-8}$ in ($10^{-14}$~erg~cm$^{-2}$~s$^{-1}$) (left\textendash PN; right\textendash MOS) plotted against the surface temperature determined from the 0.3 to 2 keV data in the same exposure.}
\label{fig:1856_flare_and_count}
\end{figure*}

The XMM satellite is subject to periods of considerable soft proton (SP) flaring when the count rate increases. Although we filter out periods of strong SP flaring, there is still residual flaring in the remaining data that can potentially affect the results. In the left panel of Fig.~\ref{fig:1856_flare_and_count}, we plot $I_{2-8}$ in individual exposures against the count rate in the source region during the times excised from the data due to flaring concerns. There is no trend in the data that would indicate that the observed excess is due to SP flaring. Note that it is well-known that the PN flares can be more intense than the MOS flares.

\subsubsection{Pileup}

Pileup of low-energy X-rays may generate spurious high-energy signatures if not accounted for. Pileup refers to the phenomenon in a CCD detector in which more than one photon arrives in a single frame time in the same region. The detector cannot distinguish the events; it reconstructs them as a single event with energy approximately equal to the sum of the individual photon energies. There are two major effects on a spectrum associated with pileup: event loss and spectral hardening. The former occurs for multiple reasons: first, a multiphoton event is detected as a single photon; second, the event energy may exceed the onboard energy threshold and be rejected; third, the deposited charge-cloud shape (known as grade for Chandra or as pattern for XMM) may become inconsistent with an X-ray photon. The spectral hardening occurs because there is a loss of low-energy events along with an increase in high-energy events. Although the amount of pileup in all of the observations analyzed in this work is relatively low, the observed tail is potentially susceptible to influence from pileup. 
 For this reason, it is necessary to verify that the hard X-ray excess is not due to pileup, and also to verify that, if a hard X-ray excess were present, its observed features would not be biased by pileup effects.
 
The amount of pileup directly depends on the count rate\textemdash the number of photons per CCD readout frame per image region.
  If the hard X-ray tail is due to pileup effects, we expect to see an increase in the count rate of hard source photons with increased total count rates as we vary over exposures. In the right panel of Fig.~\ref{fig:1856_flare_and_count}, we plot $I_{2-8}$ against the count rate in the central pixel from 0 to 2 keV for individual exposures. The wide variance in count rates is due primarily to the fact that the three different cameras have different frame times: PN $\sim$tens of milliseconds, MOS $\sim$1 s, and Chandra 3.2 s, depending on the observation submode. PN does have a higher effective area than the other two cameras, which somewhat increases the count rate. However, the XMM cameras have a much larger PSF-to-pixel-size ratio than Chandra, which further reduces the XMM count rate. For these reasons, PN is expected to be least affected by pileup while Chandra is the most affected. In Fig.~\ref{fig:1856_flare_and_count}, the Chandra count rates are similar to MOS because the Chandra exposures are in a mode that reduces the frame time. For the other NSs, Chandra is in the 3.2 s frame time mode and the count rates are higher than for the XMM exposures. In any case, we observe no significant correlation between the count rate and the reconstructed $I_{2-8}$ in individual exposures, suggesting that pileup does not strongly influence our results for RX J1856.6-3754. 

As mentioned above, we are able to generate a forward modeling matrix including pileup for the PN observations, which are also the ones that should be least affected by pileup. We find, as seen in Fig.~\ref{fig:hard1856}, that pileup is not responsible for the observed excess from 2 to 8 keV.  On the other hand, MOS and Chandra are expected to be more affected by pileup than PN.  Later in this section, we show results for Chandra simulations that include the effect of pileup, and in this case we also find that pileup of the thermal spectrum is not able to generate the observed excess.  Since MOS is expected to be less affected by pileup than Chandra, we believe that the MOS high-energy spectrum is also likely not due to pileup effects. To that end, we have used the results of~\cite{Jethwa}, which provide a method to estimate pileup in the XMM cameras. The results, presented in Tab.~\ref{tab:obs1856}, support the claim that pileup is unlikely to explain the observed hard X-ray flux in RX J1856.6-3754 for both MOS and PN.

\subsubsection{Surface Temperature}

We now investigate whether the observed excess is related to spectral variability of the surface emission of RX J1856.6-3754. Note that previous studies have found little-to-no variability in RX J1856.6-3754~\citep{Sartore_2012}.  In the bottom panel of Fig.~\ref{fig:1856_flare_and_count}, we plot $I_{2-8}$ in individual exposures against the surface temperature of the NS found in that observation. To obtain the surface temperature, we fit the 0.3-2 keV data from individual observations in XSPEC~\citep{1996ASPC..101...17A} with an absorbed thermal model with an additional 1.5\% systematic included to account for instrumental systematics such as detector location. We see no indication of a correlation between the surface temperature and the hard X-ray excess. Note that, when comparing fluxes between MOS and PN, there is an additional $\sim 3$\% systematic uncertainty in the 0.5-1 keV range, coming from cross-calibration uncertainties~\citep{Read:2014vga}. Note also that this systematic is energy-dependent and varies between $\sim 1$ and $5$\% over that range, and thus the systematic uncertainty on the temperatures themselves is likely larger.

 \subsubsection{ Chandra pileup simulation}
 \label{sec:1856-pileup}
 
  To assess the effect of pileup on the high-energy excess observed for RX J1856.6-3754 in Chandra, we perform MARX simulations \citep{2012SPIE.8443E..1AD} for each observation of this source, under two assumptions for the underlying spectrum of the source.  Our MARX simulation procedure is described in Sec.~\ref{sec:marx}.
   In both cases, we use the best-fit thermal spectrum at low energies, but in one case we also include a constant spectrum $dF/dE = 10^{-15}$~erg~cm$^{-2}$~s$^{-1}$~keV$^{-1}$.  In order to separate systematic effects that may be due to pileup from statistical fluctuations, we artificially increase the exposure time to 10 Ms.  We then pass the simulated data through the same analysis pipeline used on the real data.  
 
 It is important to clarify the limitations of the MARX software with regards to simulating pileup effects on a hard X-ray tail. MARX implements the John Davis pileup model ~\citep{Davis2001}, a probabilistic model that uses Poisson statistics to describe the probability that pileup occurs in a given frame and the probability that, in the event of pileup, the piled event will be registered as an X-ray photon (due either to energy or grade migration). However, these probabilities are generally difficult to estimate, due to the fact that many high-grade events are thrown out in-flight, in addition to the lack of a detailed photon-silicon interaction model. The latter probability, in particular, is entirely uncalibrated.

 It is unlikely that the statistical model used here can describe the data at the accuracy level required to definitively conclude that the observed hard tail is not due to pileup. Furthermore, due to these limitations, the MARX software does not assess pileup involving background photons, which could more significantly boost the event energies than the soft thermal photons. Nevertheless, the MARX simulations estimate the basic effects of pileup on the NS spectrum.
 
\begin{figure}[htb]
\begin{center}
\includegraphics[width = 0.5\textwidth]{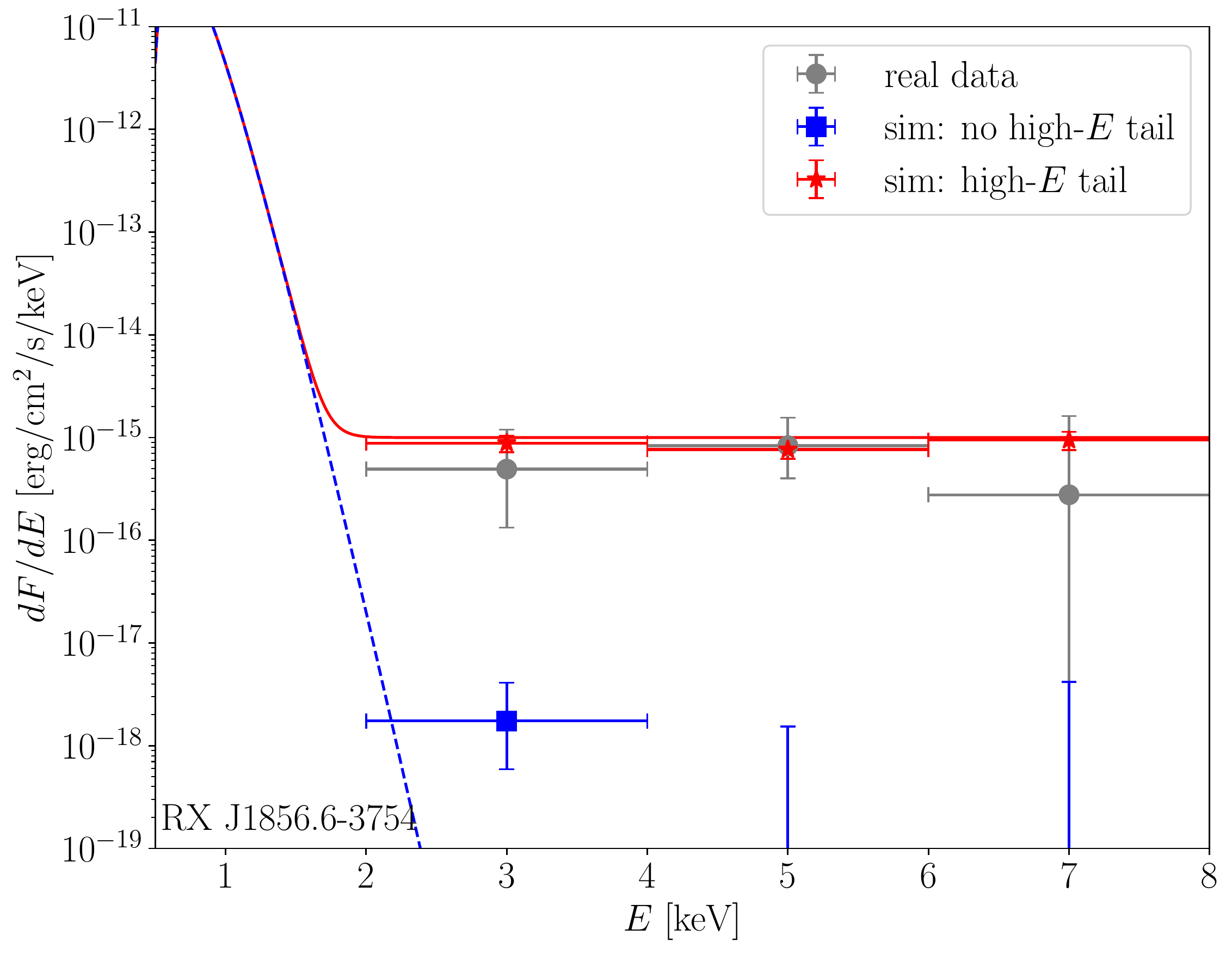}
\end{center}
\caption{MARX simulation results compared to the real Chandra data, shown in gray. Red curve shows the spectrum input to MARX with the additional flux $dF/dE = 10^{-15}$~erg~cm$^{-2}$~s$^{-1}$~keV$^{-1}$, from which we recover the red data points. Blue curve (with recovered blue data points) does not include this additional flux. Pileup of the soft emission does not appear to significantly impact the detection of the hard flux in this case, as we accurately recover it even with reduced statistical errors by inflating the exposure time to 10 Ms. When we input the spectrum with no high-energy tail, we again recover the input spectrum, as shown in blue.
 Pileup is unable to artificially reproduce the observed hard X-ray excess.
}
\label{fig:hard1856_sim}
\end{figure}

 The results of these simulations are shown in Fig.~\ref{fig:hard1856_sim}. 
In that figure, we show the spectrum measured in the real data, from 2 to 8 keV, in gray.  The red data points show the spectrum that we extract from the simulation that includes the high-energy tail.  The simulated spectrum in this case is shown as the solid red curve.  We emphasize that this simulation includes the effects of pileup.  The recovered spectrum is able to accurately describe the true underlying spectrum, which gives confidence that pileup does not affect our ability to measure a high-energy excess for this NS.  As a second cross-check, we also perform a simulation without the high-energy tail.  In this case, the recovered spectrum is shown by the blue data points.  This clearly shows that an artificial high-energy tail is not generated by pileup, at least as modeled by the MARX simulation framework.

In Sec.~\ref{sec:marx-sims} we show results of the same tests on the remaining Chandra XDINSs observations, and we find that for some NSs the effects of pileup are much more pronounced for some NSs.

\subsubsection{Variability}

\begin{figure*}[htb!]
\begin{center}
\includegraphics[width = 0.48\textwidth]{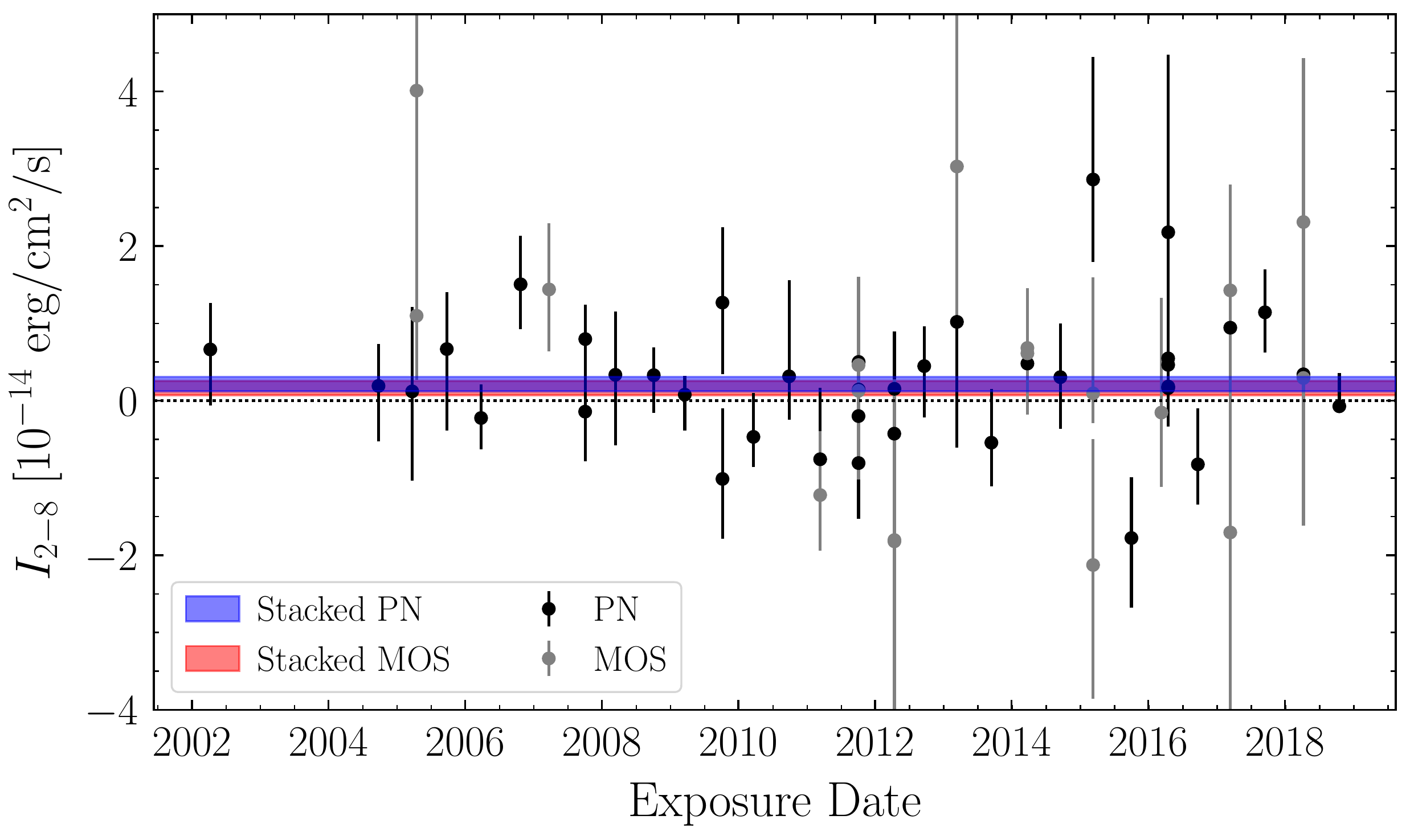} \includegraphics[width = 0.48\textwidth]{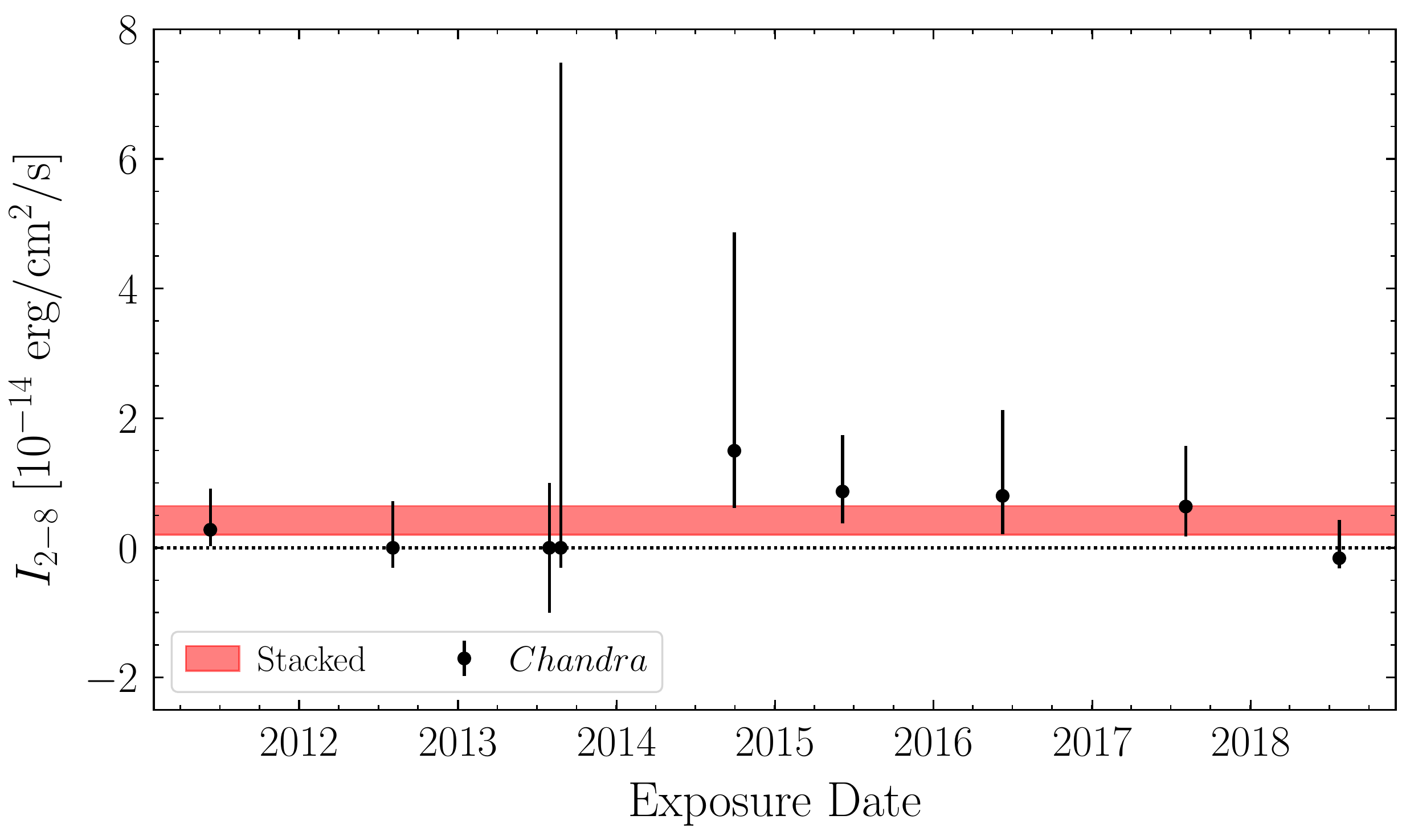}
\end{center}
\caption{Best-fit 2-8 keV intensities for RX J1856.6-3754 in (erg~cm$^{-2}$~s$^{-1}$) in the PN and MOS cameras fit from the individual exposures. Bands cover the 1$\sigma$ confidence intervals for the joint datasets.
(\textit{Left}) The PN results for the 40 individual exposures used in our analysis, and the MOS results for the 18 individual exposures used in our analysis. For PN, there appears to be no evidence for variability on the timescale shown. Between approximately 2008 and 2010, the $\sim$five $I_{2-8}$ values are low by approximately 1$\sigma$, but this may simply be a statistical fluctuation.  It could also be due to the flaring of a source in the background region.
 (\textit{Right}) The Chandra results for the nine individual exposures used in our analysis. Limits are highly one-sided due to the low number of counts. 
}
\label{fig:hard1856_obs}
\end{figure*}

It is possible that the hard X-ray signal is strongly variable in time, which would constrain the possible production mechanisms for the excess. In order to search for signs of strong variability, we analyze the individual exposure images independently, instead of working with the combined image. We stress that this search will be most sensitive to variations on timescales of years; given that both instruments were launched in 1999, our data have been taken over nearly 20 yr. We leave searches for variability on the timescale of the NS period, which is difficult due to the low number of signal counts, to future work. 
  In the left panel of Fig.~\ref{fig:hard1856_obs} we show the $I_{2-8}$ recovered from the individual exposures versus time for PN and MOS, with the analogous result for Chandra shown in the right panel. In the Chandra case, the uncertainties are strongly one-sided because the numbers of signal and background counts tend to be quite low (often as low as zero). 

As before, we determine the $I_{2-8}$ intensities by fitting the 2-8 keV spectra to a power law.  The bands in Fig.~\ref{fig:hard1856_obs} show the best-fit intensities from the analyses on the joint images over all exposures. In the PN and Chandra data, we do not observe any individual exposures with a reconstructed intensity in tension with that found in the joint image analysis. We do observe that one significant intensity deficit appears in an MOS exposure at modest global significance, although this could be due to systematic effects, such as pileup, in that particular MOS exposure. In this exposure, we find a soft 2-8 keV spectrum in the signal region with a typical (among other MOS exposures) spectrum in the background region. This might be expected if pileup heavily affects the observation, where the counts in the signal region are suppressed at high energies by energy or grade migration while the background region is unaffected. In fact, inspection of the \texttt{epatplot} results suggests that pileup affected the 0-2 keV spectrum of the observation, but there were not enough counts above 2 keV to make a definitive determination on whether pileup affected the hard spectrum. Overall, the evidence does not suggest that the hard X-ray excess in RX J1856.6-3754 is highly variable. 

\section{Search for hard X-ray excesses in the XDINS\lowercase{s}}  
\label{Sec:4}

In Sec.~\ref{1856}, we analyzed in detail the hard X-ray excess in RX J1856.6-3754.  We found evidence for such an excess in Chandra and PN data and a hint for the excess also in MOS data.  In this section, we investigate to what extent similar excesses exist in the rest of the XDINSs.  However, it should be noted that RX J1856.6-3754 is special in that it has, by far, the most exposure time across all of the X-ray cameras that we consider.
 \begin{figure}[htb]
\begin{center}
\includegraphics[width = 0.5\textwidth]{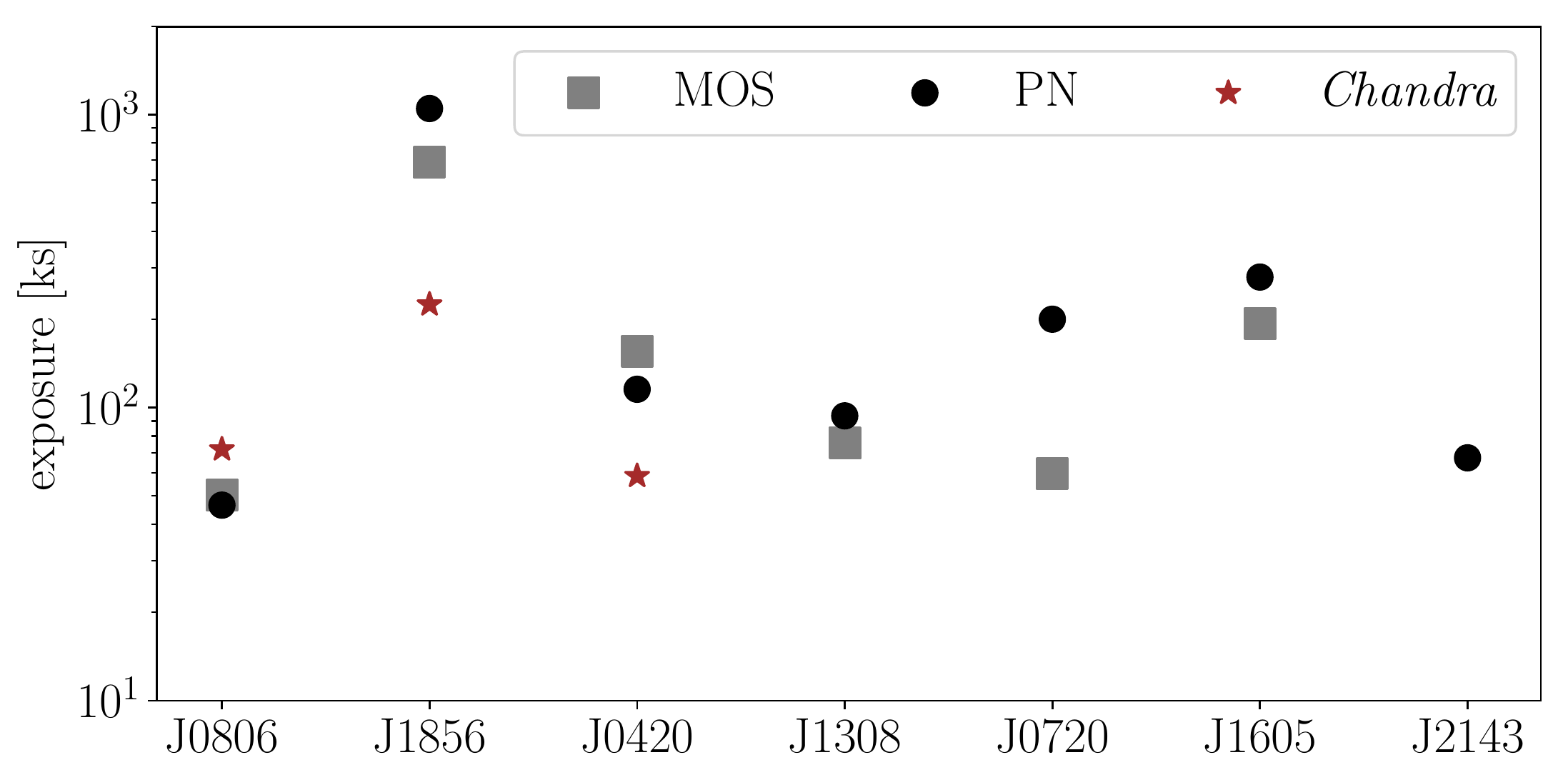} 
\end{center}
\caption{ Summed exposure times in each camera for each NS in our analysis. We have chosen not to analyze Chandra data from NSs RX J1308.6+2127, RX J0720.4-3125, and RX J1605.3+3249 due to pileup concerns.  Note that, for RX J2143.0+0654 no MOS data are available that both pass our SP flaring cut and fully contain the signal and background regions in the images. 
}
\label{fig:exposures}
\end{figure}
The total exposure times that we use for each of the XDINS are shown in Fig.~\ref{fig:exposures}.  Note that Chandra data are available for RX J1308.6+2127, RX J0720.4-3125, and RX J1605.3+3249, but as we show in the next subsection, we believe that these observations are too severely affected by pileup to reliably make a statement about the presence of a hard X-ray excess.  On the other hand, we show that none of the PN observations should be limited by pileup.  For MOS, the situation is less clear, as no pileup simulation framework is readily available, and while these observations should be less subject to pileup than Chandra, they should be more affected by pileup than the PN observations. To estimate the level of pileup in MOS, we use the results of~\cite{Jethwa}, which also apply to PN. The results for each of the XDINSs are shown in Appendix~\ref{app:IDs}. We find that pileup is unlikely to explain the observed hard X-ray excesses in any case.

\subsection{ Chandra pileup simulations}
\label{sec:marx-sims}

In Sec.~\ref{sec:1856-pileup}, we showed that pileup likely does not affect the high-energy tail observed for RX J1856.6-3754 with Chandra data.  In this section, we repeat this exercise for the other XDINSs that have Chandra observations.  To perform these simulations, we first fit the thermal model to the low-energy data (0.5-1 keV).  We then generate simulated datasets using this thermal spectrum, as in Sec.~\ref{sec:1856-pileup}, that do and do not include a possible high-energy tail.  As for RX J1856.6-3754, we model the high-energy tail as $dF/dE = 10^{-15}$~erg~cm$^{-2}$~s$^{-1}$~keV$^{-1}$ over all energies. 

\begin{figure*}[htb]
\begin{center}
\includegraphics[width = 0.48\textwidth]{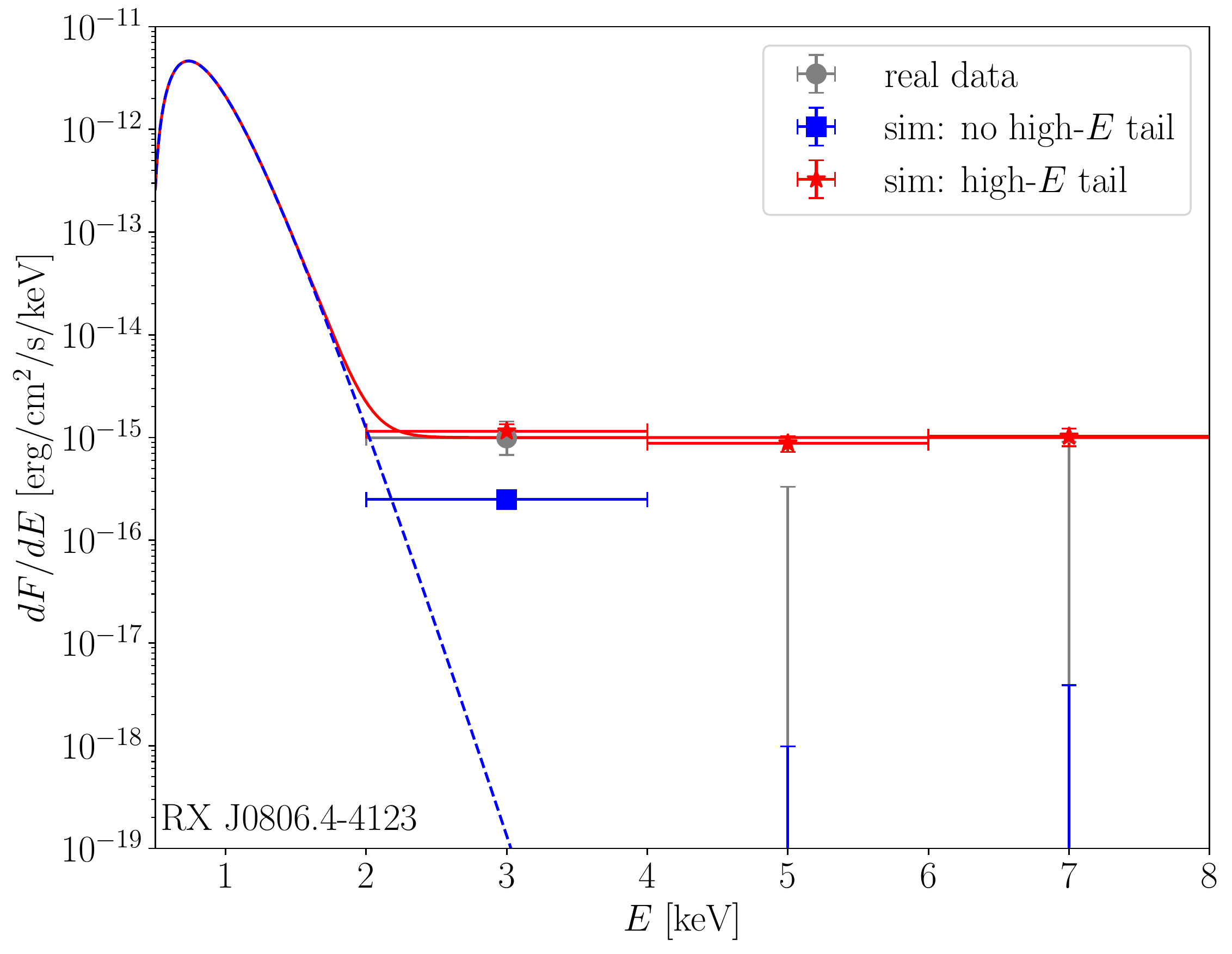} \includegraphics[width = 0.48\textwidth]{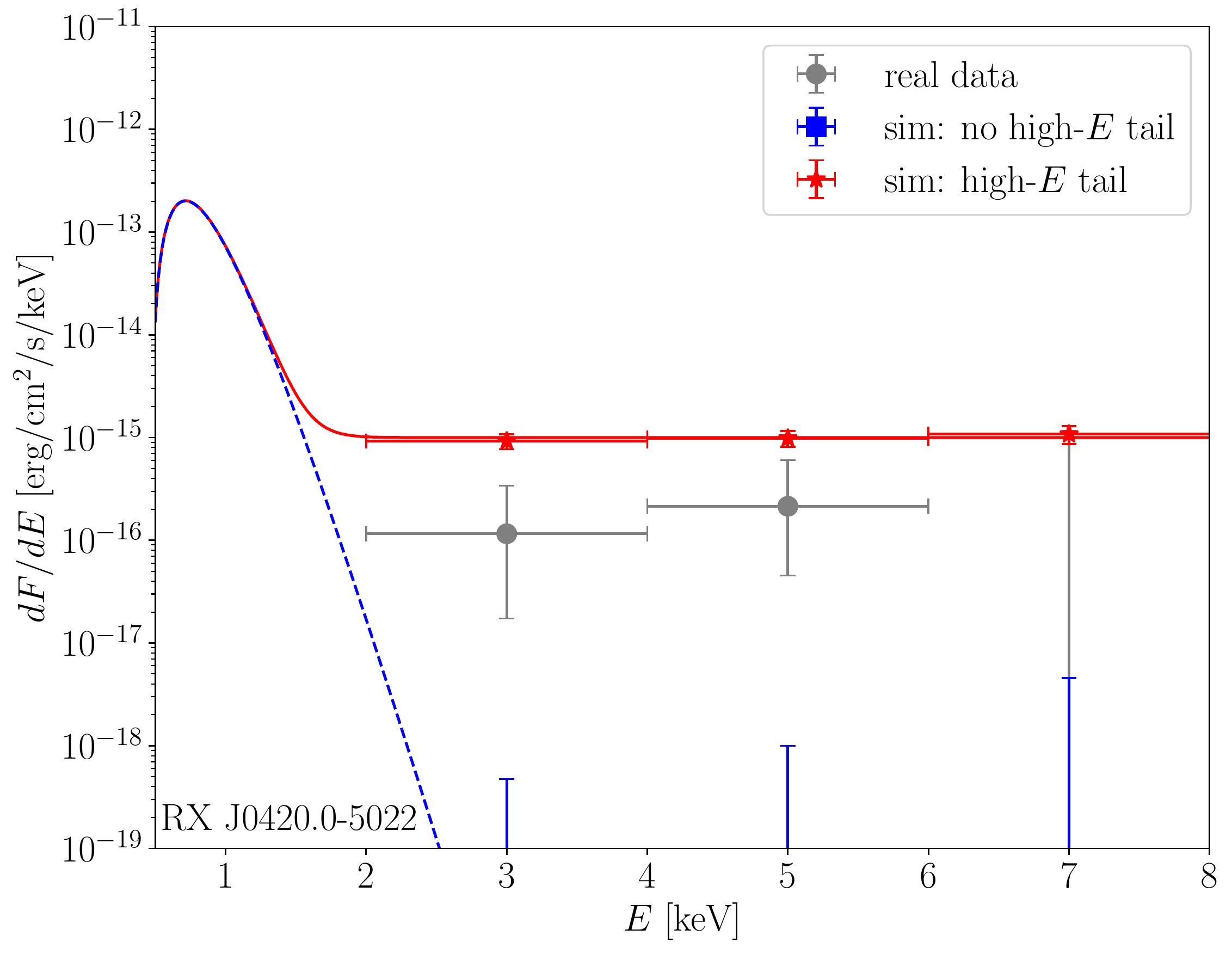}
\end{center}
\caption{As in Fig.~\ref{fig:hard1856_sim}.
({\it Left}) The MARX simulation results for RX J0806.4-4123. We see that the simulation correctly recovers the true flux when the high energy tail is input into the spectrum; however, when there is no high-energy tail, pileup generates slightly more flux in the 2-4 keV bin than expected. This energy bin is excluded from our analysis, though, because of concerns about contamination from thermal emission from the NS surface.
({\it Right}) The MARX simulation results for RX J0420.0-5022. In this case, our analysis of both the simulation results recovers the input flux. We include all three high-energy bins in our analysis of this NS. Pileup is less of a concern for this NS because of the low thermal flux. 
}
\label{fig:pileup-good}
\end{figure*}

In Fig.~\ref{fig:pileup-good} we show the results of the pileup simulations for RX J0806.4-4123 and RX J0420.0-5022.  As in Sec.~\ref{sec:1856-pileup}, we artificially increase the exposure time in the simulations to 10 Ms.
The NS RX J0420.0-5022, which is shown in the right panel, is the NS with the lowest 0.5-1 keV flux of all the XDINSs.  This NS is, correspondingly, the least affected by pileup.  The pileup simulation, clearly shows that, when no high-energy tail is included (blue), then no high-energy flux is recovered, and when the high-energy tail is included (red), the correct flux is recovered.  The same is also true in the left panel for RX J0806.4-4123, though pileup does have a small effect on the flux in the 2-4 keV energy bin.  As we further discuss later, this energy bin is excluded from the analysis for this NS because of concerns about contamination from thermal surface emission.

\begin{figure}[htb]
\begin{center}
\includegraphics[width = 0.48\textwidth]{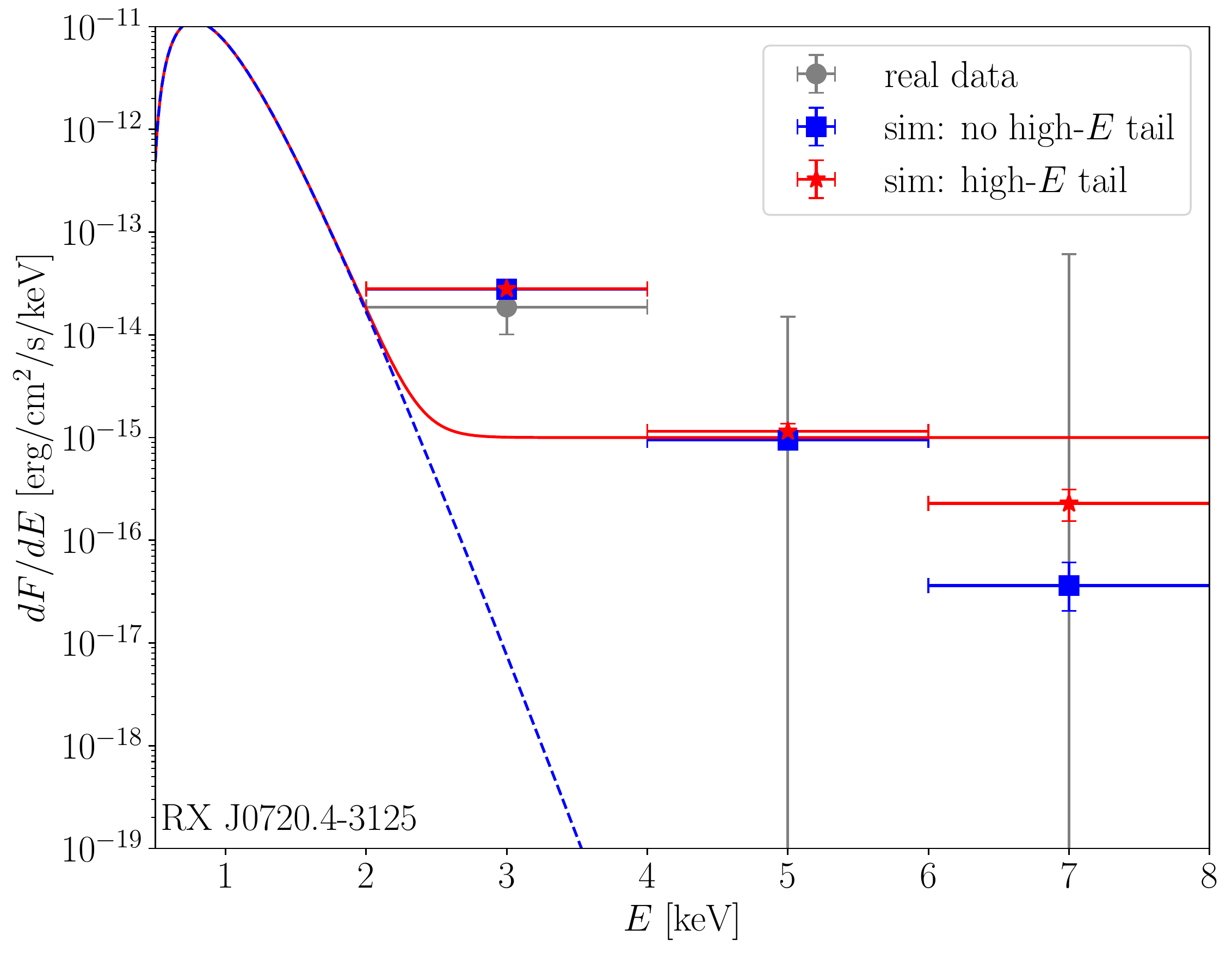} 
\end{center}
\caption{As in Fig.~\ref{fig:hard1856_sim} but for RX J0720.4-3125. In this case, the MARX simulations indicate that pileup can generate a significant excess in the 2-4 keV bin, well above the input spectrum, regardless of the existence of a hard X-ray tail. The same is true in the 4-6 and the 6-8 keV bins, so we completely remove this NS from the Chandra analyses. We find similar results for MARX simulations of RX J1308.7+2127 and RX J1605.3+3249 and exclude these NSs from the Chandra analyses as well.
}
\label{fig:pileup-bad}
\end{figure}

The simulations shown in Fig.~\ref{fig:pileup-good} should be contrasted with those in Fig.~\ref{fig:pileup-bad}, which show simulation results for the NS RX J0720.4-3125.  This NS is significantly affected by pileup.
Pileup generates an artificial, though rather soft, high-energy spectrum in the scenario where the true spectrum has no high-energy tail.  When the high-energy tail is present in the simulation, pileup actually suppresses the flux in the energy bin from 6 to 8 keV.  This likely arises from low-energy photons hitting the CCD in coincidence with true high-energy photons and then those photon pairs being rejected.  For this reason, we are unable to use the RX J0720.4-3125 Chandra data for a high-energy search.  The situation for RX J1308.6+2127 and RX J1605.3+3249 is similar; therefore, out of caution, we do not analyze the Chandra data from any of these NSs.

\subsection{NS surface modeling}
\label{sec:atmo}

In our fiducial analyses, we assume that the 0-2 keV NS spectra are blackbody in order to verify that its extrapolation does not produce the observed 2-8 keV excesses. 
However, at least some of the XDINSs likely have a thin ($\sim$1 cm) atmosphere, leading to a modified spectrum; for a comprehensive review, see~\cite{Zavlin2009} or~\cite{Potekhin:2016gnz}. The surface composition is unknown, although due to the high surface gravity, a hydrogen atmosphere is expected if hydrogen is present on the surface, usually due to accretion at formation. Moreover, the strong surface magnetic field significantly complicates the spectrum. The atomic binding energies increase and cause the absorption lines observed in some of the XDINSs. For the XDINSs' surface temperatures, hydrogen is expected to be partially ionized. Additionally, photons propagate preferentially along the field lines. Finally, the field will induce temperature inhomogeneities across the NS surface by suppressing the thermal conductivity perpendicular to the field. In general, the NS atmosphere can significantly harden the spectrum~\citep{Yakovlev:1999sk}. 

If no accretion occurred after the NS formation, a heavy-element atmosphere or bare surface may exist instead. This may be the case for RX J1856.6-3754 and possibly RX J0720.4-3125 and RX J1308.6+2127~\citep{Zavlin2009}, in which case a condensed iron surface model is appropriate. These models predict a blackbody-like spectrum with most of the deviations at low energies, and thus the hard X-ray spectrum is similar to the blackbody extrapolation. This is also the case for the thin hydrogen atmosphere model in~\cite{Ho:2006uk} that accurately reproduces the RX J1856.6-3754 spectrum.

In this subsection, we investigate the expected contribution of the NS atmosphere spectra to the 2-4 keV bin in our analysis. We use NS magnetic atmosphere models accounting for the effects discussed above, \texttt{NSMAXG}~\citep{Mori:2006jd,Ho:2008bq,Ho:2013pfa}, to fit the 0.5-1 keV spectra jointly to the phase-averaged PN spectra for each NS with the X-ray fitting software XSPEC~\citep{1996ASPC..101...17A}. Note that this procedure accounts for pileup through the PN response matrix. We account for the uncertainty in the surface composition by fitting four models: the hydrogen atmosphere model (\texttt{HB1300Thm90g1420} in XSPEC, hereafter referred to as model {\bf H90}), the carbon atmosphere model \texttt{CB1300ThB00g1438}, the oxygen atmosphere model \texttt{OB1300ThB00g1438}, and the neon atmosphere model \texttt{NeB1300ThB00g1438}. Each model assumes a dipolar magnetic field of $10^{13}$ G, although only model {\bf H90} includes the anisotropic temperature surface distribution. Model {\bf H90} assumes that the angle between the direction to Earth and the magnetic axis is $90^\circ$; to estimate the uncertainty associated with this assumption, we also fit a model \texttt{HB1300Thm00g1420}l where this angle is taken to be $0^\circ$. Finally, we fit hydrogen model \texttt{HB1350ThB00g1438} where the magnetic field strength is taken to be $3 \times 10^{13}$ G, since the XDINSs typically have larger magnetic fields than assumed in the previous models. However, this model does not account for the surface temperature and magnetic field distributions. If any of these models predict a 2-4 keV intensity $I_{2-4}$ greater than $10^{-16}$~erg~cm$^{-2}$~s$^{-1}$ we exclude that bin from further analysis in each camera for that NS. 

In practice, we find that model {\bf H90} consistently suggests the highest 2-4 keV intensity $I_{2-4}$ for each NS, so we report only these flux values. This is consistent with the fact that the mid-Z element atmospheres are known to be softer than their hydrogen counterparts~\citep{Mori:2006jd}. In Tab.~\ref{tab:atm}, we show the results of the predicted maximum fluxes in the 2-4 keV energy bin for each NS.
  We also computed the 4-6 keV intensity, but in no case was it larger than $10^{-19}$~erg~cm$^{-2}$~s$^{-1}$, and so we did not remove any higher-energy bins from the analysis. At high energies, the condensed iron atmospheres are similar to the blackbody spectra; therefore we do not expect that these models would suggest $I_{2-4} \geq 10^{-16}$~erg~cm$^{-2}$~s$^{-1}$.

\begin{table}[]
\centering
\begin{tabular}{|c|c|}
\hline
XDINS    & $I_{2-4}$ ($10^{-16}$~erg~cm$^{-2}$~s$^{-1}$) \\ \hline\hline
RX J1308.6+2127 & $4.0$        \\ \hline
RX J0420.0-5022 & $0.008$                 \\ \hline
RX J0720.4-3125 & $11.8$               \\ \hline 
RX J1605.3+3249 & $17.8$        \\ \hline 
RX J0806.4-4123 & $4.0$           \\ \hline 
RX J2143.0+0654 & $7.4$         \\ \hline 
RX J1856.6-3754 & $0.01$	 \\ \hline
\end{tabular}
\caption{\label{tab:atm} $I_{2-4}$ for each of the NSs from 2-4 keV under model {\bf H90}. If $I_{2-4} \geq 10^{-16}$~erg~cm$^{-2}$~s$^{-1}$, the 2-4 bin is discarded for the remainder of the analysis for all cameras from that NS. As such, we only analyze the 2-4 keV bin for RX J1856.6-3754 and RX J0420.0-5022.
}
\end{table}

\subsection{Characterization of the XDINSs' high-energy excess}

\begin{figure*}[htb]
\begin{center}
\includegraphics[width = 0.959\textwidth]{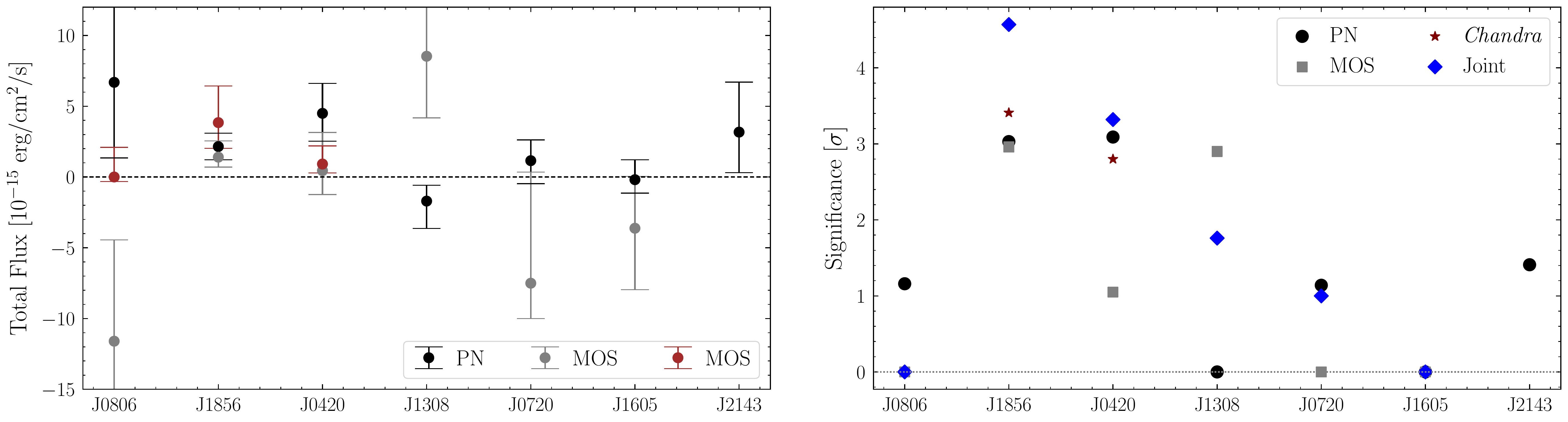} 
\end{center}
\caption{Summary of the results for all of the XDINSs.
(\textit{Left}) The total intensity in (erg~cm$^{-2}$~s$^{-1}$) recovered from the power-law fits to each of the XDINSs for the indicated instruments.  
 For RX J1856.6-3754 and RX J0420.0-5022 we fit the model to data between 2 and 8 keV and so report $I_{2-8}$. For all other NSs, we only use data between 4 and 8 keV and so report $I_{4-8}$.  In all cases, we show the best-fit intensities and the 68\% confidence intervals.
(\textit{Right}) Significances of any intensity excesses, determined through the procedure in Sec.~\ref{sec:stats}.   We also quote the significance of the joint fit across all three instruments for each NS.  RX J1856.6-3754 and RX J0420.0-5022 are the two NSs where we find significant hard X-ray excesses. 
}
\label{fig:summary}
\end{figure*}

\begin{figure*}[htb]
\begin{center}
\includegraphics[width = 1.0\textwidth]{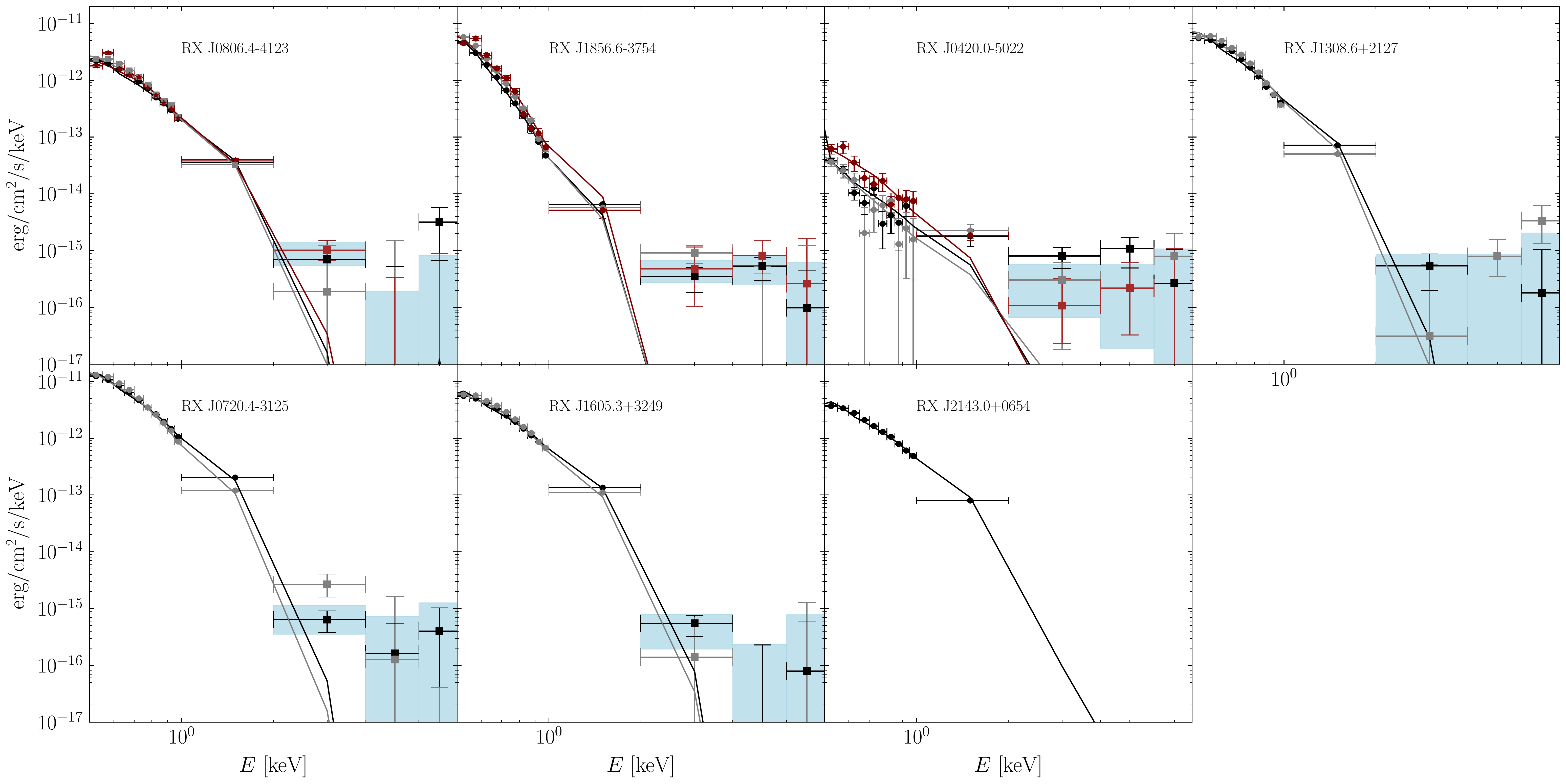}
\end{center}
\caption{As in Fig.~\ref{fig:hard1856}, but for all XDINSs. Black, grey, and red curves show the fits of the blackbody models to the low-energy (0.5-1 keV) data from PN, MOS, and Chandra, respectively, and extrapolated to higher energies.  For the PN data only the extrapolations also include pileup.  We find significant evidence for hard X-ray excesses from RX J1856.6-3754 ($\sim$4.5 $\sigma$) and RX J0420.0-5022 ($\sim$2.5 $\sigma$). Note that the joint spectra, determined from combining the data from all three cameras, are shown when more than one dataset is available.  Our hard X-ray searches use either the 2-8 keV or 4-8 keV energy ranges, depending on the NS.   We include the 2-4 keV energy bin for RX J1856.6-3754 and RX J0420.0-5022 but not for the other NSs, because of concerns about contamination to this bin from NS atmosphere emission (see Sec.~\ref{sec:atmo}).  However, the evidence for hard X-ray flux from RX J1856.6-3754 and RX J0420.0-5022 remains robust even without this energy bin.
}
\label{fig:spectrum}
\end{figure*}

We follow the same analysis procedure used for RX J1856.6-3754 to analyze the PN, MOS, and Chandra data from all of the XDINSs.  A summary of the results of these analyses is shown in Fig.~\ref{fig:summary}. 
In the left panel, we show the best-fit intensities from the fits of the spectra to the power-law model.  For RX J1856.6-3754 and RX J0420.0-5022 we show $I_{2-8}$ because we include the 2-4 keV energy bins for these analyses, while for the other five NSs, we show $I_{4-8}$.   The significances of these detections, determined through Monte Carlo simulations as described in Sec.~\ref{sec:stats}, are given in the right panel.  The spectra, along with the fits to the low-energy thermal models, are shown in Fig.~\ref{fig:spectrum}.  It is worth noting that, in Fig.~\ref{fig:spectrum}, only the PN thermal model has pileup accounted for in the blackbody spectra extrapolations.  

\begin{table}[]
\centering
\begin{tabular}{| c | c | c | c | c |}
\hline
XDINS    & PN & MOS & Chandra & Joint  \\ \hline\hline
RX J0806.4-4123 & 1.16& 0 & 0 & 0 \\ \hline
RX J1856.6-3754 & 3.03 & 2.96 & 3.41 & 4.57 \\ \hline
RX J0420.0-5022 & 3.09 & 1.05 & 2.80 & 3.32 \\ \hline
RX J1308.6+2127 &  0.0 & 2.90 & N/A  & 1.76\\ \hline
RX J0720.4-3125 & 1.14 & 0.0 & N/A & 1.0 \\ \hline
RX J1605.3+3249 & 0.0 & 0.0 & N/A & 0.0 \\ \hline
RX J2143.0+0654 & 1.41 & N/A & N/A & N/A \\ \hline
\end{tabular}
\caption{The joint and individual instrument discovery significances.}
\label{tab:SigTable}
\end{table}

Nontrivial hard X-ray flux is observed from RX J1856.6-3754 at $4.5\sigma$ significance in the joint power-law model fit over all data sets and at $2.5 \sigma$ significance from RX J0420.0-5022. 
Below, we elaborate on the observations for each of the XDINSs, setting aside RX J1856.6-3754 which was discussed in the previous section.  We also note that extended systematic tests and analysis results for each of the XDINSs are provided in the appendices~\ref{app:syst-maps},~\ref{app:syst-spec}, and~\ref{app:8-10}.
 
{\bf  RX J0806.4-4123.} 
There is no evidence for an anomalous hard X-ray excess from this NS in the 4-8 keV energy range analyzed.  As seen in Fig.~\ref{fig:summary} (with corresponding data presented in Tab.~\ref{tab:SigTable}), there is modest ($<$$1\sigma$) evidence for an excess in the PN data but no such evidence in the Chandra and MOS data.  The PN and Chandra data intensities are consistent, though the MOS intensity is recovered to be negative at marginal significance.  This is the result of the negative 6-8 keV energy bin seen in Fig.~\ref{fig:spectrum} for MOS.  Since pileup has a larger impact on the MOS spectrum, the recovered MOS spectrum in this bin may be a result of energy or grade migration.
 There is a somewhat nearby point source, but the point source mask, which we do not apply in our fiducial analysis but do apply in Appendix~\ref{app:syst-spec}, only narrowly intersects the background extraction region\textemdash and so its application does not affect our results. 

{\bf RX J0420.0-5022.} This NS is expected to be the least affected by pileup, considering that it has by far the lowest-intensity thermal flux. 
Varying the surface model shows that the presence of an atmosphere would not account for the observed emission in the 2-4 keV bin, and so this bin is included in the analysis. 
The hard X-ray excess is detected from this NS from all cameras, as seen in Fig.~\ref{fig:summary}. The best-fit spectral index for RX J0420.0-5022 combining all datasets is $n = -0.61^{+1.6}_{-2.1}$, which also suggests a hard spectrum like in RX J1856.6-3754.
 It is also interesting to note that the 1-2 keV datapoint for RX J0420.0-5022 is above the thermal model prediction for all three cameras, though we find that some of the mid-Z atmosphere models, particularly the oxygen atmosphere, can come close to explaining this datapoint.
  No nearby point source was detected for this NS.

{\bf RX J1308.6+2127.} We cut the Chandra data due to concerns about pileup arising from the MARX simulations. We additionally cut the 2-4 keV bin in the XMM data due to concerns about emission from the NS atmosphere. We observe no significant excess in the remaining bins in PN, while the MOS excess is approximately $\sim$2$\sigma$ in significance.  The joint intensity over PN and MOS data is $I_{4-8} = 2.3_{-1.7}^{+1.7} \times 10^{-15}$~erg~cm$^{-2}$~s$^{-1}$. We detect a nearby point source, but not near enough to require any masking of the extraction regions in the masked analysis.

{\bf RX J0720.4-3125.} 
We mask the 2-4 keV bin in our analysis and only consider PN data.
Although the atmosphere models do not explain the entire flux in the 2-4 keV bin, there are other systematics to consider. It is well-established that the surface temperature of RX J0720 changes on the timescale of years from around 85 to 94 eV~\citep{deVries:2004jf,Hohle:2008jv,2012MNRAS.423.1194H}. Because we jointly fit the spectra with the surface models, our procedure does not capture this time-dependence. The hotter observations may contribute the majority of the observed flux in this bin. On the other hand, RX J0720.4-3125 has been previously suggested to have a condensed surface, where the NS atmosphere models do not apply. We find no evidence for a hard X-ray excess. We detect a nearby point source, but not near enough to require any masking of the extraction regions in the masked analysis.

{\bf RX J1605.3+3249.} 
The NS atmosphere models are consistent with the entire 2-4 keV flux as observed by PN and MOS, so we mask this bin in our analysis. We find no significant hard X-ray excess. We detect a nearby point source, but not near enough to require any masking of the extraction regions in the masked analysis.

{\bf RX J2143.0+0654.} 
Since hydrogen atmosphere models suggested a large thermal flux in the 2-4 keV bin, we eliminate this bin from our analysis despite seeing no significant excess. In the remaining two bins, we find $\mathbf{I_{4-8} = 3.2_{-3.4}^{+3.0}} \times 10^{-15}$~erg~cm$^{-2}$~s$^{-1}$ from the PN data. This NS has the least exposure time; accumulating more would help us to understand the nature of the excess, if any exists. 
We detect no nearby point sources.

Altogether, the ensemble of evidence presented strongly suggests that it is likely that at least some of the XDINSs, namely RX J1856.6-3754 and RX J0420.0-5022, produce hard X-ray flux in the energy range from 2 to 8 keV through a mechanism independent of the thermal surface emission.  In the next section, we discuss various possibilities for the source of this flux.   

\section{Possible origins of the XDINS hard X-ray excess}
\label{Sec:5}

In this section, we discuss possible production mechanisms for hard X-ray flux from the XDINSs consistent with the X-ray observations presented above. We consider only the 4-8 keV flux in all NSs for simplicity.  Many pulsars are in fact observed to have two-component X-ray spectra, consisting of low-energy thermal emission from the surface and then a second, harder, nonthermal powerlaw component~\citep{1988ApJ...332..199S}.  The nonthermal emission is commonly accepted to be rotation-powered.  Indeed, a tight correlation is observed between the spin-down luminosity of pulsars and the hard X-ray luminosity (see, e.g.,~\citep{Possenti:2001uu}), although no pulsars in the sample had spin-down luminosities less than $10^{32}$~erg~s$^{-1}$. This relation includes the hard X-ray emission from a possible pulsar wind nebula~\citep{Cheng:2004yh,Posselt:2018ndf}. The emission mechanism may, for example, be synchrotron emission from accelerated charge particles in the outer regions of the magnetosphere~\citep{Abdo:2008ef}.  With that said, radio emission, which is also beamed, typically accompanies nonthermal X-ray emission.  No radio emission has been conclusively observed from the XDINSs~\citep{Kondratiev:2009my}.  Under the hypothesis that the XDINSs are normal pulsars whose radio emission is not observed because we are not in the line of sight of the beam, then it would also be expected that no nonthermal X-ray emission would be observed. This is supported by estimates of the viewing angles of RX J1308.6+2127~\citep{2011AA...534A..74H} and RX J0720.4-3125~\citep{Hambaryan:2017wvm}. Still, it is worth asking the question of whether the energetics of the hard X-ray emission observed from the XDINSs are consistent with a rotation-powered origin.

\begin{figure}[htb]
\begin{center}
\includegraphics[width = 0.49\textwidth]{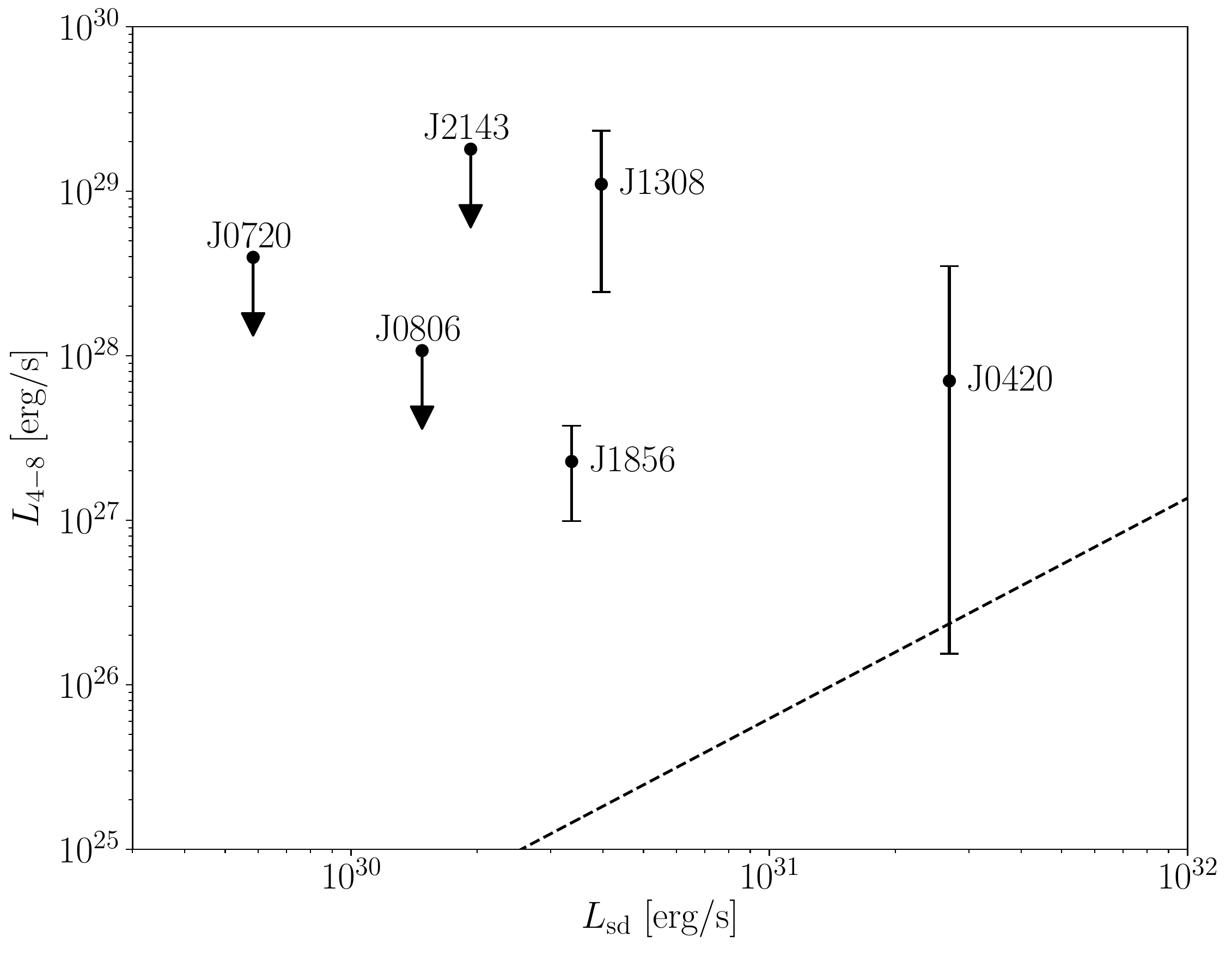}
\end{center}
\caption{Spin-down luminosities $L_{\rm sd} = 4 \pi^2 I \dot P / P^3$, calculated using Tab.~\ref{tab:M7}, plotted against the joint observed 4-8 keV luminosities $L_{4-8}$ in (erg~s$^{-1}$). Dotted line indicates the correlation observed in~\cite{Possenti:2001uu}.  Note that we do not show RX J1605.3+3249 because its period is unknown.
}
\label{fig:Lumi}
\end{figure}

In Fig.~\ref{fig:Lumi} we show the spin-down luminosity $L_{\rm sd}$ of the XDINSs versus their observed luminosities $L_{4-8}$ between 4 and 8 keV from this work. To calculate the luminosities, we use the hard X-ray intensities from joint fits over the available PN, MOS, and Chandra data.  The X-ray luminosities $L_{4-8}$ are then calculated using the observed intensities and the distances in Tab.~\ref{tab:M7}. Where our lower 1$\sigma$ bound is above the lower limit of the plot, we show the 1$\sigma$ confidence interval on the luminosity as determined by Monte Carlo; in the others, we plot the 1$\sigma$ upper limit.
The spin-down luminosities are calculated by $L_{\rm sd} = 4 \pi^2 I \dot P / P^3$, where $I$ is the moment of inertia of the NS, assumed to be $10^{45}$ g cm$^2$, and $P$ ($\dot P$) is the period (period derivative).  A summary of the XDINSs' properties is shown in Tab.~\ref{tab:M7}. Note that since there is no known spin period for RX J1605.3+3249, we do not include it in Fig.~\ref{fig:Lumi}.
\begin{table}[htb]
\centering
\begin{tabular}{|c || c | c |c |}
\hline
 XDINS & $P\ (s)$ & $\dot P$ & $d\ (pc)$ \\ \hline
 J0806 & 11.37 & $5.5 \times 10^{-14}$ & $240 \pm 25$ \\
 J1856 & 7.06 & $3 \times 10^{-14}$ & $123_{-15}^{+11}$ \\ 
 J0420 & 3.45 & $2.8 \times 10^{-14}$ & $345 \pm 200$ \\ 
 J1308 & 10.31 & $1.1 \times 10^{-13}$ & $663 \pm 137$ \\ 
 J0720 & 16.8 & $7 \times 10^{-14}$ & $361_{-88}^{+172}$ \\
 J1605 & \textemdash & \textemdash & $393 \pm 219$ \\ 
 J2143 & 9.43 & $4.1 \times 10^{-14}$ & $430 \pm 200$ \\ 
 \hline
\end{tabular}
\caption{\label{tab:M7} The period, period derivative, and distance to each XDINS, which are used to compute the spin-down and 2-8 keV luminosities. There are no known pulsations in RX J1605.3+3249. Note that the distance measures for RX J0420.0-5022, RX J1308.6+2127, and RX J2143.0+0654 are uncertain from existing observations and we have estimated large errors to be maximally conservative. The data was compiled from~\citep{2009ApJ...705..798K,Posselt:2006ud,vanKerkwijk:2007jp,2010ApJ...724..669W,Kaplan:2011xd,Kaplan:2005zr,2009AA...497..423M,Kaplan:2005fy,Hambaryan:2017wvm,Kaplan:2007qx,refId0,Malacaria:2019zqr,2009ApJ...692L..62K}.
}
\end{table}

Fig.~\ref{fig:Lumi} suggests that the hard X-ray excesses likely do not have nonthermal rotation-powered origins. The best-fit correlation between the spin-down and X-ray luminosity from~\cite{Possenti:2001uu} is shown as the dashed line.  In that work, it was shown that the 2-10 keV luminosities of pulsars (we have converted to 4-8 keV luminosities assuming a typical spectral index from that paper) typically correlate with the spin-down luminosities by that relation, with pulsars scattered typically around an order of magnitude above and below the line in $L_{4-8}$.  At least three of the XDINSs (RX J1856.6-3754, RX J1308.6+2127, and RX J0420.0-5022) show large deviations, at greater than 1$\sigma$, from this best-fit correlation, and we stress that RX J1856.6-3754 and RX J0420.0-5022 are high-significance ($\gtrsim$2.5$\sigma$) detections,  again suggesting that the hard X-ray excesses are not rotation-powered.

\begin{figure}[htb]
\begin{center}
\includegraphics[width = 0.49\textwidth]{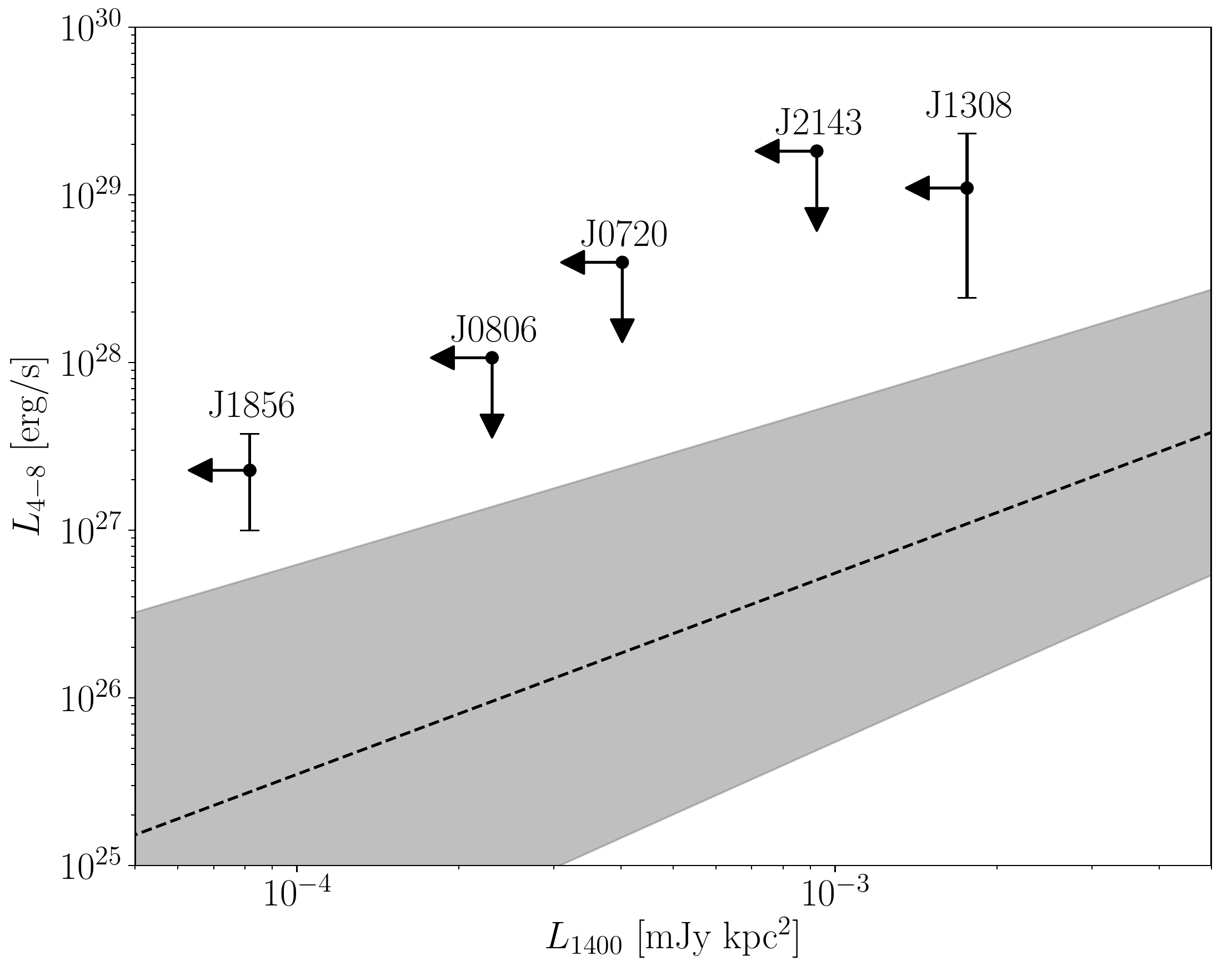}
\end{center}
\caption{Radio luminosity limits for the XDINSs in (mJy kpc$^2$) plotted against the observed joint 2-8 keV luminosities $L_{2-8}$ in (erg~s$^{-1}$) in this work. Dotted line indicates the correlation observed in~\cite{Kondratiev:2009my}. Shaded region indicates the 1$\sigma$ uncertainty on this relation.  Note that we omit RX J1605.3+3249 because its hard X-ray luminosity is negative at over 1$\sigma$, and we omit RX J0420.0-5022 because there are no radio measurements for this NS.
}
\label{fig:radio}
\end{figure}

Because the 4-8 keV emission observed from the XDINSs is very small compared to that from the typical pulsar, we might expect that we see no radio signal because it is also small.~\cite{Malov:2018uez} have observed a correlation between the 1400 MHz luminosity $L_{1400}$ and the 2-10 keV X-ray luminosity of radio pulsars (again, here we convert to 4-8 keV luminosities), albeit with large scatter. In Fig.~\ref{fig:radio} we show the radio limits for all of the XDINSs~\citep{Kondratiev:2009my} except RX J0420.0-5022 (because it has no radio luminosity measurement) and RX J1605.3+3249 (because its hard X-ray luminosity is negative at 1$\sigma$) against their measured 4-8 keV X-ray luminosities $L_{4-8}$. We have rescaled the limits in~\cite{Kondratiev:2009my} to their values assuming the distances in Tab.~\ref{tab:M7}. We see that the radio limits on the XDINSs would imply smaller 4-8 keV X-ray luminosities (for at least RX J1856.6-3754 and RX J1308.6+2127) than those observed.
 This is true in particular for the NS with the highest-significance hard X-ray detection, RX J1856.6-3754. This suggests that the XDINS hard X-ray excess is likely not due to magnetospheric emission with a corresponding radio counterpart.

Excesses of a factor 5-50 above the Rayleigh-Jeans tail of the thermal surface emission have previously been observed in the optical and UV in all of the XDINSs~\citep{Kaplan:2002hb,Kaplan:2003ae,Kaplan:2003hj,Kaplan:2011ay}. One plausible explanation of both the optical/UV and X-ray spectra is that there is an inhomogeneous temperature distribution on the surface such that cold spots explain the optical/UV emission. However, power-law fits to the optical/UV spectra deviate from the expected thermal slope, which suggests the existence of a nonthermal component.~\citep{Kaplan:2011ay} notes that the extrapolation of the optical/UV data to the X-ray band, assuming a pure power law, could potentially produce similar to those observed here. That reference comes to the same conclusion that such luminosities are unlikely to have the magnetospheric origin common in pulsars and that there is no motivated model at present that would produce such a power law nonthermal flux.  
Additionally, such power-law models are in tension with phase-resolved spectra and the absorption features; magnetized NS atmosphere models can potentially account for both the optical/UV excess and the X-ray blackbody (see, {\it e.g.}~\citep{Ho:2006uk}), although this subject is still an area of debate.

\begin{figure}[htb]
\begin{center}
\includegraphics[width = 0.49\textwidth]{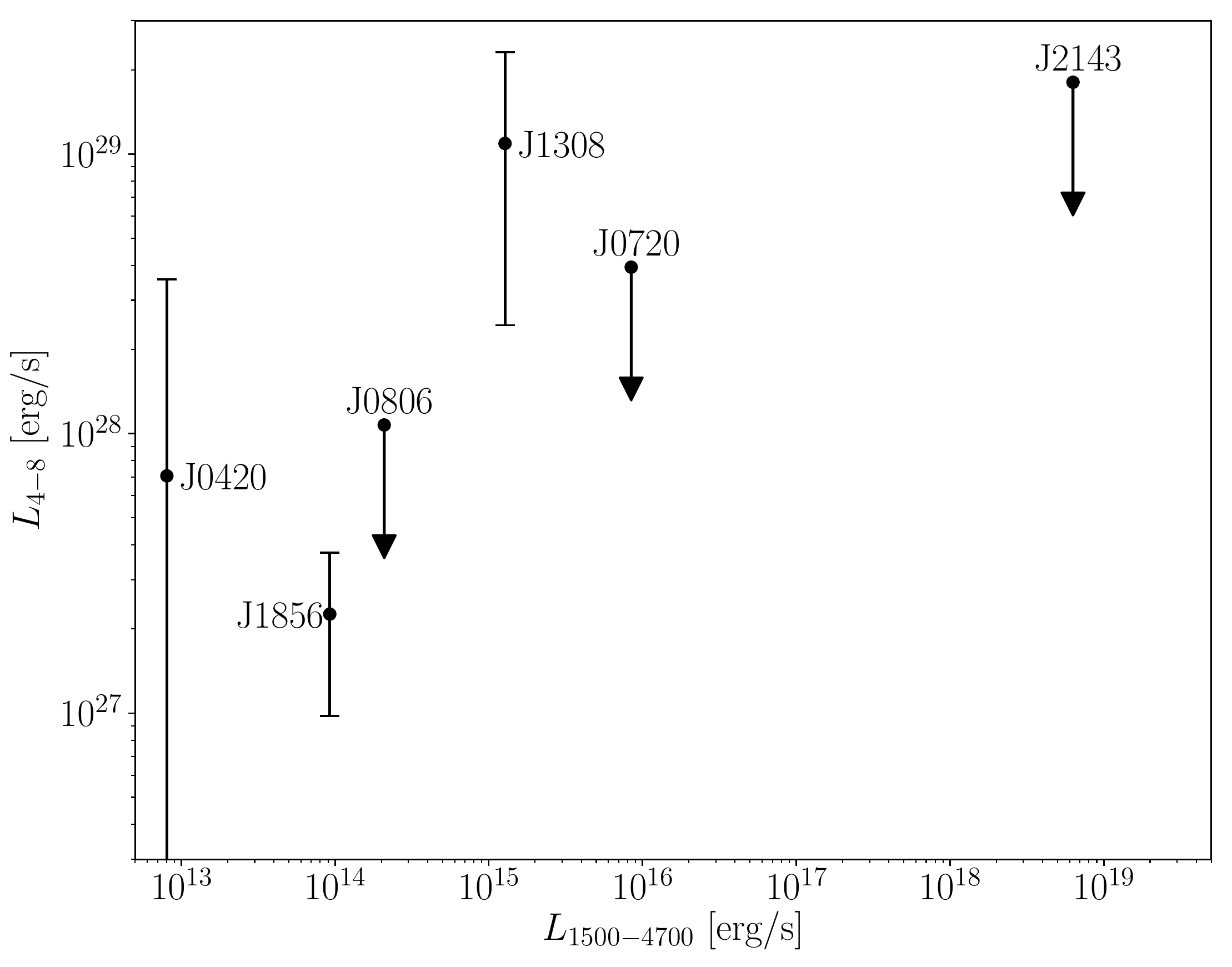}
\end{center}
\caption{Best-fit optical/UV luminosities $L_{1500-4700}$ in~\cite{Kaplan:2011ay} integrated from 1500 - 4700 $\mathring{\mathrm{A}}$ plotted against the $L_{4-8}$ found in this work. There is no observable correlation.  This is perhaps not surprising considering that likely at least some of the optical/UV excess can be explained by NS atmosphere models.  We note that RX J1605.3+3249 has been left out because it has a negative reconstructed hard X-ray luminosity at over 1$\sigma$.
}
\label{fig:optical}
\end{figure}

In Fig.~\ref{fig:optical}, we illustrate the optical luminosities integrated from 1500 - 4700 $\mathring{\mathrm{A}}$ with the best-fit fluxes and spectral indices from~\cite{Kaplan:2011ay}, $L_{1500-4700}$, against the 4-8 keV luminosities $L_{4-8}$.  Given the low sample size and nondetections in five of the seven, it is difficult to determine any possible correlation between the two. However, improved future measurements may reveal a connection, which would point to a unified emission mechanism. Again, we stress that the optical/UV excesses could possibly be explained by an NS atmosphere model.

Another possible source of X-ray flux, apart from thermal surface emission and nonthermal rotation-powered emission, is X-ray emission from accretion of the interstellar medium (see, {\it e.g.},~\citep{Treves:1999ne}).  The typical luminosities expected from accretion of the interstellar medium, assuming that the NS is in the accretion phase, which is itself nontrivial to achieve, are $\lesssim$$10^{31}$~erg~s$^{-1}$~\citep{Treves:1999ne}.  The emission is expected to be nearly thermal at a temperature $\sim$$40$ - $400$ eV, depending on the luminosity, the magnetic field, and the accretion rate.  If the temperature is on the higher side of this interval and the accretion luminosity is near $10^{31}$~erg~s$^{-1}$, then the accretion emission could potentially contribute to the hard X-ray observations from some of the XDINSs.  On the other hand, the low expected temperatures mean that the flux would, at best, be falling exponentially in the 2-8 keV energy range and only significantly contribute in the 2-4 keV energy bin.  These expectations appear inconsistent with the rather hard spectra observed from e.g.~RX J1856.6-3754. Furthermore, the high proper motions of the XDINSs make accretion unlikely to occur~\citep{2009AA...497..423M}.

\section{Discussion}
\label{Sec:6}

In this paper, we use data from XMM and Chandra to provide evidence for hard X-ray emission from some of the XDINSs in the energy band from 2 to 8 keV.  It is possible to extend the spectral analyses to 10 keV for Chandra and XMM-Newton (see Appendix~\ref{app:8-10}), though we have not included the 8-10 keV bin because of concerns about modeling the detector responses and backgrounds at these energies.  
Previously, the only X-ray emission seen from these NSs was at lower energies and consistent with thermal emission from the NS surfaces.  No radio or hard X-ray emission has previously been observed.
 Our results suggest that at least RX J1856.6-3754 and RX J0420.0-5022 produce hard X-rays by some means other than thermal surface emission.  The hard X-ray excess observed from RX J1856.6-3754 is the most significant and is seen with the PN, MOS, and Chandra cameras.  It has a hard spectral index that appears inconsistent with, e.g., being the tail of the thermal surface emission.  The excess appears, as far as we are able to test, robust from pileup effects with Chandra and point sources with PN and MOS, though each of these concerns is real and may have a larger effect than we are able to account for in this work.  

If the XDINS hard X-ray excesses survive further scrutiny, there appears to be no compelling astrophysical explanation for their existence at present.  Rotation-powered nonthermal emission scenarios fail to explain the observed relation, or lack thereof, between the hard X-ray luminosity and the spin-down luminosity.  Moreover, no radio signal has been observed from the XDINSs, which suggests that if the NSs are producing rotation-powered nonthermal emission, this emission is not beamed toward Earth. Furthermore, the hard X-ray signal observed in this work is large enough that if it was rotation-powered non-thermal emission and being beamed towards Earth, a radio signal should have been observed in some of the XDINSs. The XDINSs have previously been discussed in the literature as being candidates to observe emission from accretion of the interstellar medium, but the predicted spectra from this emission is thought to be too soft to contribute substantially in the 2-8 keV energy range, especially with the spectral index observed from, e.g., RX J1856.6-3754. In addition, the XDINSs are thought to have proper motions too large for significant accretion.

One possible exotic origin for the hard X-ray flux is the emission of hypothetical particles called axions within the NS cores and the subsequent conversion of these axions into hard X-rays in the magnetosphere.  The predicted spectrum from this scenario is hard and consistent with the index observed from, e.g., RX J1856.6-3754.  This possibility was the original motivation for the analyses described in this work, and is discussed in more depth in the companion paper~\citep{companion}.  On the other hand, this scenario is by far the most drastic, as it requires the existence of a new fundamental particle of nature. 

Additional data would be useful to help verify or better understand the XDINS hard X-ray excess.  For example, a long exposure by NuSTAR toward, e.g., RX J1856.6-3754 could both confirm the excess below $\sim$10 keV and determine if the excess continues above 10 keV.  Additional Chandra data from, e.g., RX J0806.4-4123, RX J1856.6-3754, and RX J0420.0-5022 would also be useful to gather additional statistics on the hard X-ray spectra in the 2-8 keV energy ranges for these NSs.  

\section{Acknowledgments}

We are grateful to M. Buschmann, R. Co, H. Moritz, M. Reynolds, and D. Yakovlev for useful discussions and comments.  Further, we thank the members of the XMM-Newton Helpdesk, members of the Chandra X-ray Center Helpdesk, and members of the MARX Helpdesk for assistance with their respective software.  This work was supported in part by the DOE Early Career Grant DE-SC0019225 and through computational resources and services provided by Advanced Research Computing at the University of Michigan, Ann Arbor. C. D. was partially supported by the Leinweber Graduate Fellowship at the University of Michigan, Ann Arbor. This work was performed in part at the Aspen Center for Physics, which is supported by the National Science Foundation grant PHY-1607611, and also in part at the Mainz Institute for Theoretical Physics (MITP) of the Cluster of Excellence PRISMA+ (Project ID 39083149).  We also thank the Munich Institute for Astro- and Particle Physics (MIAPP) of the DFG Excellence Cluster Origins along with the CERN Theory department for hospitality during this work. 

\newpage

\bibliographystyle{aasjournal}
\bibliography{refs_NS_hard_excess}

\newpage

\clearpage
\newpage
\mbox{}
\clearpage
\onecolumngrid
\appendix

\section{Observations used in the analyses}
\label{app:IDs}
Here, we list the observation identification numbers, by NS and instrument, which are used in the analyses presented in this work. In addition, we show the instrument used to make the observation, including the grating for Chandra, and the mode the instrument was in at observation time. For MOS, the possible modes are Full Frame (FF) and Large Window (LW), while for PN, we have also included Small Window (SW) data. For Chandra, the mode records the subarray the observation was taken in (FF, 1/2, 1/4, or 1/8). We also show the exposure time of the observation $t_{exp}$ in ks and the date it was taken. For the XMM observations, we additionally show several estimates of pileup. The singles and doubles columns indicate the ratio of observed to expected events in the 0.5-2 keV range with singles and doubles patterns, respectively. Deviations from 1.0 indicate possible pileup. We additionally show estimates of the spectral distortion (SD) and flux loss (FL) in percent due to pileup~\citep{Jethwa}.

The former pileup estimate, SD, is particularly important for our case. For the two NSs in which we find an excess, RX J1856.6-3754 and RX J0420.0-5022, we find that the SD is much lower than would be required for the thermal flux to produce the observed hard X-ray emission. This is not so surprising, as the count rates are around two orders of magnitude lower than the conservative limits on the count rate from~\cite{Jethwa}. In particular, we compute the required SD to produce the hard X-ray emission via pileup for each NS and camera. Here, we assume that every piled-up photon below 2 keV contributes to the flux above 2 keV. Note that this is significantly conservative, as most of these photons will contribute only to the flux below 2 keV. Nevertheless, we find that the SD required to reproduce the RX J1856.6-3754 hard flux is 0.78\% (1.00\%) for MOS (PN), \textbf{while the observed values are $\sim$0.04\% ($\sim$0.07\%)}. To reproduce the RX J0420.0-5022 hard flux, we find 20.8\% (32.7\%) for MOS (PN), \textbf{while the observed values are $\sim$0.01\% ($\sim$0.00\%)}. The RX J0420.0-5022 values are larger because the count rate is lower. Therefore, we conclude that pileup is not significantly contributing to the observed hard X-ray emission in these NSs.

\begin{table}[htb]
\centering
\textbf{RX J0806.4-4123} \vspace{1.5ex} \\
\begin{tabular}{|l|l|l|c|c|c|c|c|c|}
\hline
     Exposure ID &   Instrument & Mode & $t_{exp}$ &      Date &            Singles &            Doubles &    SD &    FL \\
\hline
   0106260201PNS001 &           PN &   FF &       4.1 &  11/08/00 &  $0.974 \pm 0.036$ &  $1.099 \pm 0.061$ &  0.19 &  0.48 \\\hline
   0141750501PNU002 &           PN &   FF &      12.4 &  04/24/03 &  $0.967 \pm 0.020$ &  $1.130 \pm 0.035$ &  0.20 &  0.51 \\\hline
   0552210201PNS003 &           PN &   SW &       5.9 &  05/11/08 &  $0.996 \pm 0.031$ &  $1.014 \pm 0.046$ &  0.03 &  0.04 \\\hline
   0552210401PNS003 &           PN &   SW &       3.7 &  05/29/08 &  $0.998 \pm 0.039$ &  $0.998 \pm 0.058$ &  0.03 &  0.04 \\\hline
   0552210901PNS003 &           PN &   SW &       3.7 &  11/04/08 &  $1.039 \pm 0.040$ &  $0.876 \pm 0.055$ &  0.03 &  0.04 \\\hline
   0552211001PNS003 &           PN &   SW &       6.4 &  12/10/08 &  $0.995 \pm 0.030$ &  $1.017 \pm 0.046$ &  0.03 &  0.04 \\\hline
   0552211101PNS003 &           PN &   SW &       5.4 &  03/31/09 &  $1.011 \pm 0.032$ &  $0.981 \pm 0.048$ &  0.03 &  0.04 \\\hline
   0552211601PNS003 &           PN &   SW &       3.2 &  04/11/09 &  $1.005 \pm 0.043$ &  $0.975 \pm 0.064$ &  0.03 &  0.04 \\\hline
   0672980201PNS001 &           PN &   SW &       5.4 &  05/02/11 &  $1.008 \pm 0.033$ &  $0.983 \pm 0.048$ &  0.03 &  0.04 \\\hline
   0672980301PNS001 &           PN &   SW &       3.8 &  04/20/12 &  $1.040 \pm 0.040$ &  $0.893 \pm 0.055$ &  0.03 &  0.04 \\\hline
 0106260201MOS1S002 &         MOS1 &   FF &       8.3 &  08/11/00 &  $1.053 \pm 0.052$ &  $0.817 \pm 0.081$ &  0.15 &  0.42 \\\hline
 0106260201MOS2S003 &         MOS2 &   FF &       8.7 &  08/11/00 &  $1.053 \pm 0.051$ &  $0.818 \pm 0.079$ &  0.15 &  0.44 \\\hline
 0141750501MOS1U002 &         MOS1 &   FF &      16.6 &  24/04/03 &  $1.057 \pm 0.038$ &  $0.801 \pm 0.058$ &  0.14 &  0.39 \\\hline
 0141750501MOS2U002 &         MOS2 &   FF &      16.9 &  24/04/03 &  $1.057 \pm 0.037$ &  $0.802 \pm 0.057$ &  0.15 &  0.43 \\\hline
               2789 &  ACIS-I/NONE &   FF &      17.7 &  21/02/02 &{\textemdash} &{\textemdash} &   {\textemdash} &   {\textemdash} \\\hline
               5540 &  ACIS-I/NONE &   FF &      19.7 &  18/02/05 &{\textemdash} &{\textemdash} &   {\textemdash} &   {\textemdash} \\\hline
              16953 &  ACIS-I/NONE &   FF &      34.7 &  20/02/15 &{\textemdash} &{\textemdash} &   {\textemdash} &   {\textemdash} \\
\hline
\end{tabular} \\
\caption{\label{tab:obs0806}  The exposures (Exposure ID) used in our fiducial analyses for RX J0806.4-4123 along with supplementary data.  We list the instrument and its operating mode, the length of the observation $t_{exp}$ in ks, the date of the observation, along with four pileup metrics. The singles and doubles columns refer to the observed to expected events with singles and doubles fractions in the 0.5-2 keV range. The spectral distortion (SD) and flux loss (FL) [\%] are additional metrics for pileup described in~\citep{Jethwa}. Note we do not show these metrics for Chandra as in that case we perform dedicated simulations.
}
\end{table}

{\centering
\textbf{RX J1856.6-3754} \vspace{-0.5ex} \\}
\begin{longtable}{|l|l|l|c|c|c|c|c|c|}
\hline
     Exposure ID &   Instrument & Mode & $t_{exp}$ &      Date &            Singles &            Doubles &    SD &    FL \\
\hline
   0791580301PNS001 &           PN &   SW &       4.0 &  04/17/16 &  $0.967 \pm 0.033$ &  $1.116 \pm 0.055$ &  0.04 &  0.05 \\\hline
   0791580501PNS001 &           PN &   SW &       4.3 &  04/17/16 &  $0.983 \pm 0.034$ &  $1.066 \pm 0.055$ &  0.03 &  0.04 \\\hline
   0791580201PNS001 &           PN &   SW &       6.3 &  04/16/16 &  $0.989 \pm 0.028$ &  $1.030 \pm 0.044$ &  0.03 &  0.04 \\\hline
   0791580601PNS001 &           PN &   SW &       9.5 &  04/17/16 &  $0.979 \pm 0.024$ &  $1.086 \pm 0.039$ &  0.04 &  0.05 \\\hline
   0791580101PNS001 &           PN &   SW &      12.5 &  04/16/16 &  $0.998 \pm 0.030$ &  $1.016 \pm 0.047$ &  0.03 &  0.03 \\\hline
   0791580401PNS001 &           PN &   SW &      12.7 &  04/17/16 &  $0.993 \pm 0.023$ &  $1.030 \pm 0.036$ &  0.03 &  0.03 \\\hline
   0165971601PNS003 &           PN &   SW &      21.6 &  09/24/04 &  $0.982 \pm 0.014$ &  $1.067 \pm 0.023$ &  0.04 &  0.05 \\\hline
   0165971901PNS003 &           PN &   SW &      10.9 &  03/23/05 &  $0.977 \pm 0.022$ &  $1.090 \pm 0.036$ &  0.04 &  0.05 \\\hline
   0165972001PNS003 &           PN &   SW &      17.9 &  09/24/05 &  $0.985 \pm 0.017$ &  $1.057 \pm 0.027$ &  0.04 &  0.05 \\\hline
   0106260101PNS001 &           PN &   SW &      38.3 &  04/08/02 &  $0.986 \pm 0.011$ &  $1.054 \pm 0.018$ &  0.04 &  0.05 \\\hline
   0412600601PNU002 &           PN &   SW &      41.1 &  10/05/08 &  $0.985 \pm 0.011$ &  $1.057 \pm 0.017$ &  0.04 &  0.05 \\\hline
   0412600701PNS003 &           PN &   SW &      45.4 &  03/19/09 &  $0.970 \pm 0.011$ &  $1.107 \pm 0.018$ &  0.04 &  0.05 \\\hline
   0412601101PNU002 &           PN &   SW &      43.8 &  09/28/10 &  $0.982 \pm 0.011$ &  $1.065 \pm 0.018$ &  0.03 &  0.04 \\\hline
   0165972101PNS003 &           PN &   SW &      47.0 &  03/26/06 &  $0.981 \pm 0.011$ &  $1.067 \pm 0.017$ &  0.04 &  0.05 \\\hline
   0412600301PNS003 &           PN &   SW &      14.7 &  10/04/07 &  $0.991 \pm 0.018$ &  $1.029 \pm 0.029$ &  0.04 &  0.05 \\\hline
   0412600301PNU002 &           PN &   SW &       9.2 &  10/04/07 &  $0.983 \pm 0.024$ &  $1.066 \pm 0.038$ &  0.04 &  0.05 \\\hline
   0727761301PNS001 &           PN &   SW &      20.4 &  04/10/18 &  $0.982 \pm 0.018$ &  $1.065 \pm 0.028$ &  0.04 &  0.05 \\\hline
   0727761101PNS001 &           PN &   SW &      31.9 &  03/15/17 &  $0.979 \pm 0.015$ &  $1.075 \pm 0.024$ &  0.04 &  0.04 \\\hline
   0727761001PNS001 &           PN &   SW &      45.5 &  09/23/16 &  $0.974 \pm 0.011$ &  $1.093 \pm 0.018$ &  0.04 &  0.05 \\\hline
   0727760101PNS001 &           PN &   SW &      43.5 &  09/14/13 &  $0.972 \pm 0.011$ &  $1.099 \pm 0.018$ &  0.04 &  0.05 \\\hline
   0412600101PNS003 &           PN &   SW &      47.9 &  10/24/06 &  $0.984 \pm 0.010$ &  $1.062 \pm 0.016$ &  0.04 &  0.05 \\\hline
   0412600901PNS003 &           PN &   SW &      30.5 &  03/22/10 &  $0.986 \pm 0.013$ &  $1.057 \pm 0.021$ &  0.04 &  0.05 \\\hline
   0412602201PNS003 &           PN &   SW &      46.5 &  03/14/13 &  $0.960 \pm 0.011$ &  $1.138 \pm 0.018$ &  0.04 &  0.05 \\\hline
   0412600401PNU002 &           PN &   SW &      25.1 &  03/13/08 &  $0.983 \pm 0.014$ &  $1.066 \pm 0.023$ &  0.04 &  0.05 \\\hline
   0810840101PNS001 &           PN &   SW &      35.2 &  10/19/18 &  $0.982 \pm 0.012$ &  $1.064 \pm 0.020$ &  0.04 &  0.04 \\\hline
   0727760201PNS001 &           PN &   SW &      33.5 &  03/26/14 &  $0.985 \pm 0.014$ &  $1.058 \pm 0.021$ &  0.04 &  0.05 \\\hline
   0727760401PNS001 &           PN &   SW &      33.9 &  03/12/15 &  $0.958 \pm 0.013$ &  $1.148 \pm 0.023$ &  0.04 &  0.05 \\\hline
   0727761201PNS001 &           PN &   SW &      26.6 &  09/16/17 &  $0.958 \pm 0.013$ &  $1.145 \pm 0.023$ &  0.04 &  0.05 \\\hline
   0412601401PNS003 &           PN &   SW &      20.5 &  04/13/12 &  $0.970 \pm 0.017$ &  $1.111 \pm 0.027$ &  0.04 &  0.05 \\\hline
   0412601401PNU002 &           PN &   SW &       4.2 &  04/14/12 &  $0.994 \pm 0.035$ &  $1.016 \pm 0.053$ &  0.04 &  0.05 \\\hline
   0727760301PNS001 &           PN &   SW &      48.2 &  09/18/14 &  $0.973 \pm 0.009$ &  $1.098 \pm 0.016$ &  0.04 &  0.05 \\\hline
   0412602301PNS003 &           PN &   SW &      47.0 &  09/20/12 &  $0.965 \pm 0.010$ &  $1.125 \pm 0.017$ &  0.04 &  0.05 \\\hline
   0412600801PNS003 &           PN &   SW &       4.2 &  10/07/09 &  $1.007 \pm 0.036$ &  $0.969 \pm 0.054$ &  0.04 &  0.05 \\\hline
   0412600801PNU002 &           PN &   SW &      34.0 &  10/07/09 &  $0.984 \pm 0.012$ &  $1.058 \pm 0.019$ &  0.04 &  0.05 \\\hline
   0727760501PNS001 &           PN &   SW &      42.6 &  10/03/15 &  $0.972 \pm 0.011$ &  $1.103 \pm 0.018$ &  0.04 &  0.05 \\\hline
   0412601301PNS003 &           PN &   SW &      28.6 &  03/14/11 &  $0.986 \pm 0.014$ &  $1.059 \pm 0.022$ &  0.04 &  0.05 \\\hline
   0412601501PNS600 &           PN &   SW &      16.9 &  10/05/11 &  $0.982 \pm 0.018$ &  $1.070 \pm 0.029$ &  0.04 &  0.05 \\\hline
   0412601501PNS601 &           PN &   SW &      15.9 &  10/05/11 &  $0.977 \pm 0.019$ &  $1.085 \pm 0.031$ &  0.04 &  0.05 \\\hline
   0412601501PNS602 &           PN &   SW &      13.8 &  10/05/11 &  $0.958 \pm 0.019$ &  $1.148 \pm 0.033$ &  0.04 &  0.05 \\\hline
   0412601501PNS603 &           PN &   SW &      14.2 &  10/06/11 &  $0.966 \pm 0.019$ &  $1.113 \pm 0.031$ &  0.04 &  0.05 \\\hline
 0213080101MOS1U002 &         MOS1 &   LW &       2.8 &  15/04/05 &  $0.974 \pm 0.082$ &  $1.199 \pm 0.170$ &  0.06 &  0.17 \\\hline
 0213080101MOS2U002 &         MOS2 &   LW &       1.9 &  15/04/05 &  $0.971 \pm 0.091$ &  $1.230 \pm 0.194$ &  0.07 &  0.20 \\\hline
 0415180101MOS1S002 &         MOS1 &   LW &      19.6 &  25/03/07 &  $0.971 \pm 0.032$ &  $1.220 \pm 0.068$ &  0.06 &  0.18 \\\hline
 0727761301MOS1S002 &         MOS1 &   LW &      29.5 &  10/04/18 &  $0.955 \pm 0.033$ &  $1.282 \pm 0.071$ &  0.06 &  0.17 \\\hline
 0727761301MOS2S003 &         MOS2 &   LW &      25.0 &  10/04/18 &  $0.951 \pm 0.039$ &  $1.282 \pm 0.086$ &  0.05 &  0.14 \\\hline
 0727761101MOS1S002 &         MOS1 &   LW &      48.7 &  15/03/17 &  $0.965 \pm 0.025$ &  $1.210 \pm 0.052$ &  0.06 &  0.16 \\\hline
 0727761101MOS2S003 &         MOS2 &   LW &      54.0 &  15/03/17 &  $0.949 \pm 0.026$ &  $1.320 \pm 0.060$ &  0.05 &  0.14 \\\hline
 0412602201MOS1S001 &         MOS1 &   LW &      63.4 &  14/03/13 &  $0.949 \pm 0.018$ &  $1.309 \pm 0.040$ &  0.07 &  0.19 \\\hline
 0727760201MOS1S002 &         MOS1 &   LW &      44.0 &  26/03/14 &  $0.971 \pm 0.025$ &  $1.193 \pm 0.052$ &  0.06 &  0.16 \\\hline
 0727760201MOS2S003 &         MOS2 &   LW &      43.5 &  26/03/14 &  $0.959 \pm 0.025$ &  $1.256 \pm 0.055$ &  0.05 &  0.15 \\\hline
 0727760401MOS1S002 &         MOS1 &   LW &      47.7 &  12/03/15 &  $0.969 \pm 0.025$ &  $1.196 \pm 0.052$ &  0.06 &  0.16 \\\hline
 0727760401MOS2S003 &         MOS2 &   LW &      51.4 &  12/03/15 &  $0.990 \pm 0.042$ &  $1.100 \pm 0.082$ &  0.05 &  0.15 \\\hline
 0412601401MOS1U002 &         MOS1 &   LW &       5.4 &  14/04/12 &  $0.959 \pm 0.058$ &  $1.279 \pm 0.127$ &  0.06 &  0.17 \\\hline
 0412601401MOS2U002 &         MOS2 &   LW &       5.8 &  14/04/12 &  $0.972 \pm 0.061$ &  $1.197 \pm 0.128$ &  0.05 &  0.15 \\\hline
 0727760601MOS1S002 &         MOS1 &   LW &      32.6 &  11/03/16 &  $0.991 \pm 0.038$ &  $1.083 \pm 0.074$ &  0.06 &  0.16 \\\hline
 0412601301MOS1S001 &         MOS1 &   LW &      39.1 &  14/03/11 &  $0.952 \pm 0.037$ &  $1.322 \pm 0.083$ &  0.06 &  0.16 \\\hline
 0412601501MOS1S001 &         MOS1 &   LW &      83.8 &  05/10/11 &  $0.963 \pm 0.016$ &  $1.245 \pm 0.034$ &  0.06 &  0.18 \\\hline
 0412601501MOS2S002 &         MOS2 &   LW &      90.2 &  05/10/11 &  $0.958 \pm 0.017$ &  $1.266 \pm 0.038$ &  0.06 &  0.17 \\\hline
              13198 &  ACIS-S/LETG &  1/8 &      28.4 &  10/06/11 &{\textemdash} &{\textemdash} &   {\textemdash} &   {\textemdash}  \\\hline
              18416 &  ACIS-S/LETG &  1/4 &      28.8 &  08/06/16 &{\textemdash} &{\textemdash} &   {\textemdash} &   {\textemdash}  \\\hline
              19848 &  ACIS-S/LETG &  1/8 &      28.4 &  03/08/17 &{\textemdash} &{\textemdash} &   {\textemdash} &   {\textemdash}  \\\hline
              20718 &  ACIS-S/LETG &  1/8 &      28.2 &  23/07/18 &{\textemdash} &{\textemdash} &   {\textemdash} &   {\textemdash}  \\\hline
              14267 &  ACIS-S/LETG &  1/8 &      28.4 &  03/08/12 &{\textemdash} &{\textemdash} &   {\textemdash} &   {\textemdash}  \\\hline
              15474 &  ACIS-S/LETG &  1/8 &      11.7 &  30/07/13 &{\textemdash} &{\textemdash} &   {\textemdash} &   {\textemdash}  \\\hline
              16265 &  ACIS-S/LETG &  1/8 &      16.1 &  25/08/13 &{\textemdash} &{\textemdash} &   {\textemdash} &   {\textemdash}  \\\hline
              16422 &  ACIS-S/LETG &  1/8 &      26.0 &  29/09/14 &{\textemdash} &{\textemdash} &   {\textemdash} &   {\textemdash}  \\\hline
              17394 &  ACIS-S/LETG &  1/8 &      29.1 &  05/06/15 &{\textemdash} &{\textemdash} &   {\textemdash} &   {\textemdash}  \\
\hline
\caption{\label{tab:obs1856} As in Tab.~\ref{tab:obs0806} but for RX J1856.6-3754.
}\\
\end{longtable}

{\centering
\textbf{RX J0420.0-5022} \vspace{-0.5ex} \\}
\begin{longtable}{|l|l|l|c|c|c|c|c|c|}
\hline
     Exposure ID &   Instrument & Mode & $t_{exp}$ &      Date &            Singles &            Doubles &    SD &    FL \\
\hline
   0141750101PNS003 &           PN &   FF &      17.3 &  12/30/02 &  $1.021 \pm 0.138$ &  $0.888 \pm 0.198$ &  0.01 &  0.01 \\\hline
   0141751001PNS003 &           PN &   FF &       9.8 &  12/31/02 &  $0.984 \pm 0.135$ &  $1.060 \pm 0.216$ &  0.01 &  0.01 \\\hline
   0141751101PNS003 &           PN &   FF &      14.2 &  01/19/03 &  $0.955 \pm 0.121$ &  $1.075 \pm 0.197$ &  0.01 &  0.01 \\\hline
   0141751201PNS003 &           PN &   FF &      17.6 &  07/25/03 &  $0.968 \pm 0.129$ &  $1.116 \pm 0.223$ &  0.01 &  0.01 \\\hline
   0651470201PNS003 &           PN &   SW &       2.9 &  03/30/10 &  $1.184 \pm 0.290$ &  $0.429 \pm 0.226$ &  0.00 &  0.00 \\\hline
   0651470501PNS003 &           PN &   SW &       2.4 &  05/21/10 &  $0.967 \pm 0.307$ &  $1.146 \pm 0.492$ &  0.00 &  0.00 \\\hline
   0651470601PNS003 &           PN &   SW &       4.4 &  07/29/10 &  $0.956 \pm 0.197$ &  $1.112 \pm 0.307$ &  0.00 &  0.00 \\\hline
   0651470701PNS003 &           PN &   SW &       6.8 &  09/21/10 &  $0.999 \pm 0.157$ &  $0.894 \pm 0.211$ &  0.00 &  0.00 \\\hline
   0651470801PNS003 &           PN &   SW &       8.1 &  10/02/10 &  $1.010 \pm 0.143$ &  $0.893 \pm 0.191$ &  0.00 &  0.00 \\\hline
   0651470901PNS003 &           PN &   SW &       9.0 &  10/03/10 &  $0.944 \pm 0.137$ &  $1.147 \pm 0.221$ &  0.00 &  0.00 \\\hline
   0651471001PNS003 &           PN &   SW &       5.3 &  10/04/10 &  $1.033 \pm 0.179$ &  $0.775 \pm 0.213$ &  0.00 &  0.00 \\\hline
   0651471101PNS003 &           PN &   SW &       5.6 &  10/06/10 &  $0.949 \pm 0.131$ &  $1.108 \pm 0.204$ &  0.00 &  0.00 \\\hline
   0651471201PNS003 &           PN &   SW &       3.7 &  11/26/10 &  $1.022 \pm 0.204$ &  $0.924 \pm 0.275$ &  0.00 &  0.00 \\\hline
   0651471401PNU002 &           PN &   SW &       4.8 &  03/31/11 &  $1.018 \pm 0.192$ &  $0.942 \pm 0.263$ &  0.00 &  0.00 \\\hline
   0651471501PNS003 &           PN &   SW &       3.6 &  04/11/11 &  $0.922 \pm 0.211$ &  $1.108 \pm 0.347$ &  0.00 &  0.00 \\\hline
 0141750101MOS1S001 &         MOS1 &   FF &      20.8 &  30/12/02 &  $0.998 \pm 0.223$ &  $1.076 \pm 0.438$ &  0.00 &  0.01 \\\hline
 0141750101MOS2S002 &         MOS2 &   FF &      20.9 &  30/12/02 &  $1.016 \pm 0.283$ &  $0.970 \pm 0.518$ &  0.00 &  0.01 \\\hline
 0141751001MOS1S001 &         MOS1 &   FF &      15.9 &  31/12/02 &  $0.977 \pm 0.185$ &  $1.083 \pm 0.369$ &  0.00 &  0.01 \\\hline
 0141751001MOS2S002 &         MOS2 &   FF &      16.3 &  31/12/02 &  $1.006 \pm 0.195$ &  $1.033 \pm 0.370$ &  0.00 &  0.01 \\\hline
 0141751101MOS1S001 &         MOS1 &   FF &      20.5 &  19/01/03 &  $0.948 \pm 0.155$ &  $1.279 \pm 0.349$ &  0.00 &  0.01 \\\hline
 0141751101MOS2S002 &         MOS2 &   FF &      19.7 &  19/01/03 &  $1.038 \pm 0.179$ &  $0.845 \pm 0.299$ &  0.00 &  0.01 \\\hline
 0141751201MOS1S001 &         MOS1 &   FF &      20.5 &  25/07/03 &  $1.070 \pm 0.246$ &  $0.671 \pm 0.352$ &  0.00 &  0.01 \\\hline
 0141751201MOS2S002 &         MOS2 &   FF &      21.2 &  25/07/03 &  $1.042 \pm 0.215$ &  $0.813 \pm 0.351$ &  0.00 &  0.01 \\\hline
               2788 &  ACIS-S/NONE &   FF &      19.4 &  11/11/02 &{\textemdash} &{\textemdash} &   {\textemdash} &   {\textemdash}  \\\hline
               5541 &  ACIS-S/NONE &   FF &      19.7 &  07/11/05 &{\textemdash} &{\textemdash} &   {\textemdash} &   {\textemdash}  \\\hline
              17457 &  ACIS-S/NONE &   FF &      19.4 &  11/11/15 &{\textemdash} &{\textemdash} &   {\textemdash} &   {\textemdash}  \\
\hline
\caption{\label{tab:obs0420} As in Tab.~\ref{tab:obs0806} but for RX J0420.0-5022.
}\\
\end{longtable}

\begin{table}[htb]
\centering
\textbf{RX J1308.6+2127} \vspace{1.5ex} \\
\begin{tabular}{|l|l|l|c|c|c|c|c|c|}
\hline
     Exposure ID &   Instrument & Mode & $t_{exp}$ &      Date &            Singles &            Doubles &    SD &    FL \\
\hline
   0157360101PNS005 &         PN &   FF &      23.2 &  01/01/03 &  $0.964 \pm 0.009$ &  $1.136 \pm 0.017$ &  0.45 &  1.19 \\\hline
   0163560101PNS003 &         PN &   FF &      16.5 &  12/30/03 &  $0.976 \pm 0.011$ &  $1.084 \pm 0.019$ &  0.46 &  1.22 \\\hline
   0305900201PNS003 &         PN &   FF &       2.1 &  06/25/05 &  $0.962 \pm 0.031$ &  $1.134 \pm 0.055$ &  0.46 &  1.21 \\\hline
   0305900301PNS003 &         PN &   FF &      10.9 &  06/27/05 &  $0.977 \pm 0.014$ &  $1.086 \pm 0.024$ &  0.45 &  1.21 \\\hline
   0305900401PNS003 &         PN &   FF &       9.2 &  07/15/05 &  $0.972 \pm 0.015$ &  $1.108 \pm 0.026$ &  0.45 &  1.18 \\\hline
   0305900601PNS003 &         PN &   FF &      12.8 &  01/10/06 &  $0.961 \pm 0.012$ &  $1.147 \pm 0.022$ &  0.46 &  1.23 \\\hline
   0402850301PNS003 &         PN &   LW &       3.4 &  06/08/06 &  $0.972 \pm 0.025$ &  $1.104 \pm 0.042$ &  0.31 &  0.78 \\\hline
   0402850401PNS003 &         PN &   LW &       5.6 &  06/16/06 &  $0.957 \pm 0.019$ &  $1.158 \pm 0.035$ &  0.30 &  0.76 \\\hline
   0402850501PNS003 &         PN &   LW &       2.6 &  06/27/06 &  $0.972 \pm 0.028$ &  $1.108 \pm 0.048$ &  0.31 &  0.77 \\\hline
   0402850701PNS003 &         PN &   LW &       7.3 &  12/27/06 &  $0.976 \pm 0.017$ &  $1.090 \pm 0.029$ &  0.30 &  0.77 \\\hline
   0402850901PNS003 &         PN &   LW &       2.6 &  07/05/06 &  $0.971 \pm 0.029$ &  $1.106 \pm 0.049$ &  0.30 &  0.75 \\\hline
 0157360101MOS1S003 &       MOS1 &   LW &      27.5 &  01/01/03 &  $1.083 \pm 0.020$ &  $0.711 \pm 0.027$ &  0.12 &  0.33 \\\hline
 0163560101MOS1S001 &       MOS1 &   LW &      23.9 &  30/12/03 &  $1.081 \pm 0.021$ &  $0.727 \pm 0.029$ &  0.12 &  0.33 \\\hline
 0163560101MOS2S002 &       MOS2 &   LW &      24.2 &  30/12/03 &  $1.083 \pm 0.021$ &  $0.713 \pm 0.029$ &  0.12 &  0.34 \\
\hline
\end{tabular} \\
\caption{\label{tab:obs1308} As in Tab.~\ref{tab:obs0806} but for RX J1308.6+2127.
}
\end{table}

\begin{table}[htb]
\centering
\textbf{RX J0720.4-3125} \vspace{1.5ex} \\
\begin{tabular}{|l|l|l|c|c|c|c|c|c|}
\hline
     Exposure ID &   Instrument & Mode & $t_{exp}$ &      Date &            Singles &            Doubles &    SD &    FL \\
\hline
 0156960401PNS003 &         PN &   FF &      24.3 &  11/08/02 &  $0.878 \pm 0.006$ &  $1.438 \pm 0.014$ &  0.85 &  2.14 \\\hline
 0158360201PNU002 &         PN &   SW &      37.5 &  05/02/03 &  $1.004 \pm 0.007$ &  $0.998 \pm 0.011$ &  0.09 &  0.11 \\\hline
 0161960201PNS007 &         PN &   SW &      12.2 &  10/27/03 &  $1.008 \pm 0.009$ &  $0.983 \pm 0.013$ &  0.13 &  0.23 \\\hline
 0161960201PNS008 &         PN &   SW &      16.0 &  10/27/03 &  $1.005 \pm 0.008$ &  $0.990 \pm 0.012$ &  0.13 &  0.22 \\\hline
 0164560501PNS001 &         PN &   FF &      18.7 &  05/22/04 &  $0.907 \pm 0.006$ &  $1.321 \pm 0.013$ &  1.13 &  2.87 \\\hline
 0300520201PNS003 &         PN &   FF &      18.5 &  04/28/05 &  $0.900 \pm 0.007$ &  $1.350 \pm 0.014$ &  1.09 &  2.76 \\\hline
 0300520301PNS003 &         PN &   FF &      13.0 &  09/23/05 &  $0.908 \pm 0.008$ &  $1.319 \pm 0.016$ &  1.08 &  2.75 \\\hline
 0311590101PNS003 &         PN &   FF &      30.1 &  11/12/05 &  $0.898 \pm 0.005$ &  $1.360 \pm 0.011$ &  1.08 &  2.75 \\\hline
 0400140301PNS001 &         PN &   FF &      13.6 &  05/22/06 &  $0.901 \pm 0.008$ &  $1.345 \pm 0.016$ &  1.09 &  2.76 \\\hline
 0400140401PNS001 &         PN &   FF &      16.7 &  11/05/06 &  $0.889 \pm 0.007$ &  $1.391 \pm 0.015$ &  1.08 &  2.73 \\\hline
 0502710201PNS001 &         PN &   FF &       6.0 &  05/05/07 &  $0.885 \pm 0.013$ &  $1.404 \pm 0.027$ &  1.04 &  2.63 \\\hline
 0502710301PNS001 &         PN &   FF &      19.0 &  11/17/07 &  $0.880 \pm 0.007$ &  $1.420 \pm 0.015$ &  1.02 &  2.59 \\\hline
 0554510101PNS003 &         PN &   FF &       8.9 &  03/21/09 &  $0.868 \pm 0.011$ &  $1.454 \pm 0.025$ &  1.00 &  2.54 \\\hline
 0650920101PNS003 &         PN &   FF &       8.0 &  04/11/11 &  $0.889 \pm 0.011$ &  $1.391 \pm 0.024$ &  0.94 &  2.38 \\\hline
 0670700201PNS003 &         PN &   FF &       9.0 &  05/02/11 &  $0.873 \pm 0.011$ &  $1.433 \pm 0.024$ &  0.93 &  2.36 \\\hline
 0670700301PNS003 &         PN &   FF &      21.4 &  10/01/11 &  $0.884 \pm 0.007$ &  $1.409 \pm 0.016$ &  0.92 &  2.31 \\\hline
 0690070201PNS003 &         PN &   FF &      21.5 &  09/18/12 &  $0.870 \pm 0.007$ &  $1.462 \pm 0.016$ &  0.89 &  2.24 \\
\hline
\end{tabular} \\
\caption{\label{tab:obs0720} As in Tab.~\ref{tab:obs0806} but for RX J0720.4-3125.
}
\end{table}

\begin{table}[htb]
\centering
\textbf{RX J1605.3+3249} \vspace{1.5ex} \\
\begin{tabular}{|l|l|l|c|c|c|c|c|c|}
\hline
     Exposure ID &   Instrument & Mode & $t_{exp}$ &      Date &            Singles &            Doubles &    SD &    FL \\
\hline
   0157360401PNS005 &         PN &   LW &      21.8 &  01/17/03 &  $0.931 \pm 0.009$ &  $1.238 \pm 0.017$ &  0.33 &  0.85 \\\hline
   0157360601PNS005 &         PN &   LW &       5.5 &  02/26/03 &  $0.954 \pm 0.022$ &  $1.150 \pm 0.038$ &  0.24 &  0.60 \\\hline
   0671620101PNS003 &         PN &   FF &      26.4 &  03/06/12 &  $0.921 \pm 0.008$ &  $1.278 \pm 0.016$ &  0.51 &  1.34 \\\hline
   0764460201PNS003 &         PN &   FF &      90.5 &  07/21/15 &  $0.932 \pm 0.004$ &  $1.239 \pm 0.008$ &  0.51 &  1.33 \\\hline
   0764460301PNS003 &         PN &   FF &      50.9 &  07/26/15 &  $0.930 \pm 0.006$ &  $1.251 \pm 0.011$ &  0.51 &  1.34 \\\hline
   0764460401PNS003 &         PN &   FF &      38.9 &  08/20/15 &  $0.942 \pm 0.007$ &  $1.203 \pm 0.013$ &  0.49 &  1.30 \\\hline
   0764460501PNS003 &         PN &   FF &      45.2 &  02/10/16 &  $0.930 \pm 0.006$ &  $1.251 \pm 0.012$ &  0.52 &  1.37 \\\hline
 0073140201MOS1S004 &       MOS1 &   FF &      26.2 &  15/01/02 &  $1.031 \pm 0.018$ &  $0.925 \pm 0.030$ &  0.37 &  1.09 \\\hline
 0073140201MOS2S005 &       MOS2 &   FF &      25.8 &  15/01/02 &  $1.029 \pm 0.018$ &  $0.933 \pm 0.030$ &  0.40 &  1.19 \\\hline
 0073140301MOS2S005 &       MOS2 &   FF &      16.8 &  09/01/02 &  $1.017 \pm 0.022$ &  $0.999 \pm 0.040$ &  0.40 &  1.17 \\\hline
 0073140501MOS1S004 &       MOS1 &   FF &      21.0 &  19/01/02 &  $1.026 \pm 0.020$ &  $0.954 \pm 0.035$ &  0.37 &  1.10 \\\hline
 0073140501MOS2S005 &       MOS2 &   FF &      21.0 &  19/01/02 &  $1.026 \pm 0.020$ &  $0.956 \pm 0.035$ &  0.38 &  1.14 \\\hline
 0157360401MOS2S004 &       MOS2 &   FF &      26.0 &  17/01/03 &  $1.027 \pm 0.018$ &  $0.936 \pm 0.031$ &  0.38 &  1.14 \\\hline
 0302140501MOS1S002 &       MOS1 &   FF &       3.3 &  12/02/06 &  $1.030 \pm 0.052$ &  $0.912 \pm 0.086$ &  0.34 &  1.01 \\\hline
 0302140501MOS2S003 &       MOS2 &   FF &       3.0 &  12/02/06 &  $1.033 \pm 0.055$ &  $0.892 \pm 0.090$ &  0.37 &  1.10 \\\hline
 0671620101MOS1U002 &       MOS1 &   FF &       6.9 &  07/03/12 &  $1.045 \pm 0.039$ &  $0.857 \pm 0.062$ &  0.29 &  0.86 \\\hline
 0671620101MOS2S002 &       MOS2 &   FF &      34.9 &  06/03/12 &  $1.036 \pm 0.017$ &  $0.897 \pm 0.028$ &  0.35 &  1.03 \\\hline
 0764460501MOS1S001 &       MOS1 &   LW &      58.2 &  10/02/16 &  $1.061 \pm 0.013$ &  $0.779 \pm 0.020$ &  0.11 &  0.30 \\
\hline
\end{tabular} \\
\caption{\label{tab:obs1605} As in Tab.~\ref{tab:obs0806} but for RX J1605.3+3249.
}
\end{table}

\begin{table}[htb]
\centering
\textbf{RX J2143.0+0654} \vspace{1.5ex} \\
\begin{tabular}{|l|l|l|c|c|c|c|c|c|}
\hline
     Exposure ID &   Instrument & Mode & $t_{exp}$ &      Date &            Singles &            Doubles &    SD &    FL \\
\hline
 0201150101PNS006 &         PN &   SW &      14.6 &  05/31/04 &  $1.012 \pm 0.015$ &  $0.973 \pm 0.022$ &  0.06 &  0.07 \\\hline
 0502040601PNS003 &         PN &   SW &       5.2 &  05/13/07 &  $1.004 \pm 0.024$ &  $0.988 \pm 0.036$ &  0.06 &  0.07 \\\hline
 0502040701PNS003 &         PN &   SW &       9.0 &  05/17/07 &  $1.004 \pm 0.018$ &  $0.990 \pm 0.027$ &  0.06 &  0.07 \\\hline
 0502040901PNS003 &         PN &   SW &       5.0 &  06/12/07 &  $1.013 \pm 0.025$ &  $0.967 \pm 0.036$ &  0.06 &  0.07 \\\hline
 0502041001PNS003 &         PN &   SW &       5.8 &  11/03/07 &  $1.014 \pm 0.023$ &  $0.965 \pm 0.034$ &  0.06 &  0.07 \\\hline
 0502041101PNS003 &         PN &   SW &       7.7 &  11/07/07 &  $1.006 \pm 0.020$ &  $0.995 \pm 0.030$ &  0.06 &  0.07 \\\hline
 0502041201PNS003 &         PN &   SW &       6.1 &  11/08/07 &  $1.011 \pm 0.022$ &  $0.970 \pm 0.033$ &  0.06 &  0.07 \\\hline
 0502041301PNS003 &         PN &   SW &       3.7 &  11/23/07 &  $0.986 \pm 0.029$ &  $1.055 \pm 0.045$ &  0.06 &  0.07 \\\hline
 0502041401PNS003 &         PN &   SW &       5.1 &  12/10/07 &  $0.996 \pm 0.024$ &  $1.021 \pm 0.037$ &  0.06 &  0.07 \\\hline
 0502041801PNS003 &         PN &   SW &       5.3 &  05/19/08 &  $1.011 \pm 0.024$ &  $0.972 \pm 0.035$ &  0.06 &  0.07 \\
\hline
\end{tabular} \\
\caption{\label{tab:obs2143} As in Tab.~\ref{tab:obs0806} but for RX J2143.0+0654.
}
\end{table}
\newpage

\section{Count statistics and exposures for the XDINS\lowercase{s}}
\label{sec:data}

\begin{table}[htb]
\centering
\textbf{RX J0806.4-4123} \vspace{1.5ex} \\
\begin{tabular}{| c | c | c | c | c| c | c | c | c|}
\hline
\multicolumn{9}{|c|}{\textbf{PN}} \\

\hline
Energy Range & $ \quad c_s \quad $ & $\quad c_B \quad $ & \, $\sum_{p \in R_S}\,$ & $\sum_{p \in R_B}\,$ &  $\, \bar w_S $  [cm$^2$ s] & $\, \bar w_B \,$  [cm$^2$ s] & $\qquad \chi_{S} \qquad$ & $\qquad \chi_{B} \qquad $ \\ \hline
2-4 keV & 39 &37& 2729 &3136 & $2.9 \times 10^{6}$ & $2.89 \times 10^{6}$ & $7.95 \times 10^{-1}$ & $9.72 \times 10^{-2}$\\ 
\hline
4-6 keV & 26& 35& 2729 &3136& $2.77 \times 10^{6}$ & $2.76 \times 10^{6}$ & $7.81 \times 10^{-1}$ & $9.59 \times 10^{-2}$ \\ \hline 
6-8 keV & 32& 26& 2729 &3136 & $2.16 \times 10^{6}$ & $2.16 \times 10^{6}$ & $7.62 \times 10^{-1}$ & $9.75 \times 10^{-2}$ \\ 
\hline

\multicolumn{9}{| c |}{\textbf{MOS}} \\

\hline
Energy Range & $ \quad c_s \quad $ & $\quad c_B \quad $ & \, $\sum_{p \in R_S}\,$ & $\sum_{p \in R_B}\,$ &  $\, \bar w_S $  [cm$^2$ s] & $\, \bar w_B \,$  [cm$^2$ s] & $\qquad \chi_{S} \qquad$ & $\qquad \chi_{B} \qquad $ \\ \hline
2-4 keV & 7 &7& 1086 &1218 & $2.88 \times 10^{6}$ & $2.86 \times 10^{6}$ & $7.95 \times 10^{-1}$ & $9.74 \times 10^{-2}$\\ 
\hline
4-6 keV & 5& 6& 1086 &1218& $2.52 \times 10^{6}$ & $2.51 \times 10^{6}$ & $7.81 \times 10^{-1}$ & $9.62 \times 10^{-2}$ \\ \hline 
6-8 keV & 1& 5& 1086 &1218 & $1.21 \times 10^{6}$ & $1.21 \times 10^{6}$ & $7.62 \times 10^{-1}$ & $9.8 \times 10^{-2}$ \\ 
\hline

\multicolumn{9}{|c|}{\textit{\textbf{Chandra}}} \\
\hline
Energy Range & $ \quad c_s \quad $ & $\quad c_B \quad $ & \, $\sum_{p \in R_S}\,$ & $\sum_{p \in R_B}\,$ &  $\, \bar w_S $  [cm$^2$ s] & $\, \bar w_B \,$  [cm$^2$ s] & $\qquad \chi_{S} \qquad$ & $\qquad \chi_{B} \qquad $ \\ \hline
2-4 keV & 9 &0& 51 &1029 & $8.49 \times 10^{6}$ & $8.07 \times 10^{6}$ & $9.25 \times 10^{-2}$ & $6.88 \times 10^{-1}$\\ 
\hline
4-6 keV & 0& 0& 51 &1029& $8.41 \times 10^{6}$ & $7.97 \times 10^{6}$ & $9.14 \times 10^{-2}$ & $6.89 \times 10^{-1}$ \\ \hline 
6-8 keV & 0& 2& 51 &1029 & $3.00 \times 10^{6}$ & $2.85 \times 10^{6}$ & $9.19 \times 10^{-2}$ & $6.87 \times 10^{-1}$ \\ 
\hline
\end{tabular} \\
\caption{\label{tab:data0806}  The exposure-stacked data for all cameras used in our fiducial analyses for RX J0806.4-4123.  We include the number of signal counts $c_S$, the number of background counts $c_B$, the number of pixels in the signal (background) region $\sum_{p \in R_S}\,$  ($\sum_{p \in R_B}\,$), the average exposure in the signal (background) region $\, \bar w_S $ ($\, \bar w_B $), and the fraction of source flux expected in the signal (background) region due to the PSF $\chi_{S}$ ($\chi_{B}$).  Note that the weights are reported without the keV$^{-1}$.   
}
\end{table}

\begin{table}[htb]
\centering
\vspace{1.5ex}\textbf{RX J1856.6-3754} \vspace{1.5ex} \\
\begin{tabular}{| c | c | c | c | c| c | c | c | c|}
\hline
\multicolumn{9}{|c|}{\textbf{PN}} \\

\hline
Energy Range & $ \quad c_s \quad $ & $\quad c_B \quad $ & \, $\sum_{p \in R_S}\,$ & $\sum_{p \in R_B}\,$ &  $\, \bar w_S $  [cm$^2$ s] & $\, \bar w_B \,$  [cm$^2$ s] & $\qquad \chi_{S} \qquad$ & $\qquad \chi_{B} \qquad $ \\ \hline
2-4 keV & 874 &903& 10668 &12285 & $1.91 \times 10^{7}$ & $1.91 \times 10^{7}$ & $7.95 \times 10^{-1}$ & $9.74 \times 10^{-2}$\\ 
\hline
4-6 keV & 669& 681& 10668 &12285& $1.84 \times 10^{7}$ & $1.84 \times 10^{7}$ & $7.81 \times 10^{-1}$ & $9.61 \times 10^{-2}$ \\ \hline 
6-8 keV & 462& 522& 10668 &12285 & $1.45 \times 10^{7}$ & $1.45 \times 10^{7}$ & $7.62 \times 10^{-1}$ & $9.79 \times 10^{-2}$ \\ 
\hline

\multicolumn{9}{| c |}{\textbf{MOS}} \\

\hline
Energy Range & $ \quad c_s \quad $ & $\quad c_B \quad $ & \, $\sum_{p \in R_S}\,$ & $\sum_{p \in R_B}\,$ &  $\, \bar w_S $  [cm$^2$ s] & $\, \bar w_B \,$  [cm$^2$ s] & $\qquad \chi_{S} \qquad$ & $\qquad \chi_{B} \qquad $ \\ \hline
2-4 keV & 99 &72& 3753 &4421 & $7.1 \times 10^{6}$ & $7. \times 10^{6}$ & $7.64 \times 10^{-1}$ & $1.09 \times 10^{-1}$\\ 
\hline
4-6 keV & 61& 71& 3753 &4421& $6.28 \times 10^{6}$ & $6.19 \times 10^{6}$ & $7.51 \times 10^{-1}$ & $1.07 \times 10^{-1}$ \\ \hline 
6-8 keV & 48& 57& 3753 &4421 & $3.09 \times 10^{6}$ & $3.05 \times 10^{6}$ & $7.31 \times 10^{-1}$ & $1.09 \times 10^{-1}$ \\ 
\hline

\multicolumn{9}{|c|}{\textit{\textbf{Chandra}}} \\

\hline
Energy Range & $ \quad c_s \quad $ & $\quad c_B \quad $ & \, $\sum_{p \in R_S}\,$ & $\sum_{p \in R_B}\,$ &  $\, \bar w_S $  [cm$^2$ s] & $\, \bar w_B \,$  [cm$^2$ s] & $\qquad \chi_{S} \qquad$ & $\qquad \chi_{B} \qquad $ \\ \hline
2-4 keV & 2 &13& 157 &3719 & $8.63 \times 10^{5}$ & $8.74 \times 10^{5}$ & $9.59 \times 10^{-2}$ & $6.73 \times 10^{-1}$\\ 
\hline
4-6 keV & 3& 5& 157 &3719& $1.67 \times 10^{6}$ & $1.69 \times 10^{6}$ & $9.54 \times 10^{-2}$ & $6.73 \times 10^{-1}$ \\ \hline 
6-8 keV & 1& 16& 157 &3719 & $8.66 \times 10^{5}$ & $8.8 \times 10^{5}$ & $9.49 \times 10^{-2}$ & $6.75 \times 10^{-1}$ \\ 
\hline

\end{tabular} \\
\caption{\label{tab:data1856} As in Tab.~\ref{tab:data0806} but for RX J1856.6-3754.
}
\end{table}

\begin{table}[htb]
\centering
\vspace{1.5ex}\textbf{RX J0420.0-5022} \vspace{1.5ex} \\
\begin{tabular}{| c | c | c | c | c| c | c | c | c|}
\hline
\multicolumn{9}{|c|}{\textbf{PN}} \\
\hline
Energy Range & $ \quad c_s \quad $ & $\quad c_B \quad $ & \, $\sum_{p \in R_S}\,$ & $\sum_{p \in R_B}\,$ &  $\, \bar w_S $  [cm$^2$ s] & $\, \bar w_B \,$  [cm$^2$ s] & $\qquad \chi_{S} \qquad$ & $\qquad \chi_{B} \qquad $ \\ \hline
2-4 keV & 66 &48& 4043 &4712 & $6.15 \times 10^{6}$ & $6.19 \times 10^{6}$ & $7.95 \times 10^{-1}$ & $9.78 \times 10^{-2}$\\ 
\hline
4-6 keV & 65& 54& 4043 &4712& $5.9 \times 10^{6}$ & $5.93 \times 10^{6}$ & $7.81 \times 10^{-1}$ & $9.68 \times 10^{-2}$ \\ \hline 
6-8 keV & 35& 38& 4043 &4712 & $4.64 \times 10^{6}$ & $4.66 \times 10^{6}$ & $7.62 \times 10^{-1}$ & $9.85 \times 10^{-2}$ \\ 
\hline

\multicolumn{9}{| c |}{\textbf{MOS}} \\
\hline
Energy Range & $ \quad c_s \quad $ & $\quad c_B \quad $ & \, $\sum_{p \in R_S}\,$ & $\sum_{p \in R_B}\,$ &  $\, \bar w_S $  [cm$^2$ s] & $\, \bar w_B \,$  [cm$^2$ s] & $\qquad \chi_{S} \qquad$ & $\qquad \chi_{B} \qquad $ \\ \hline
2-4 keV & 14 &13& 1739 &2426 & $6.04 \times 10^{6}$ & $6.02 \times 10^{6}$ & $7.64 \times 10^{-1}$ & $1.2 \times 10^{-1}$\\ 
\hline
4-6 keV & 5& 12& 1739 &2426& $5.29 \times 10^{6}$ & $5.28 \times 10^{6}$ & $7.52 \times 10^{-1}$ & $1.17 \times 10^{-1}$ \\ \hline 
6-8 keV & 5& 4& 1739 &2426 & $2.55 \times 10^{6}$ & $2.55 \times 10^{6}$ & $7.32 \times 10^{-1}$ & $1.2 \times 10^{-1}$ \\ 
\hline
\multicolumn{9}{|c|}{\textit{\textbf{Chandra}}} \\
\hline
Energy Range & $ \quad c_s \quad $ & $\quad c_B \quad $ & \, $\sum_{p \in R_S}\,$ & $\sum_{p \in R_B}\,$ &  $\, \bar w_S $  [cm$^2$ s] & $\, \bar w_B \,$  [cm$^2$ s] & $\qquad \chi_{S} \qquad$ & $\qquad \chi_{B} \qquad $ \\ \hline
2-4 keV & 1 &1& 44 &1057 & $7.79 \times 10^{6}$ & $7.78 \times 10^{6}$ & $9.4 \times 10^{-2}$ & $6.88 \times 10^{-1}$\\ 
\hline
4-6 keV & 1& 1& 44 &1057& $6.55 \times 10^{6}$ & $6.55 \times 10^{6}$ & $9.37 \times 10^{-2}$ & $6.89 \times 10^{-1}$ \\ \hline 
6-8 keV & 0& 3& 44 &1057 & $2.02 \times 10^{6}$ & $2.02 \times 10^{6}$ & $9.26 \times 10^{-2}$ & $6.88 \times 10^{-1}$ \\ 
\hline
\end{tabular} \\
\caption{\label{tab:data0420} As in Tab.~\ref{tab:data0806} but for RX J0420.0-5022.
}
\end{table}

\begin{table}[htb]
\centering
\textbf{RX J1308.6+2127} \vspace{1ex} \\
\begin{tabular}{| c | c | c | c | c| c | c | c | c|}
\hline
\multicolumn{9}{|c|}{\textbf{PN}} \\

\hline
Energy Range & $ \quad c_s \quad $ & $\quad c_B \quad $ & \, $\sum_{p \in R_S}\,$ & $\sum_{p \in R_B}\,$ &  $\, \bar w_S $  [cm$^2$ s] & $\, \bar w_B \,$  [cm$^2$ s] & $\qquad \chi_{S} \qquad$ & $\qquad \chi_{B} \qquad $ \\ \hline
2-4 keV & 41 &32& 3015 &3453 & $6.71 \times 10^{6}$ & $6.69 \times 10^{6}$ & $7.94 \times 10^{-1}$ & $9.73 \times 10^{-2}$\\ 
\hline
4-6 keV & 31& 50& 3015 &3453& $6.42 \times 10^{6}$ & $6.4 \times 10^{6}$ & $7.82 \times 10^{-1}$ & $9.57 \times 10^{-2}$ \\ \hline 
6-8 keV & 25& 27& 3015 &3453 & $5.07 \times 10^{6}$ & $5.05 \times 10^{6}$ & $7.63 \times 10^{-1}$ & $9.72 \times 10^{-2}$ \\ 
\hline

\multicolumn{9}{| c |}{\textbf{MOS}} \\
\hline
Energy Range & $ \quad c_s \quad $ & $\quad c_B \quad $ & \, $\sum_{p \in R_S}\,$ & $\sum_{p \in R_B}\,$ &  $\, \bar w_S $  [cm$^2$ s] & $\, \bar w_B \,$  [cm$^2$ s] & $\qquad \chi_{S} \qquad$ & $\qquad \chi_{B} \qquad $ \\ \hline
2-4 keV & 8 &9& 804 &933 & $7.37 \times 10^{6}$ & $7.35 \times 10^{6}$ & $7.95 \times 10^{-1}$ & $9.69 \times 10^{-2}$\\ 
\hline
4-6 keV & 3& 0& 804 &933& $6.44 \times 10^{6}$ & $6.42 \times 10^{6}$ & $7.81 \times 10^{-1}$ & $9.56 \times 10^{-2}$ \\ \hline 
6-8 keV & 6& 2& 804 &933 & $3.12 \times 10^{6}$ & $3.11 \times 10^{6}$ & $7.62 \times 10^{-1}$ & $9.75 \times 10^{-2}$ \\ 
\hline
\end{tabular} \\
\caption{\label{tab:data1308} As in Tab.~\ref{tab:data0806} but for RX J1308.6+2127.
}
\end{table}

\begin{table}[htb]
\centering
\vspace{1.5ex} \textbf{RX J0720.4-3125} \vspace{1.5ex} \\
\begin{tabular}{| c | c | c | c | c| c | c | c | c|}
\hline
\multicolumn{9}{|c|}{\textbf{PN}} \\
\hline
Energy Range & $ \quad c_s \quad $ & $\quad c_B \quad $ & \, $\sum_{p \in R_S}\,$ & $\sum_{p \in R_B}\,$ &  $\, \bar w_S $  [cm$^2$ s] & $\, \bar w_B \,$  [cm$^2$ s] & $\qquad \chi_{S} \qquad$ & $\qquad \chi_{B} \qquad $ \\ \hline
2-4 keV & 181 &158& 4575 &5230 & $1.19 \times 10^{7}$ & $1.18 \times 10^{7}$ & $7.95 \times 10^{-1}$ & $9.71 \times 10^{-2}$\\ 
\hline
4-6 keV & 115& 124& 4575 &5230& $1.15 \times 10^{7}$ & $1.14 \times 10^{7}$ & $7.82 \times 10^{-1}$ & $9.57 \times 10^{-2}$ \\ \hline 
6-8 keV & 98& 102& 4575 &5230 & $9.12 \times 10^{6}$ & $9.08 \times 10^{6}$ & $7.62 \times 10^{-1}$ & $9.71 \times 10^{-2}$ \\ 
\hline

\multicolumn{9}{|c|}{\textbf{MOS}} \\
\hline
Energy Range & $ \quad c_s \quad $ & $\quad c_B \quad $ & \, $\sum_{p \in R_S}\,$ & $\sum_{p \in R_B}\,$ &  $\, \bar w_S $  [cm$^2$ s] & $\, \bar w_B \,$  [cm$^2$ s] & $\qquad \chi_{S} \qquad$ & $\qquad \chi_{B} \qquad $ \\ \hline
2-4 keV & 21 &13& 1276 &1787 & $2.33 \times 10^{6}$ & $2.33 \times 10^{6}$ & $7.63 \times 10^{-1}$ & $1.20 \times 10^{-1}$\\ 
\hline
4-6 keV & 6& 8& 1276 &1787& $2.05 \times 10^{6}$ & $2.05 \times 10^{6}$ & $7.51 \times 10^{-1}$ & $1.17 \times 10^{-1}$ \\ \hline 
6-8 keV & 3& 8& 1276 &1787 & $9.96 \times 10^{5}$ & $9.97 \times 10^{5}$ & $7.31 \times 10^{-1}$ & $1.2 \times 10^{-1}$ \\ 
\hline
\end{tabular} \\
\caption{\label{tab:data0720} As in Tab.~\ref{tab:data0806} but for RX J0720.4-3125.
}
\end{table}

\begin{table}[htb]
\centering
\vspace{1.5ex}\textbf{RX J1605.3+3249} \vspace{1.5ex} \\
\begin{tabular}{| c | c | c | c | c| c | c | c | c|} 
\hline
\multicolumn{9}{|c|}{\textbf{PN}} \\
\hline
Energy Range & $ \quad c_s \quad $ & $\quad c_B \quad $ & \, $\sum_{p \in R_S}\,$ & $\sum_{p \in R_B}\,$ &  $\, \bar w_S $  [cm$^2$ s] & $\, \bar w_B \,$  [cm$^2$ s] & $\qquad \chi_{S} \qquad$ & $\qquad \chi_{B} \qquad $ \\ \hline
2-4 keV & 124 &103& 1841 &2150 & $2.92 \times 10^{7}$ & $2.93 \times 10^{7}$ & $7.85 \times 10^{-1}$ & $1.02 \times 10^{-1}$\\ 
\hline
4-6 keV & 76& 94& 1841 &2150& $2.79 \times 10^{7}$ & $2.81 \times 10^{7}$ & $7.72 \times 10^{-1}$ & $1.00 \times 10^{-1}$ \\ \hline 
6-8 keV & 57& 65& 1841 &2150 & $2.21 \times 10^{7}$ & $2.22 \times 10^{7}$ & $7.54 \times 10^{-1}$ & $1.02 \times 10^{-1}$ \\ 
\hline
\multicolumn{9}{| c |}{\textbf{MOS}} \\
\hline
Energy Range & $ \quad c_s \quad $ & $\quad c_B \quad $ & \, $\sum_{p \in R_S}\,$ & $\sum_{p \in R_B}\,$ &  $\, \bar w_S $  [cm$^2$ s] & $\, \bar w_B \,$  [cm$^2$ s] & $\qquad \chi_{S} \qquad$ & $\qquad \chi_{B} \qquad $ \\ \hline
2-4 keV & 35 &38& 2926 &3380 & $4.21 \times 10^{6}$ & $4.21 \times 10^{6}$ & $7.85 \times 10^{-1}$ & $1.02 \times 10^{-1}$\\ 
\hline
4-6 keV & 17& 28& 2926 &3380& $3.70 \times 10^{6}$ & $3.70 \times 10^{6}$ & $7.72 \times 10^{-1}$ & $1.01 \times 10^{-1}$ \\ \hline 
6-8 keV & 15& 20& 2926 &3380 & $1.81 \times 10^{6}$ & $1.81 \times 10^{6}$ & $7.53 \times 10^{-1}$ & $1.02 \times 10^{-1}$ \\ 
\hline
\end{tabular} \\
\caption{\label{tab:data1605} As in Tab.~\ref{tab:data0806} but for RX J1605.3+3249.
}
\end{table}

\begin{table}[htb]
\centering
\vspace{1.5ex}\textbf{RX J2143.0+0654}\vspace{1.5ex} \\
\begin{tabular}{| c | c | c | c | c| c | c | c | c|}
\hline
Energy Range & $ \quad c_s \quad $ & $\quad c_B \quad $ & \, $\sum_{p \in R_S}\,$ & $\sum_{p \in R_B}\,$ &  $\, \bar w_S $  [cm$^2$ s] & $\, \bar w_B \,$  [cm$^2$ s] & $\qquad \chi_{S} \qquad$ & $\qquad \chi_{B} \qquad $ \\ \hline
2-4 keV & 61 &62& 2720 &3148 & $5.06 \times 10^{6}$ & $5.05 \times 10^{6}$ & $7.95 \times 10^{-1}$ & $9.64 \times 10^{-2}$\\ 
\hline
4-6 keV & 41& 37& 2720 &3148& $4.86 \times 10^{6}$ & $4.84 \times 10^{6}$ & $7.82 \times 10^{-1}$ & $9.56 \times 10^{-2}$ \\ \hline 
6-8 keV & 33& 34& 2720 &3148 & $3.84 \times 10^{6}$ & $3.82 \times 10^{6}$ & $7.63 \times 10^{-1}$ & $9.72 \times 10^{-2}$ \\ 
\hline

\end{tabular}
\caption{\label{tab:data2143} As in Tab.~\ref{tab:data0806} but for RX J2143.0+0654.
}
\end{table}

Here we give, for each NS and instrument, the data used in our fiducial analyses after stacking all exposures together. In particular, we list: the number of counts in the signal region, $c_S$; the number of counts in the background region, $c_B$; the number of pixels included in the signal region, $\sum_{p \in R_S}$; the number of pixels included in the background region, $\sum_{p \in R_B}$; the mean pixel exposure for signal region pixels, $\bar w_S$; the mean pixel exposure for background region pixels, $\bar w_B$; the fraction of signal counts that will appear in signal region pixels due to the instrument PSF, $\chi_S$; and the fraction of signal counts that will appear in background region pixels due to the instrument PSF, $\chi_B$.  The data are provided in Tabs.~\ref{tab:data0806},~\ref{tab:data1856},~\ref{tab:data0420},~\ref{tab:data1308},~\ref{tab:data0720},~\ref{tab:data1605}, and~\ref{tab:data2143} for each NS, respectively.

\section{Test statistic maps for the XDINS\lowercase{s}}
\label{app:syst-maps}

\begin{figure*}[htb]
\begin{center}
\includegraphics[height = 0.23\textheight]{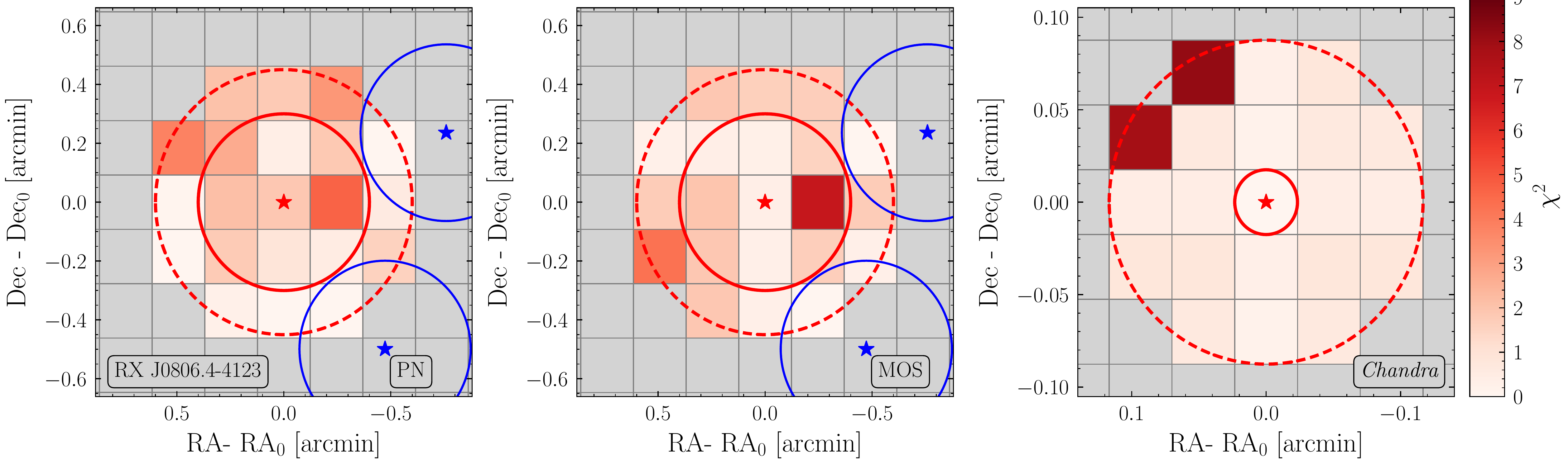}
\end{center}
\caption{$\chi^2$ maps as in Fig.~\ref{fig:hard1856_map} but for RX J0806.4-4123 in PN, MOS and Chandra computed using counts from the 4 to 8 keV energy range. No excess is observed in the signal region for any instrument. A nearby point source is found in the joint analysis of PN and MOS data; its point source mask would remove a small number of pixels from the background extraction region.}
\label{fig:hard0806_map}
\end{figure*}

\begin{figure*}[htb]
\begin{center}
\includegraphics[height = 0.23\textheight]{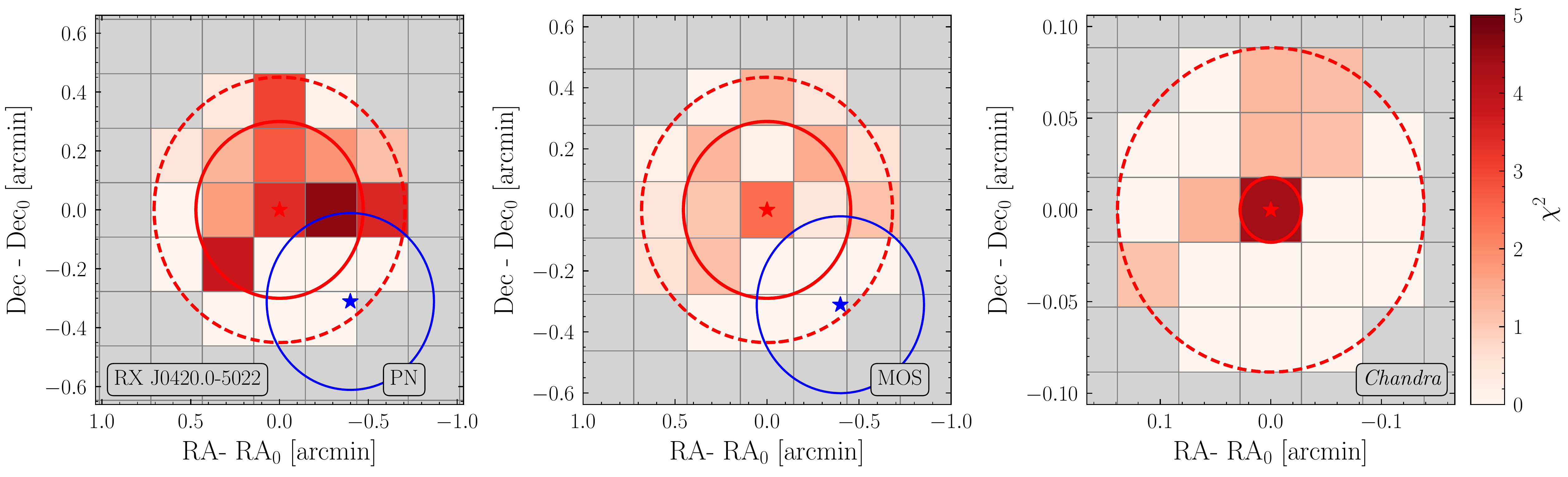}
\end{center}
\caption{The $\chi^2$ maps for RX J0420.0-5022 in PN, MOS, and Chandra computed using counts from the 2 to 8 keV energy range.  Evidence for an excess in the central pixel is observed for the Chandra and MOS maps, while the most significant excesses for the PN map is displaced from the center by one pixel.  This one-pixel displacement of the most significant pixel from the source center is consistent with the spread expected given the angular resolution and the $\sim$3$\sigma$ detection significance for PN  (see Tab.~\ref{tab:SigTable}).
    See the text for an expanded discussion.  }
\label{fig:hard0420_map}
\end{figure*}

\begin{figure*}[htb]
\begin{center}
\includegraphics[height = 0.23\textheight]{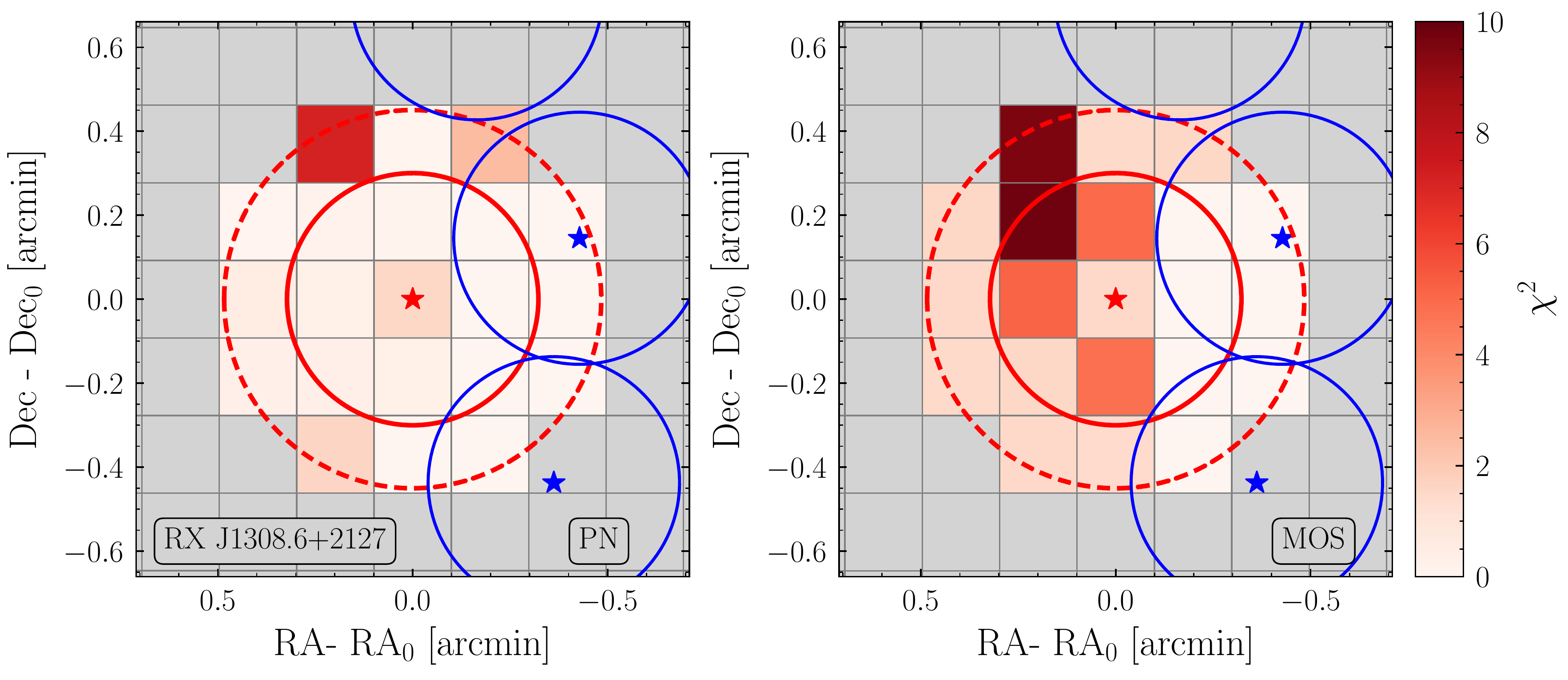}
\end{center}
\caption{The $\chi^2$ maps for RX J1308.6+2127 in PN and MOS. There is a nearby point source, but its mask does not include any of the signal or background extraction regions. There is no strong evidence for an excess in the central pixel.}
\label{fig:hard1308_map}
\end{figure*}

\begin{figure*}[htb]
\begin{center}
\includegraphics[height = 0.23\textheight]{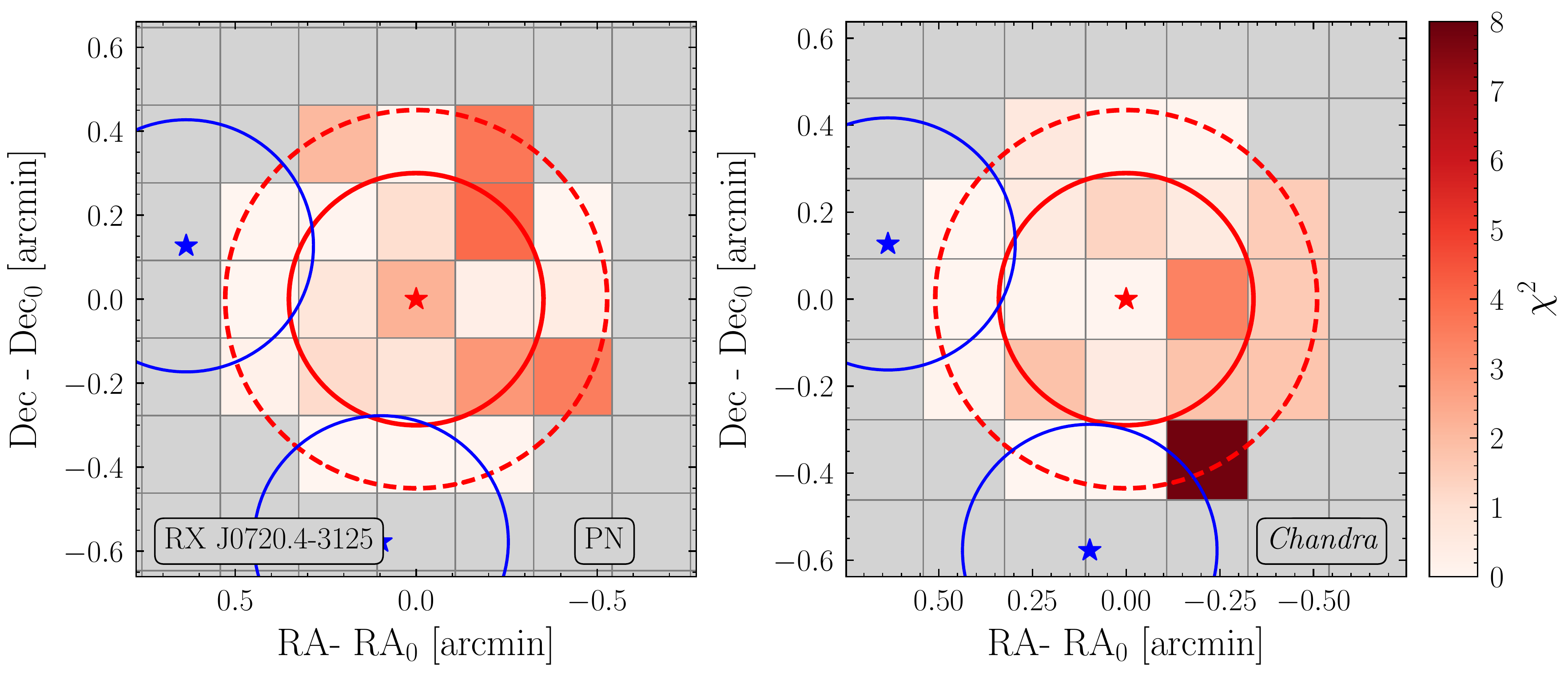}
\end{center}
\caption{The $\chi^2$ map for RX J0720.4-3125 as observed by the PN and MOS instruments. There is no evidence for a central-pixel excess. There is a somewhat nearby point source.}
\label{fig:hard0720_map}
\end{figure*}

\begin{figure*}[htb]
\begin{center}
\includegraphics[height = 0.23\textheight]{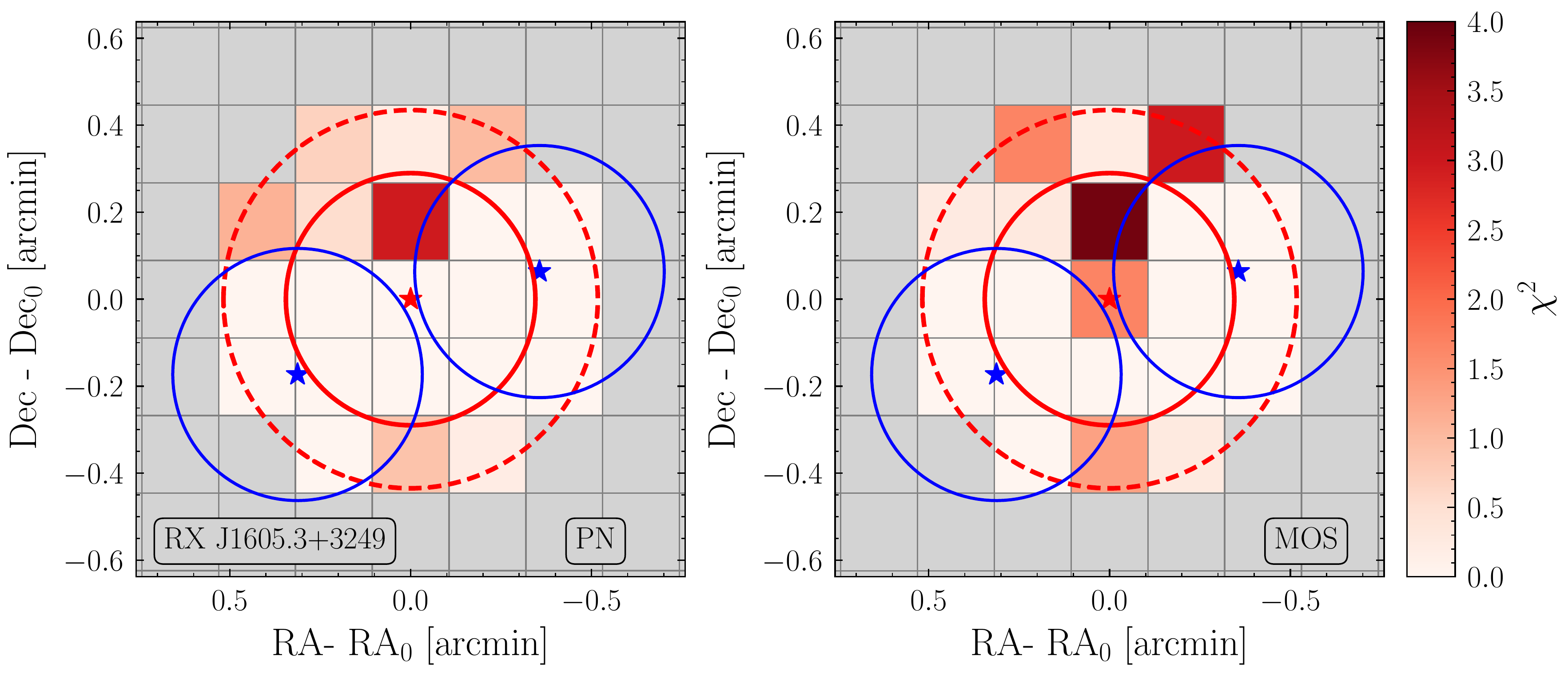}
\end{center}
\caption{The $\chi^2$ maps for RX J1605.3+3249 for observations using the PN and MOS instruments. No relevant point sources are detected, and there is no evidence for a central-pixel excess in either instrument.}
\label{fig:hard1605_map}
\end{figure*}

\begin{figure*}[htb]
\begin{center}
\includegraphics[height = 0.23\textheight]{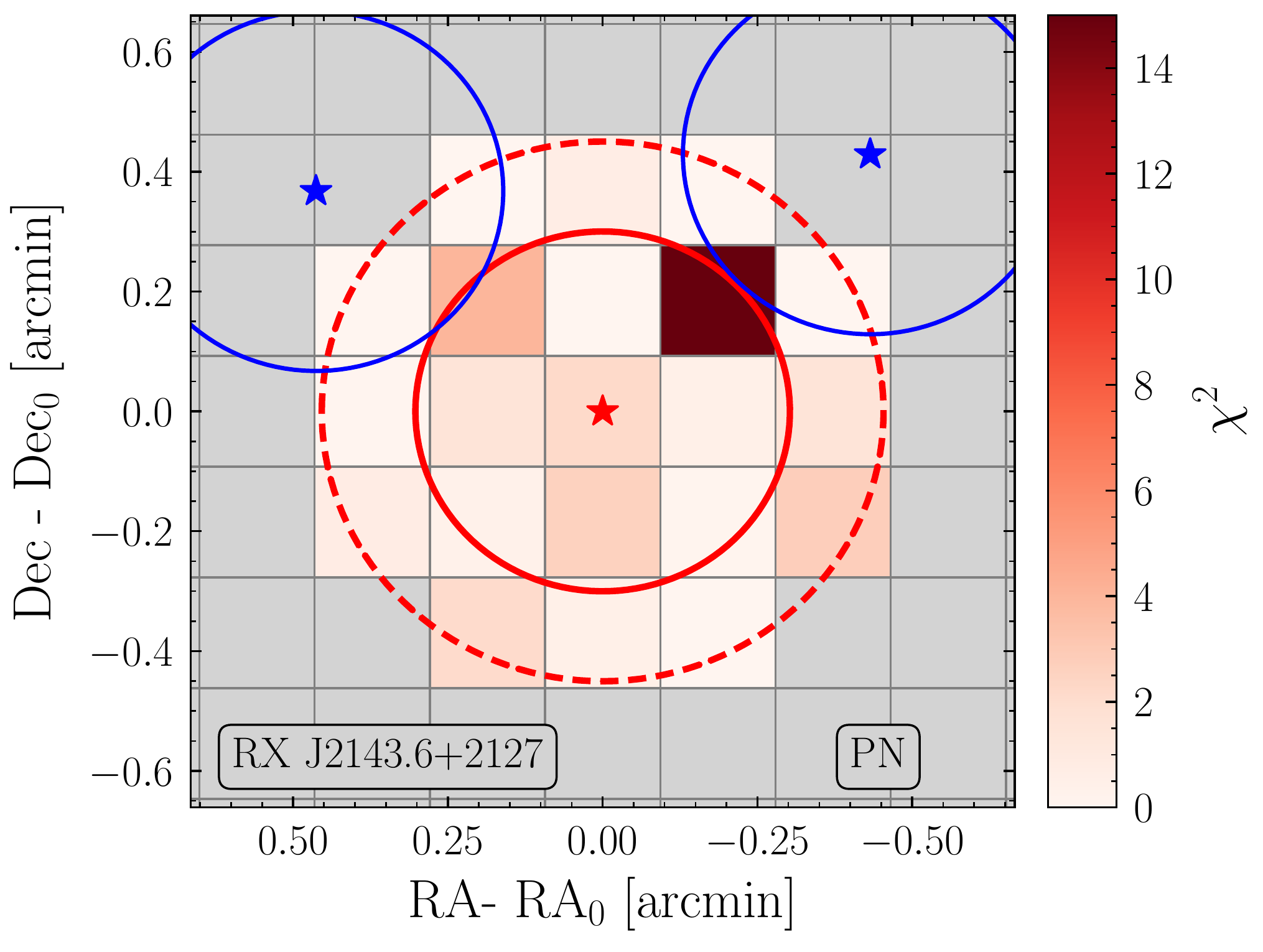}
\end{center}
\caption{The $\chi^2$ map for RX J2143.0+0654 for observations using the PN instrument. No relevant point sources are detected, and there is no significant evidence for a central-pixel excess.}
\label{fig:hard2143_map}
\end{figure*}

We present the test statistic maps for all NSs and for all instruments in which they are observed, as shown in Fig.~\ref{fig:hard1856_map} for RX J1856.6-3754. For RX J1856.6-3754 and RX J0420.0-5022, the test statistic maps are computed using flux from 2 to 8 keV. For all other NSs, the test statistic maps are computed using only the flux from 4 to 8 keV.  The maps are presented in Figs.~\ref{fig:hard0806_map},~\ref{fig:hard0420_map},~\ref{fig:hard1308_map},~\ref{fig:hard0720_map},~\ref{fig:hard1605_map}, and~\ref{fig:hard2143_map}.

It is worth pointing out that the brightest pixel in Fig.~\ref{fig:hard0420_map} is one pixel displaced from the center.  Recall that RX J0420.0-5022 was detected at $\sim$3$\sigma$ significance with the PN data using the joint likelihood.
Note also that we expect to be able to localize a 3$\sigma$ signal to roughly within the  1/3 of the 90\% EEF radius for the instrumental PSF.  In the 2-8 keV energy range, the 90\% EEF radius for PN and MOS is approximately 1\farcm\  Thus, it should not be too surprising that, in Fig.~\ref{fig:hard0420_map}, the most significant pixel for the PN map is not the central pixel but rather the pixel whose center is slightly displaced from the source.  On the other hand, the brightest pixels for the Chandra and MOS RX J0420.0-5022 maps are the central pixels. 

\section{Spectral limits and systematic tests for the XDINS\lowercase{s}}
\label{app:syst-spec}

\begin{figure*}[htb]
\begin{center}
\includegraphics[width = 0.48\textwidth]{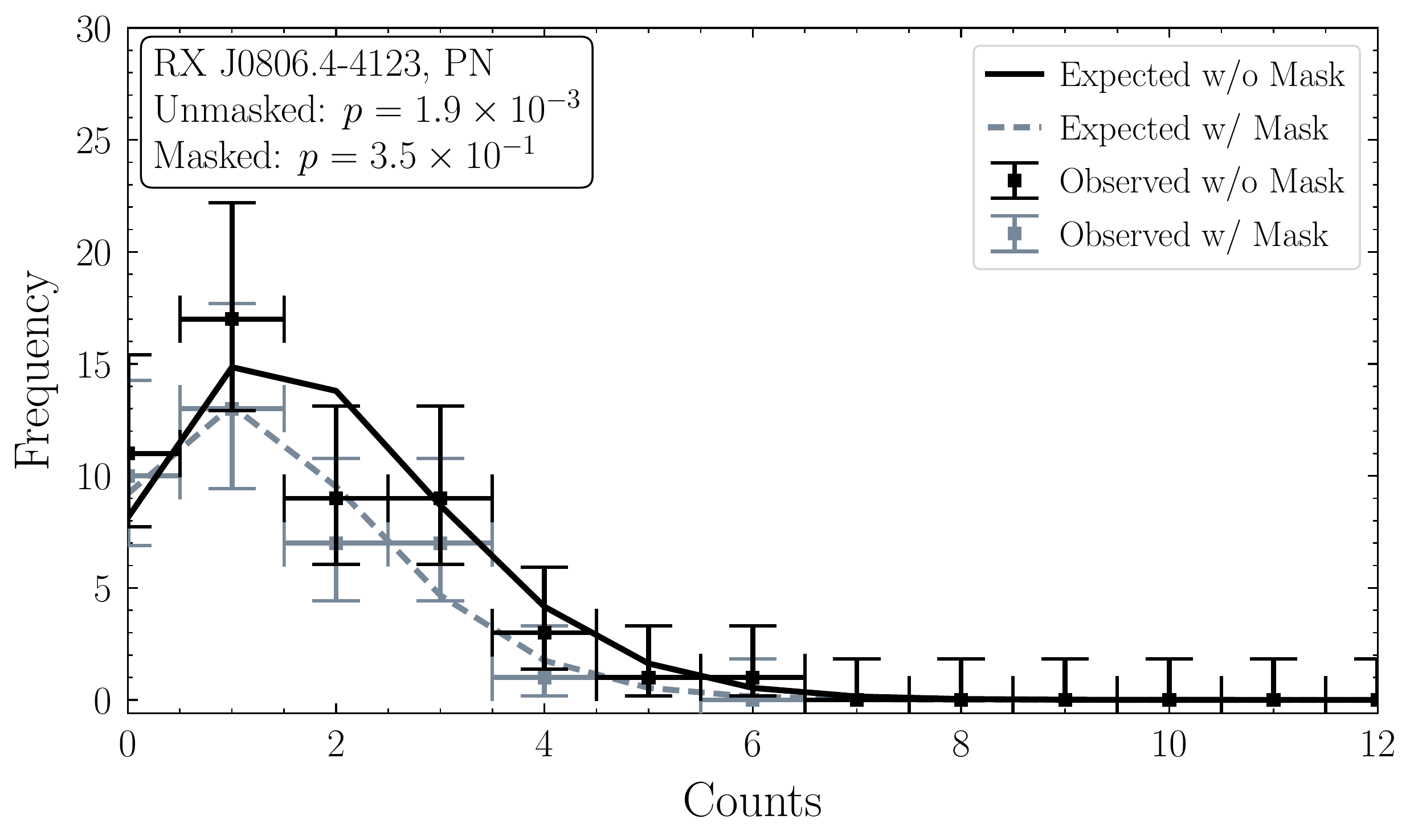} \includegraphics[width = 0.48\textwidth]{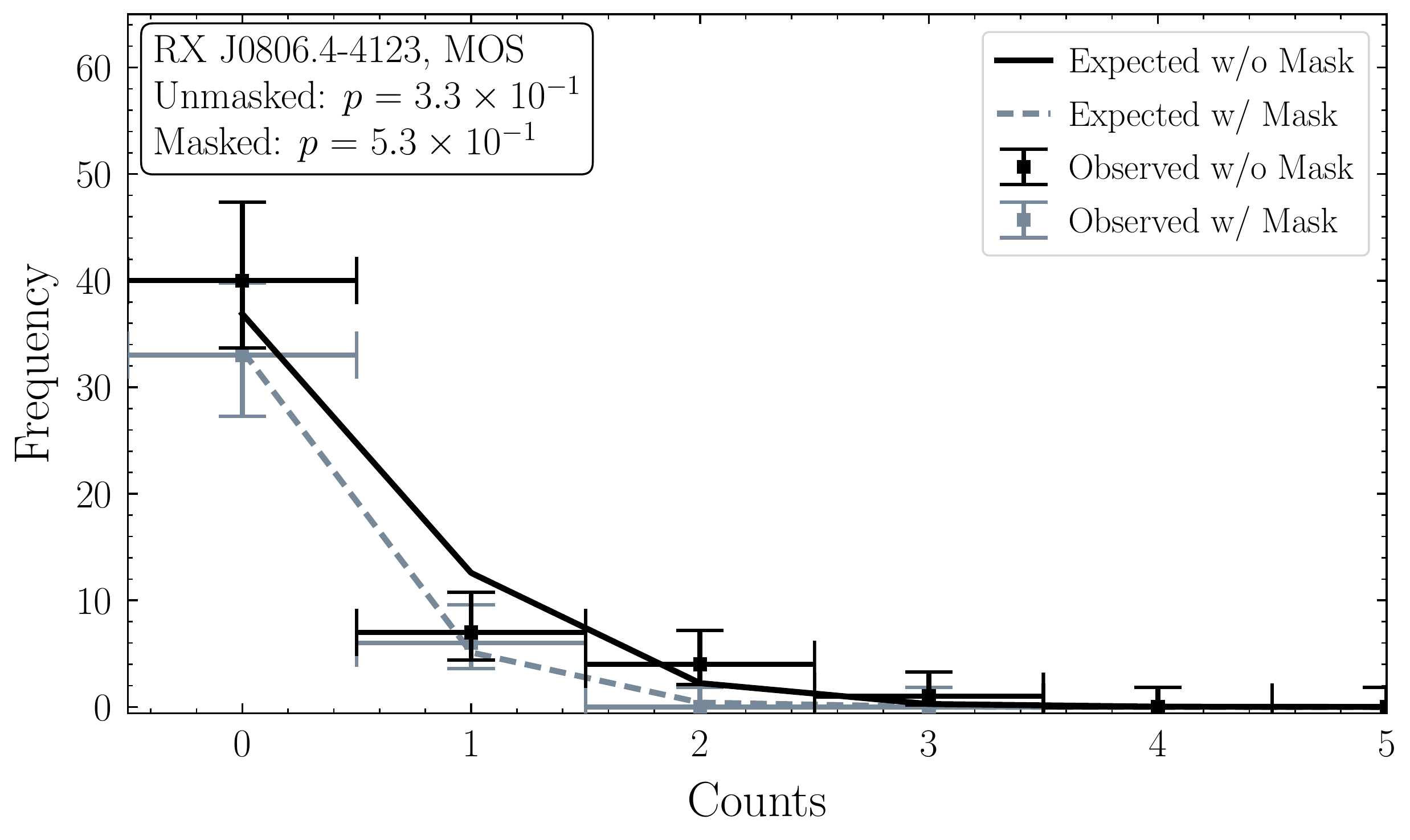}
\end{center}
\caption{
As in Fig.~\ref{fig:PoissonDistTest} but for RX J0806.4-4123.  In particular, we show the distribution of background counts by pixel for RX J0806.4-4123 with and without point source masking in both PN  (\textit{left})  and MOS  (\textit{right})  instruments. The point source mask only narrowly overlaps with the background extraction regions and therefore has marginal impact on the goodness of fit.}
\label{fig:PoissonDistTest_0806}
\end{figure*}

\begin{figure*}[htb]
\begin{center}
\includegraphics[width = .85\textwidth]{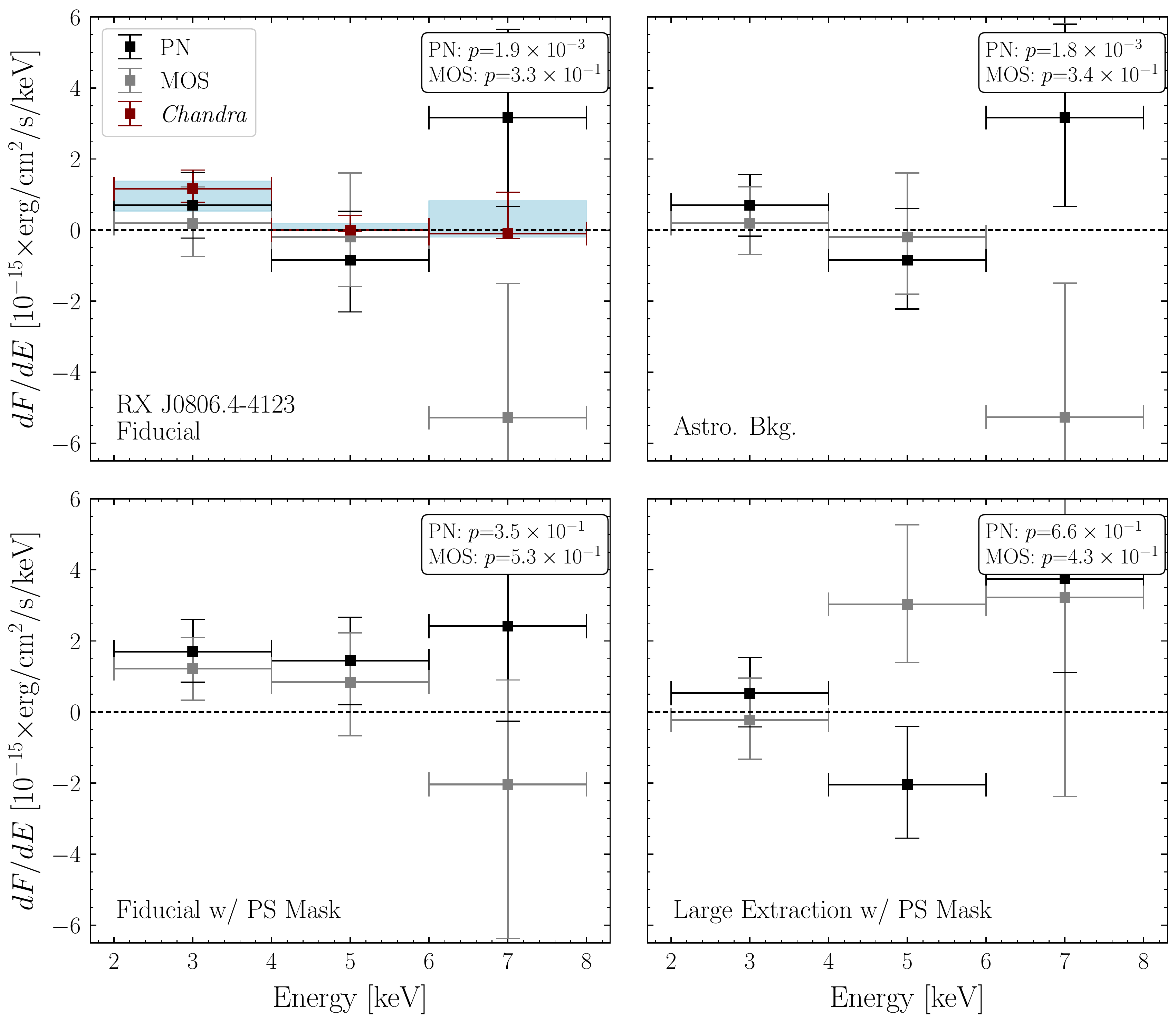}
\end{center}
\caption{As in Fig.~\ref{fig:Systematic} but for RX J0806.4-4123. Because the point source mask only narrowly overlaps with the background extraction regions, the effect of its inclusion on the reconstructed fluxes and limits is negligible.}
\label{fig:Systematic_0806}
\end{figure*}

\begin{figure*}[htb]
\begin{center}
\includegraphics[width = 0.48\textwidth]{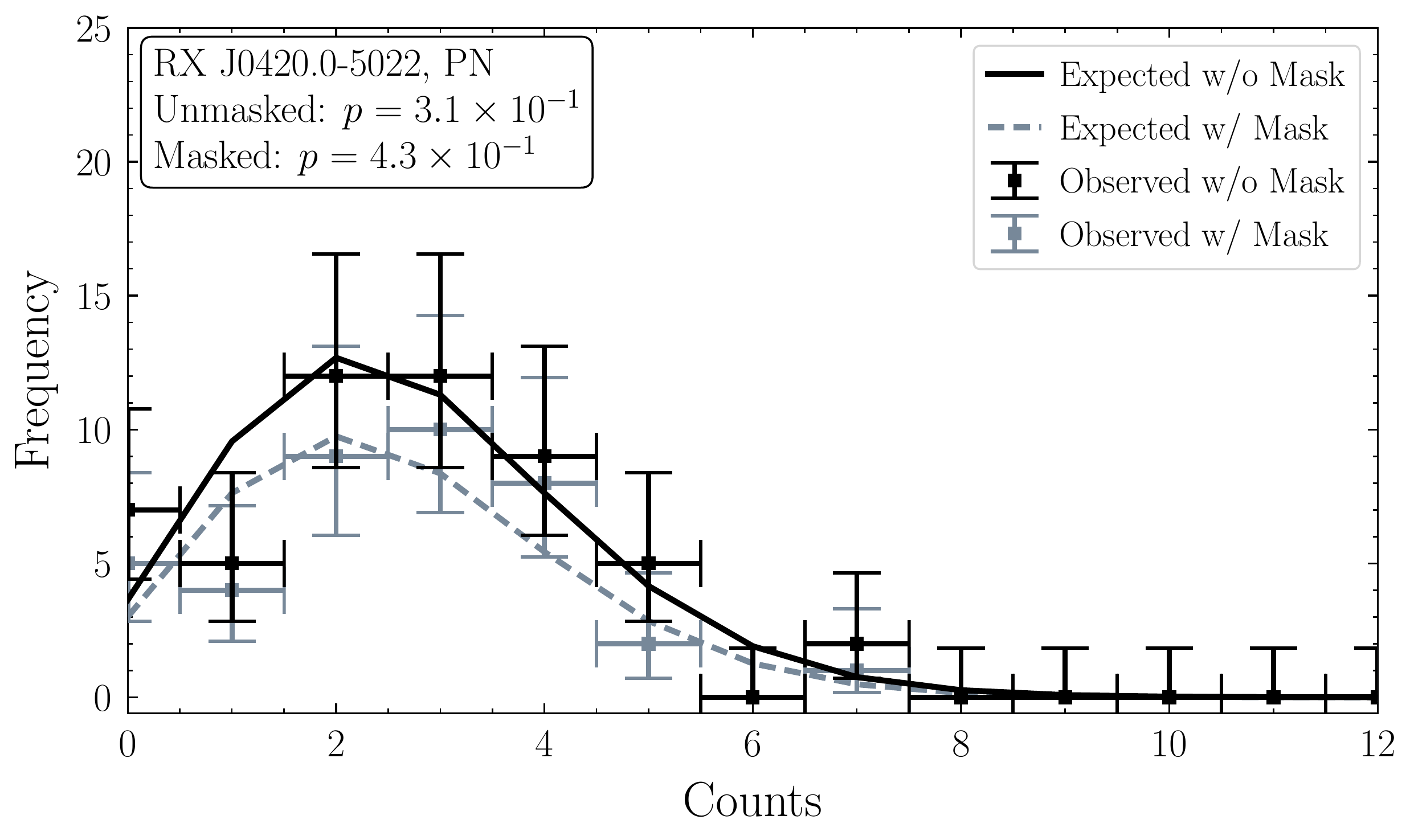} \includegraphics[width = 0.48\textwidth]{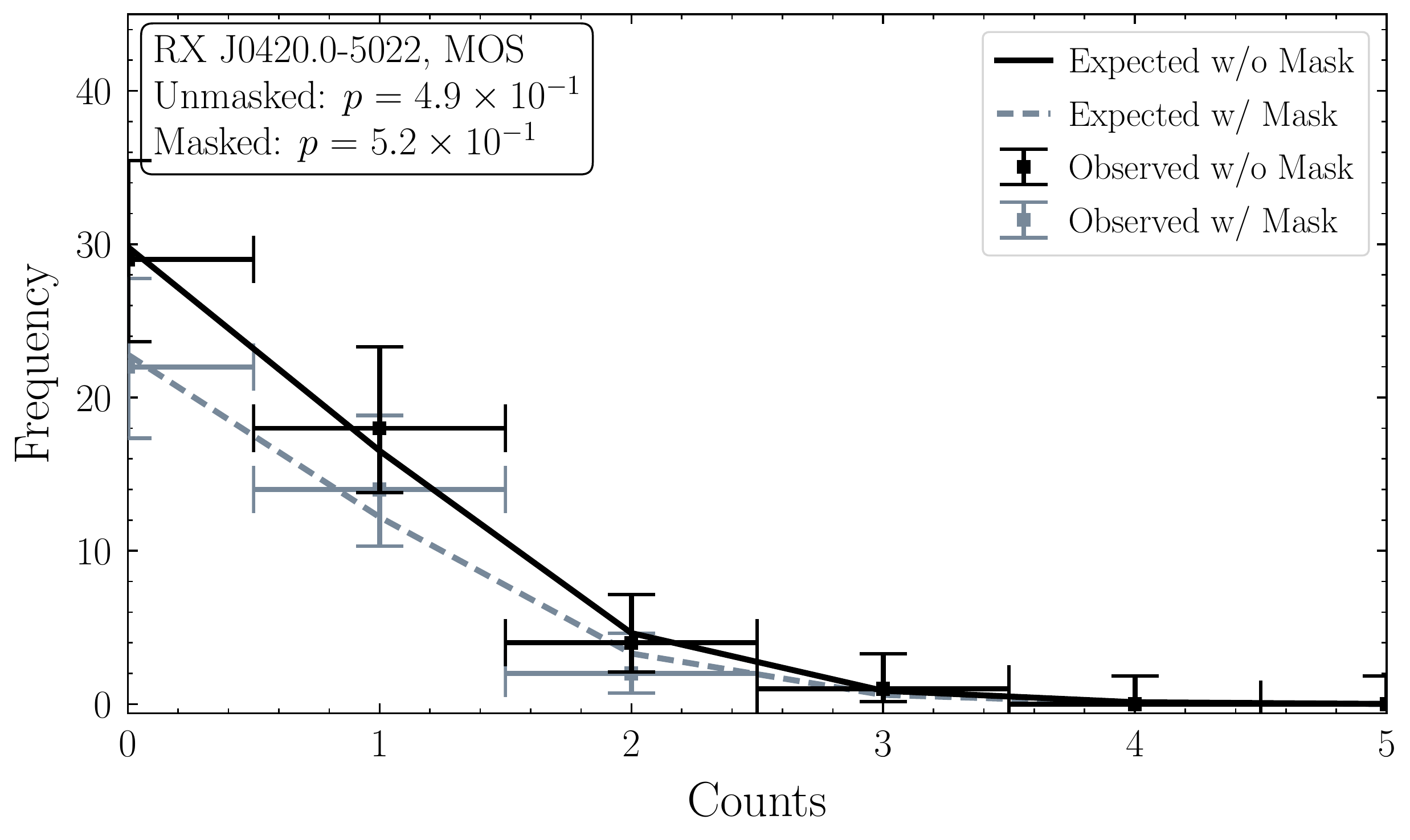}
\end{center}
\caption{The distribution of background counts by pixel for RX J0420.0-5022 in the PN  (\textit{left})  and MOS  (\textit{right}) instruments. No point source was found near enough to the signal or background extraction regions to require masking.}
\label{fig:PoissonDistTest_0420}
\end{figure*}

\begin{figure*}[htb]
\begin{center}
\includegraphics[width = .99\textwidth]{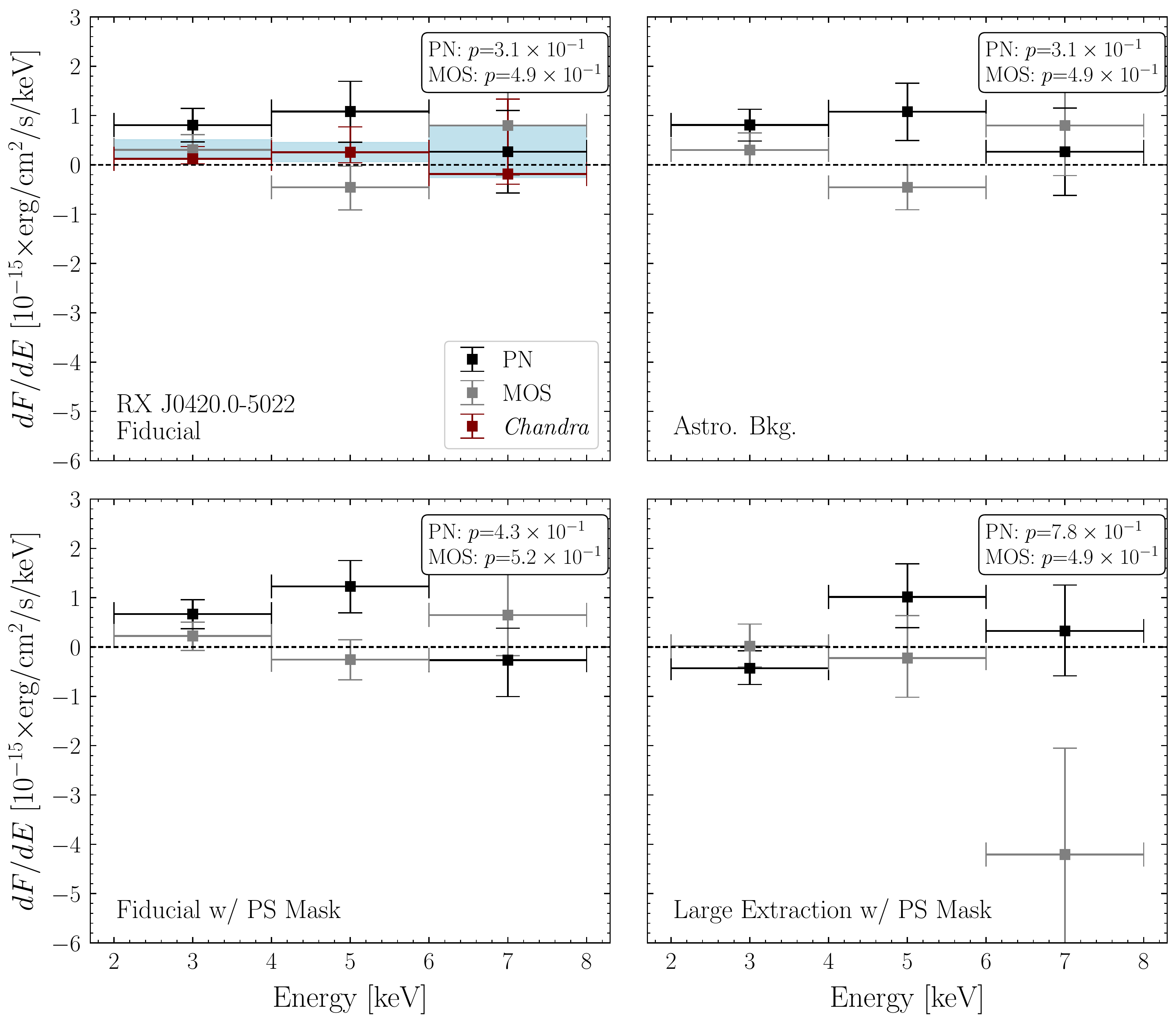}
\end{center}
\caption{Systematic variations on the analysis procedure on the reconstructed fluxes and limits at each energy bin for RX J0420.0-5022. A statistically significant excess in the power-law fit is found for this NS.}
\label{fig:Systematic_0420}
\end{figure*}

\begin{figure*}[htb]
\begin{center}
\includegraphics[width = 0.48\textwidth]{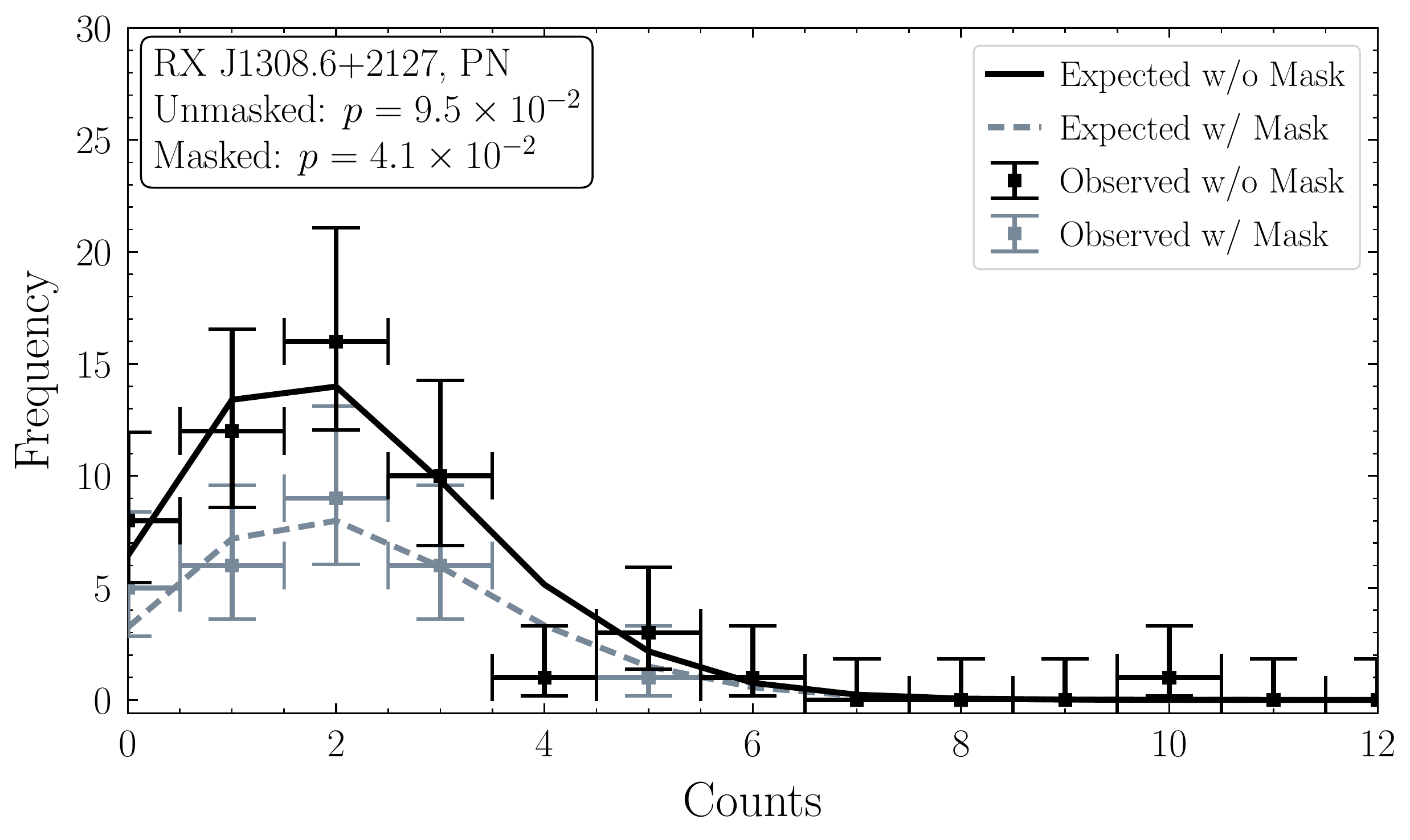} \includegraphics[width = 0.48\textwidth]{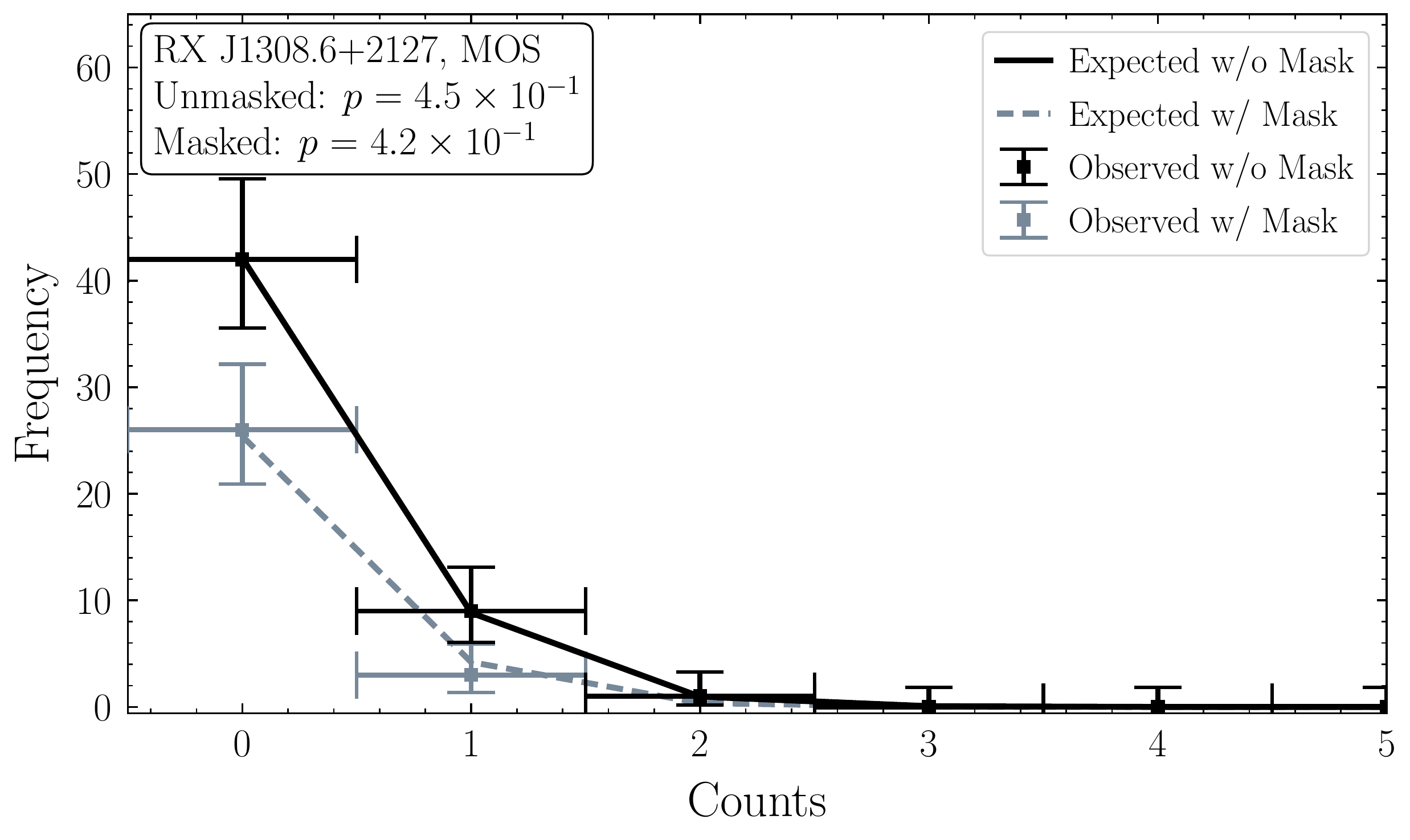}
\end{center}
\caption{The distribution of background counts by pixel for RX J1308+2127 in both PN (\textit{left}) and MOS (\textit{right}) instruments. No nearby point sources are detected that required masking.}
\label{fig:PoissonDistTest_1308}
\end{figure*}

\begin{figure*}[htb]
\begin{center}
\includegraphics[width = .99\textwidth]{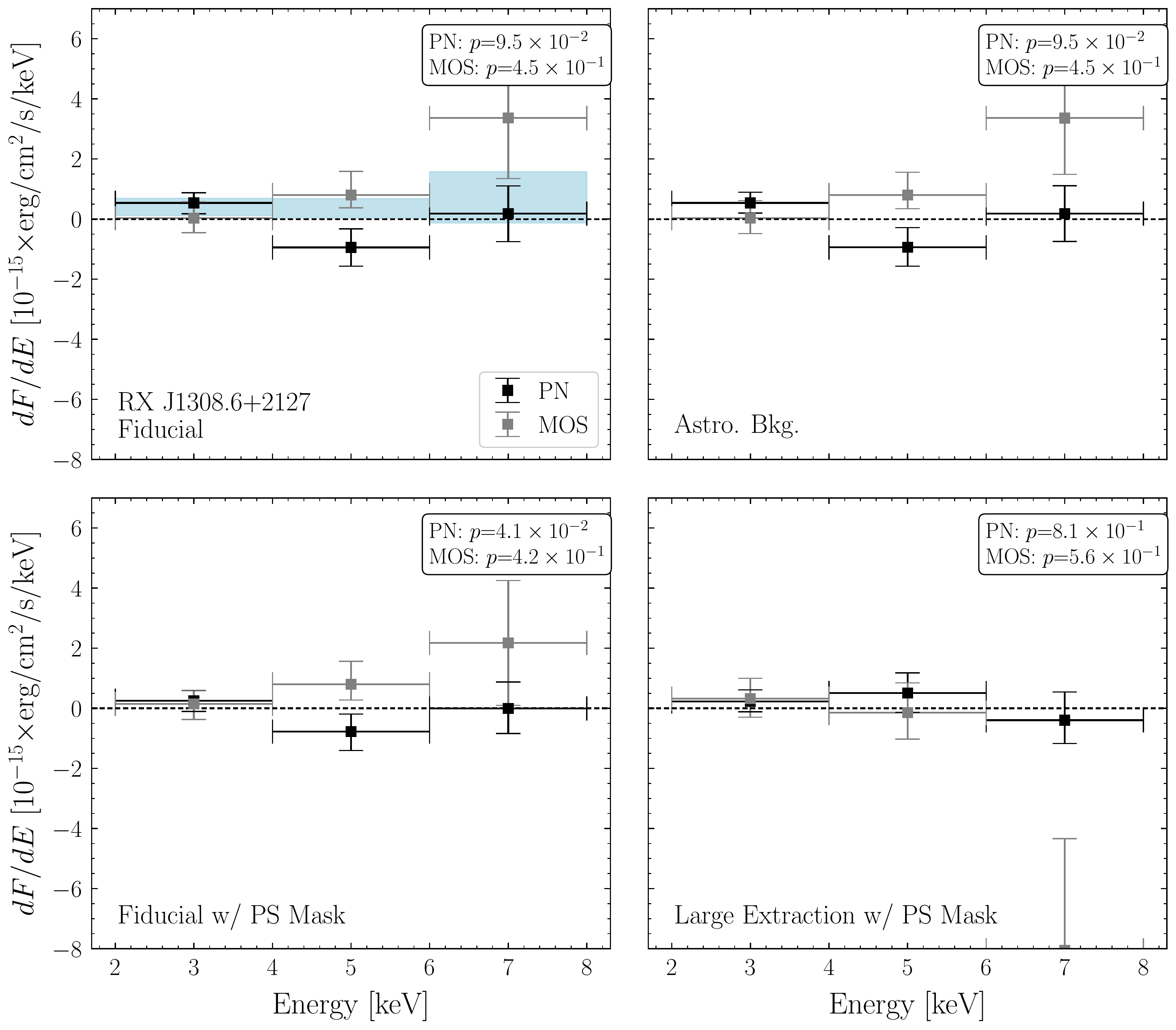}
\end{center}
\caption{Systematic variations on the analysis procedure on the reconstructed fluxes and limits at each energy bin for  RX J1308+2127.}
\label{fig:Systematic_1308}
\end{figure*}

\begin{figure*}[htb]
\begin{center}
\includegraphics[width = 0.48\textwidth]{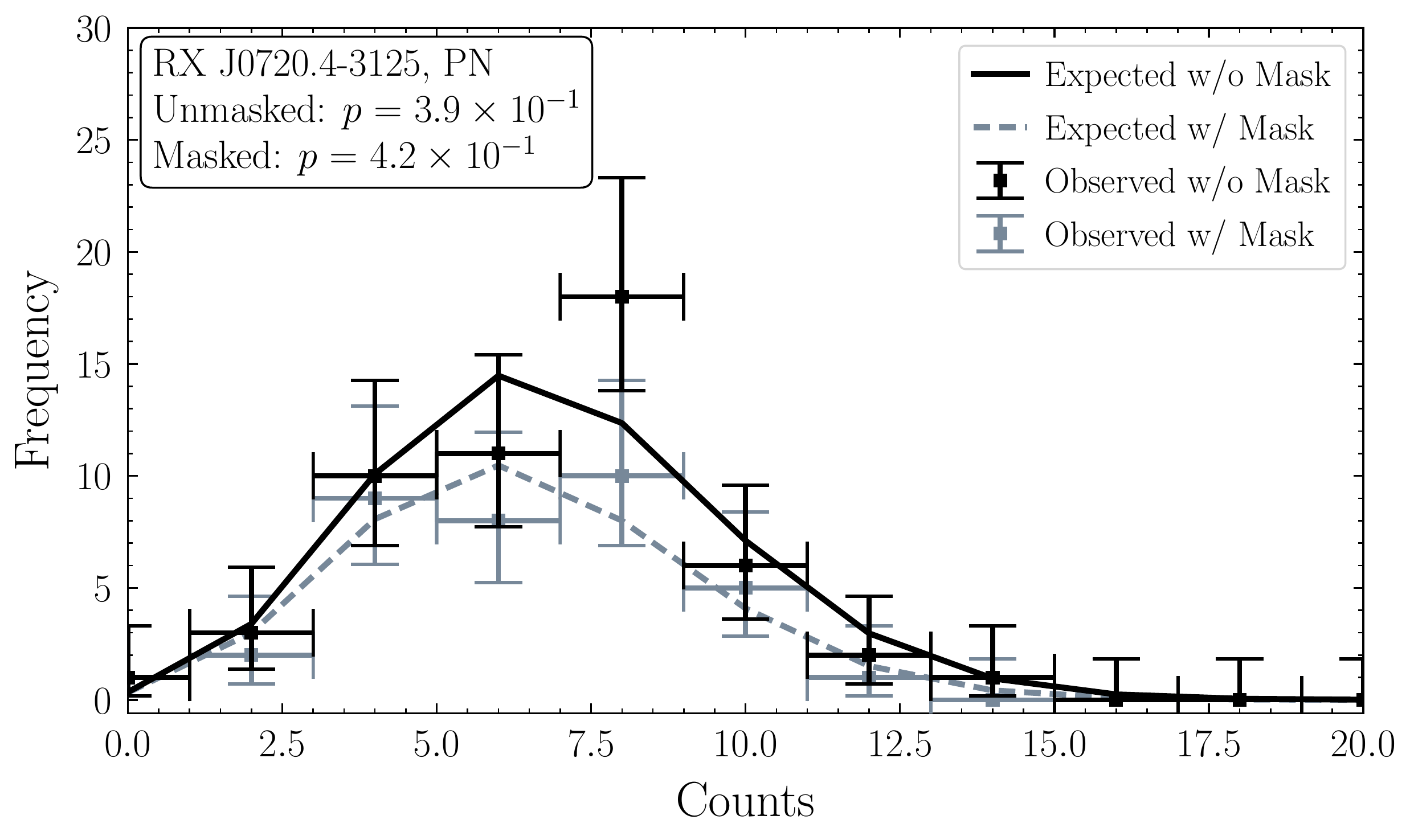} \includegraphics[width = 0.48\textwidth]{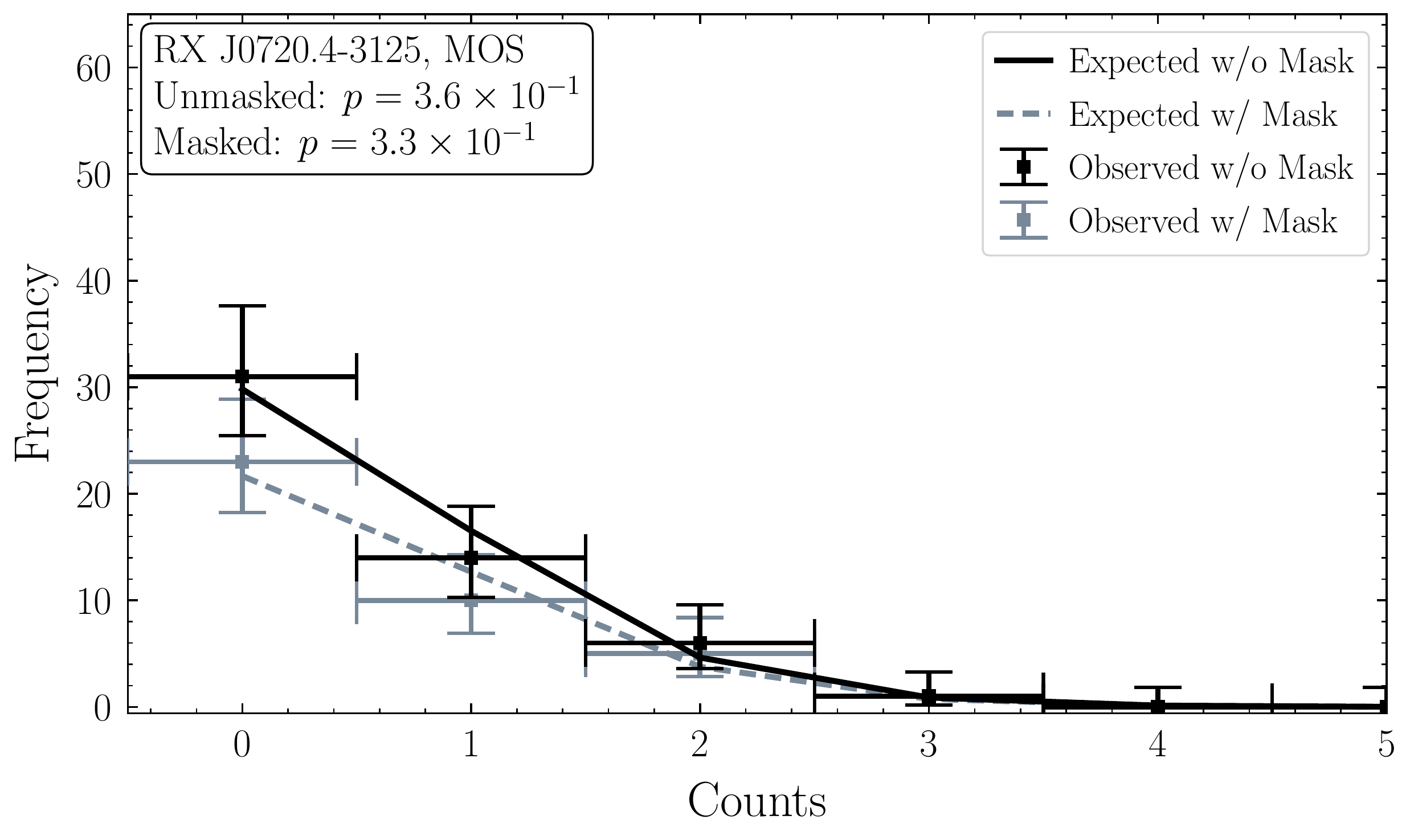} 
\end{center}
\caption{The distribution of background counts by pixel for RX J0720.4-3125 in both PN (\textit{left}) and MOS (\textit{right}). A nearby point source is detected, but masking it has marginal impact on the goodness of fit.}
\label{fig:PoissonDistTest_0720}
\end{figure*}

\begin{figure*}[htb]
\begin{center}
\includegraphics[width = .85 \textwidth]{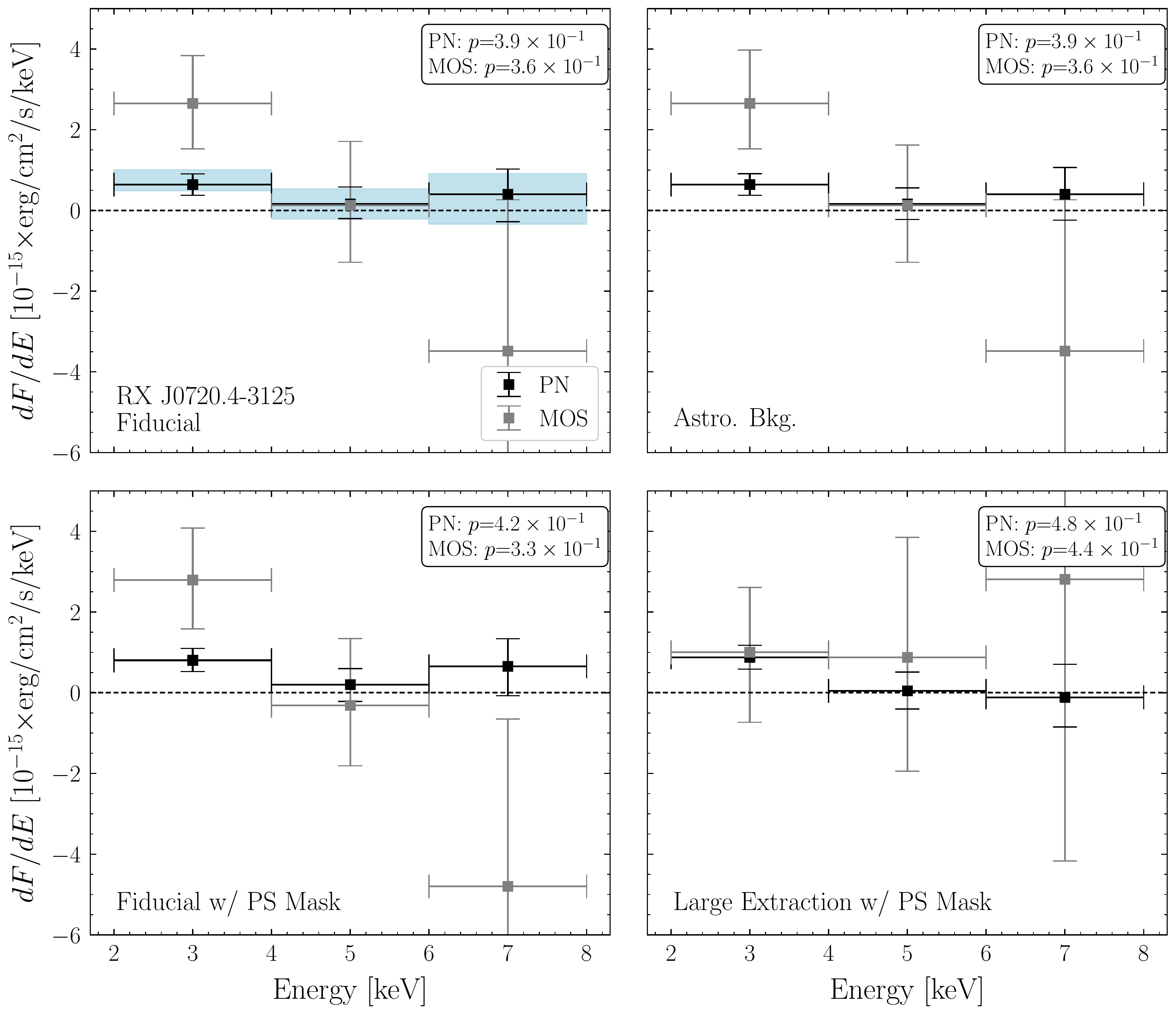}
\end{center}
\caption{Systematic variations on the analysis procedure on the reconstructed fluxes at each energy bin for  RX J0720.4-3125.}
\label{fig:Systematic_0720}
\end{figure*}

\begin{figure*}[htb]
\begin{center}
\includegraphics[width = 0.48\textwidth]{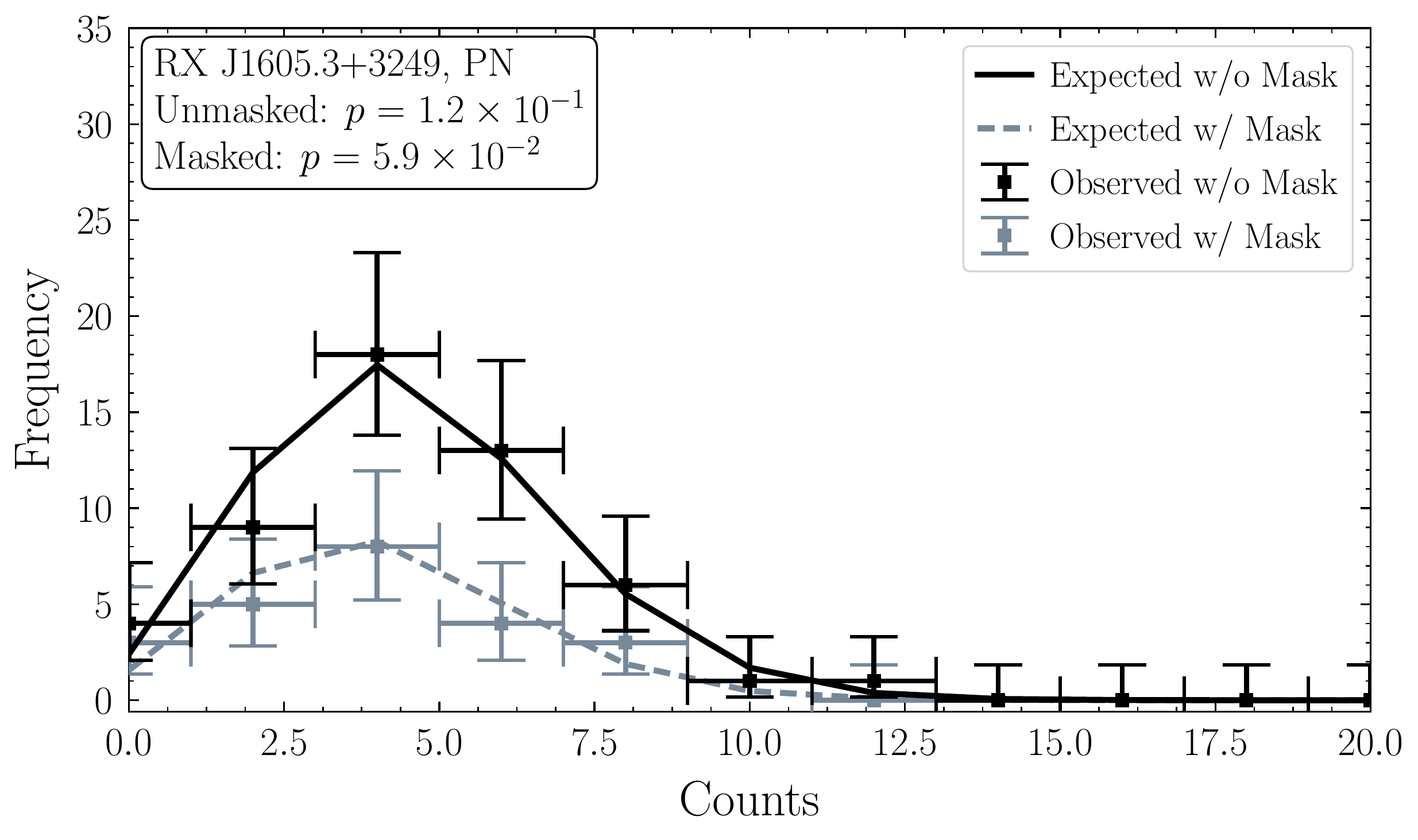} \includegraphics[width = 0.48\textwidth]{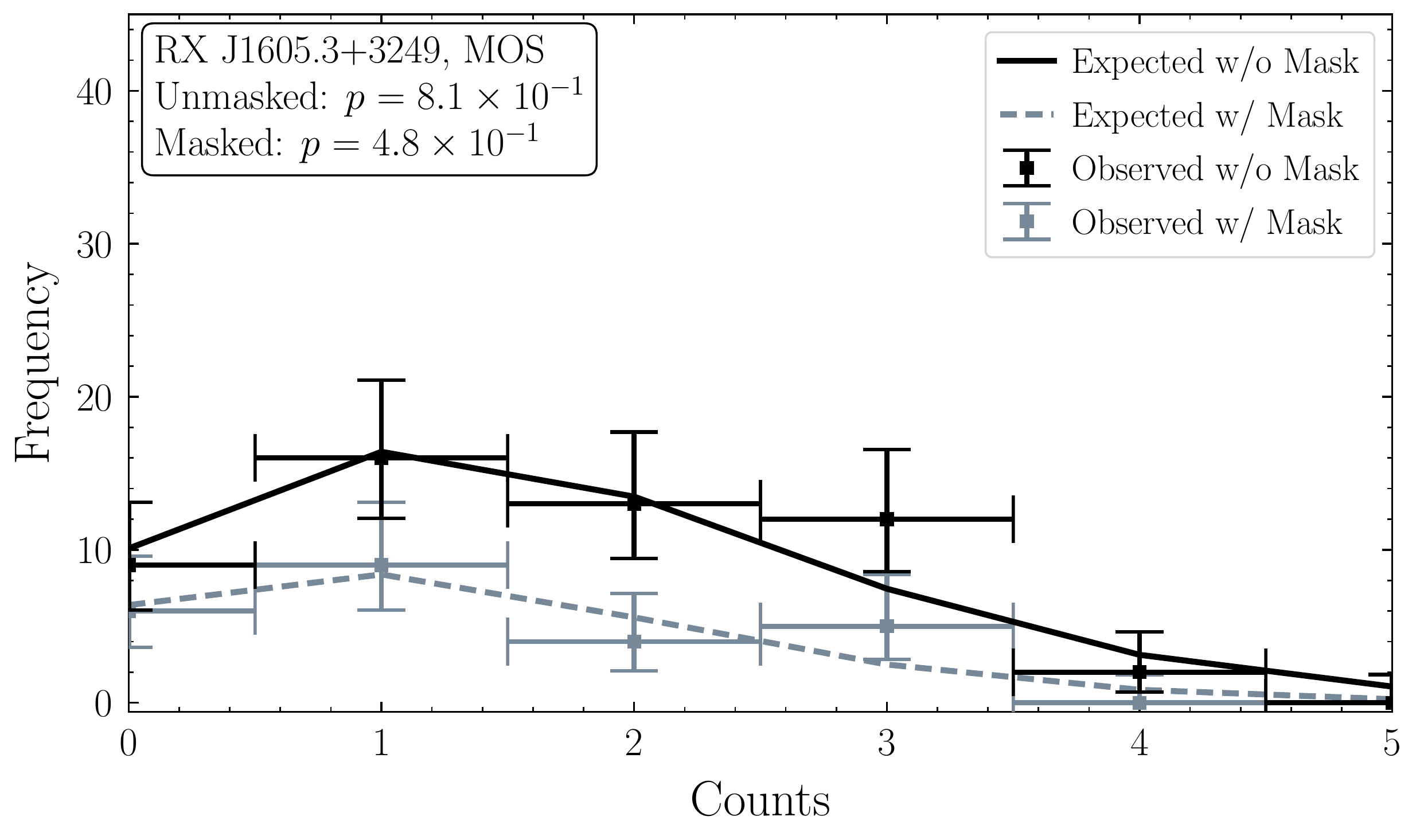}
\end{center}
\caption{The distribution of background counts by pixel for RX J1605.3+3249 in the PN (\textit{left}) and MOS (\textit{right}) instrument.}
\label{fig:PoissonDistTest_1605}
\end{figure*}

\begin{figure*}[htb]
\begin{center}
\includegraphics[width = .99\textwidth]{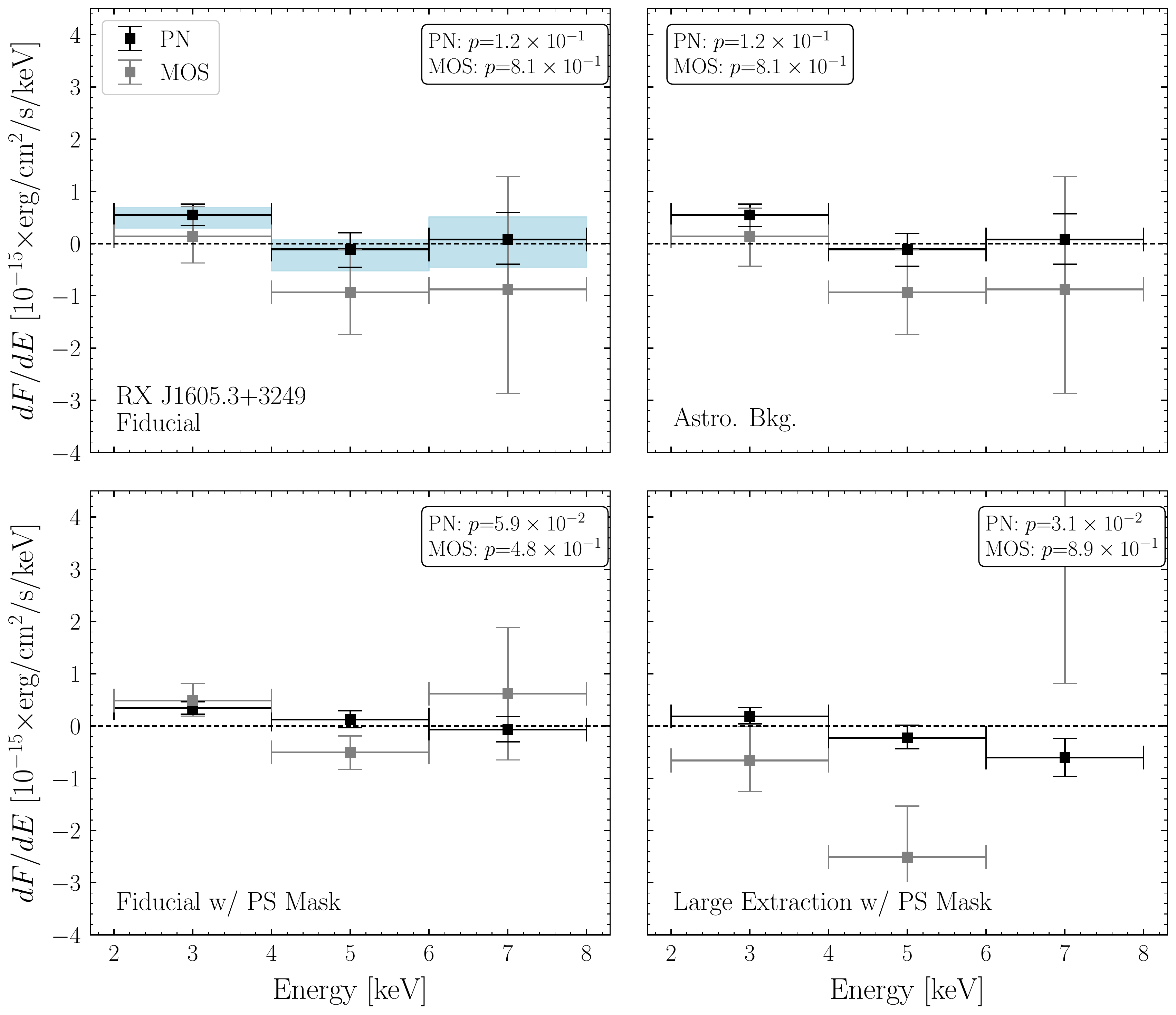}
\end{center}
\caption{Systematic variations on the analysis procedure on the reconstructed fluxes at each energy bin for    RX J1605.3+3249.}
\label{fig:Systematic_1605}
\end{figure*}

\begin{figure*}[htb]
\begin{center}
\includegraphics[width = 0.48\textwidth]{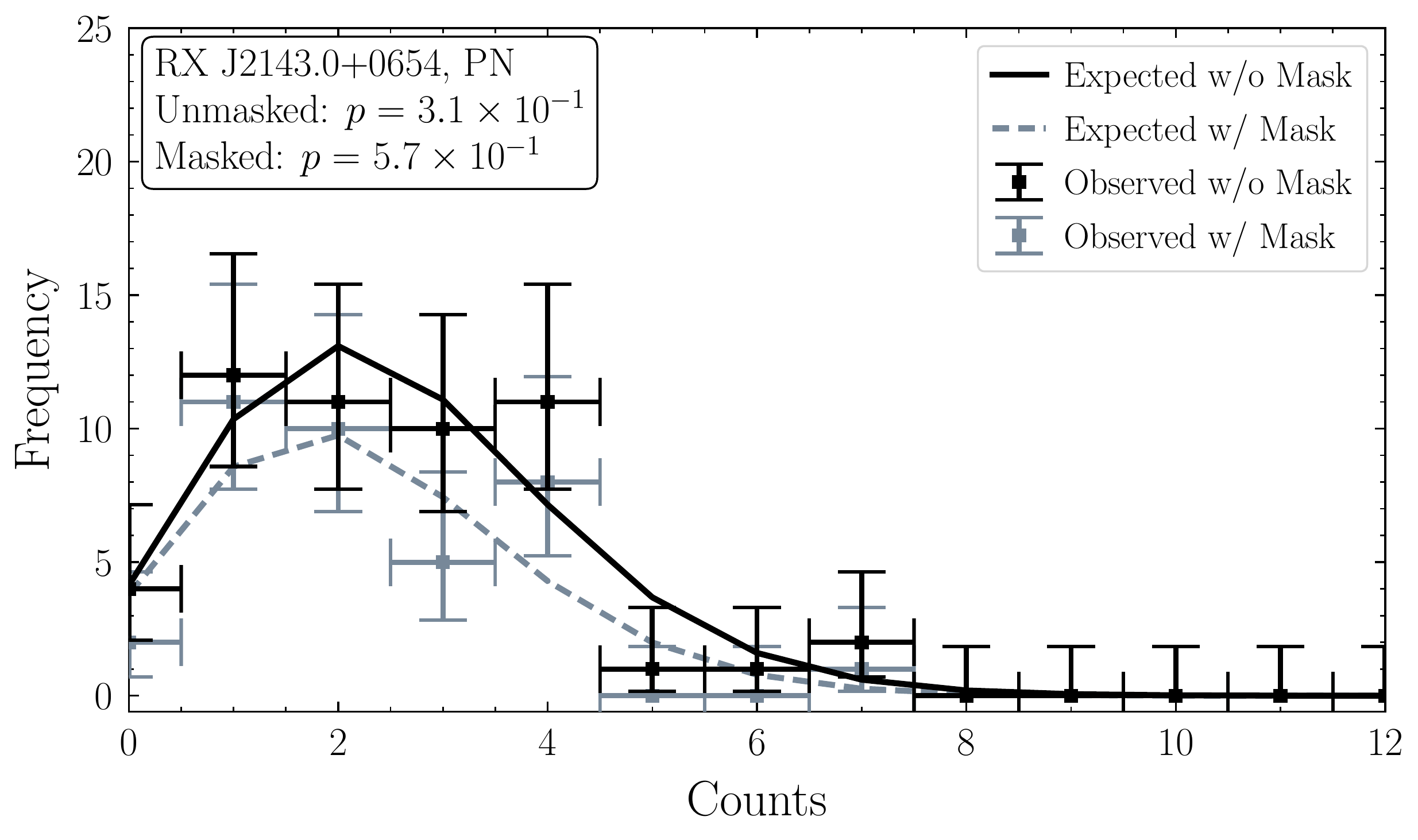} 
\end{center}
\caption{The distribution of background counts by pixel for RX J2143.0+0654 in the PN instrument. }
\label{fig:PoissonDistTest_2143}
\end{figure*}

\begin{figure*}[htb]
\begin{center}
\includegraphics[width = .99\textwidth]{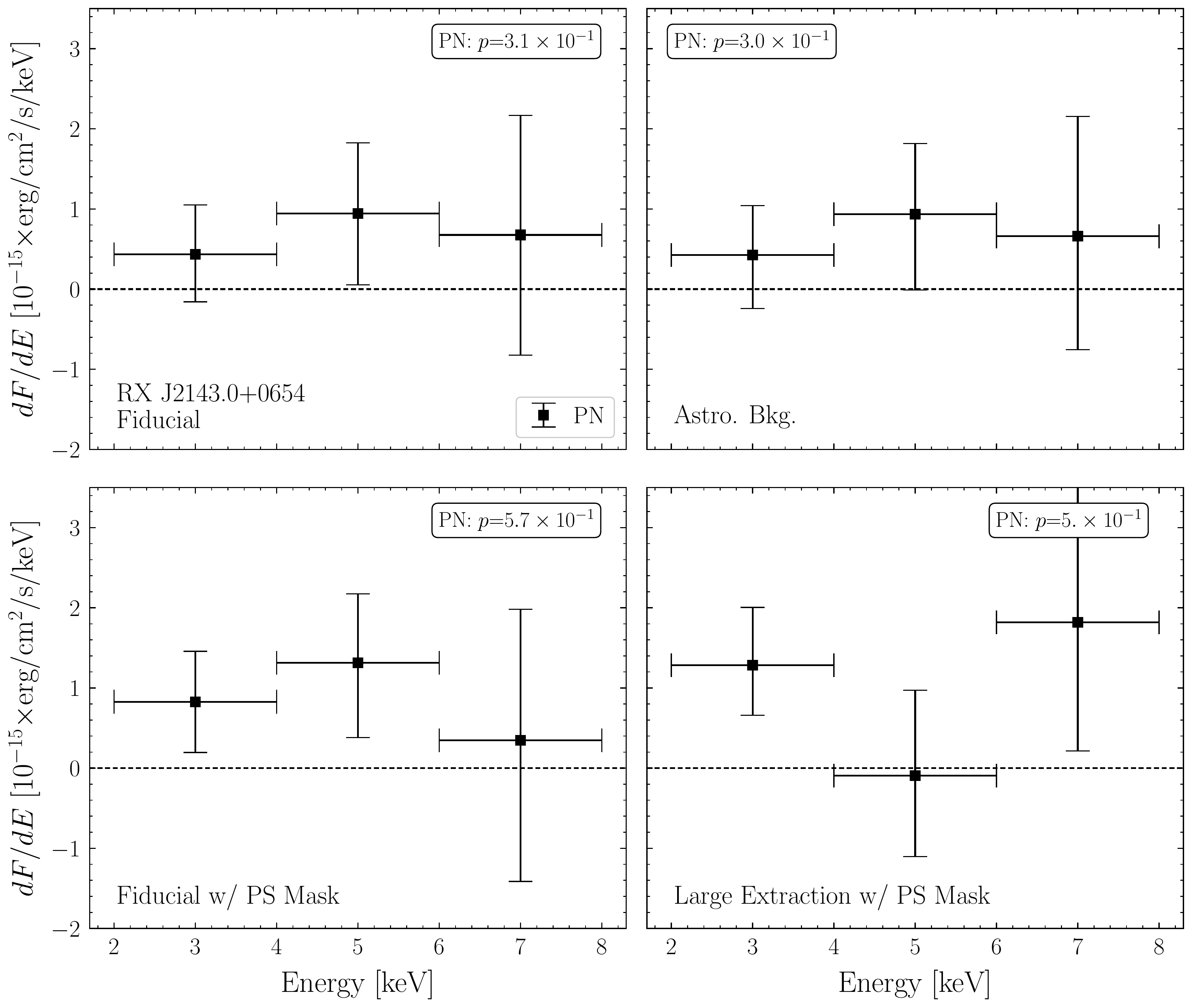}
\end{center}
\caption{Systematic variations on the analysis procedure on the reconstructed fluxes at each energy bin for    RX J2143.0+0654.}
\label{fig:Systematic_2143}
\end{figure*}

Here, we present the fiducial spectral limits on the flux in each energy bin from 2 to 8 keV for each NS in each instrument in which they are observed, along with systematic variations on our analysis procedure that test the robustness of the reconstructed flux. We also inspect the count distribution in the background extraction region compared to the one expected under the fitted background rate, and include the $p$-values for the background goodness of fit under each analysis procedure. These results are in analogy to those shown in Fig.~\ref{fig:PoissonDistTest} and Fig.~\ref{fig:Systematic} for RX J1856.6-3754.  We present the results for the other six NSs in Figs.~\ref{fig:PoissonDistTest_0806},~\ref{fig:Systematic_0806},~\ref{fig:PoissonDistTest_0420},~\ref{fig:Systematic_0420},~\ref{fig:PoissonDistTest_1308},~\ref{fig:Systematic_1308},~\ref{fig:PoissonDistTest_0720},~\ref{fig:Systematic_0720},~\ref{fig:PoissonDistTest_1605},~\ref{fig:Systematic_1605},~\ref{fig:PoissonDistTest_2143}, and~\ref{fig:Systematic_2143}.  

\clearpage

\section{Inspection of the 8-10 keV energy bin}
\label{app:8-10}

In this analysis, we have chosen to exclude the analysis of X-ray counts in the energies between 8 and 10 keV. This choice is motivated by a number of statistical and systematic technical issues. In the 8-10 keV bin, the background count rate increases substantially. For instance, for RX J1856.6-3754, in PN the effective area decreases by 45\% from the 6-8 keV bin to the 8-10 keV bin, while the absolute number of counts increases by 30\% from 6-8 to 8-10 keV. Likewise, in MOS the effective area decreases by 67\% while the absolute counts decreases by only 21\%, and in Chandra, the effective area decreases by 75\% while the absolute counts decreases by only 52\%. This reduces our overall sensitivity in the 8-10 keV bin while also rendering our analysis more susceptible to mismodeling the background, which is spatially inhomogeneous over the detector. Moreover, the calibration of the instruments becomes more uncertain at higher energies.
Additionally, pileup may significantly suppress the counts in this bin due to migrating the photon energies above the detector threshold.  Finally, the detector PSF increases with energy and our signal region can become appreciably contaminated by nearby point sources. 

\begin{figure*}[htb]
\begin{center}
\includegraphics[width = .99\textwidth]{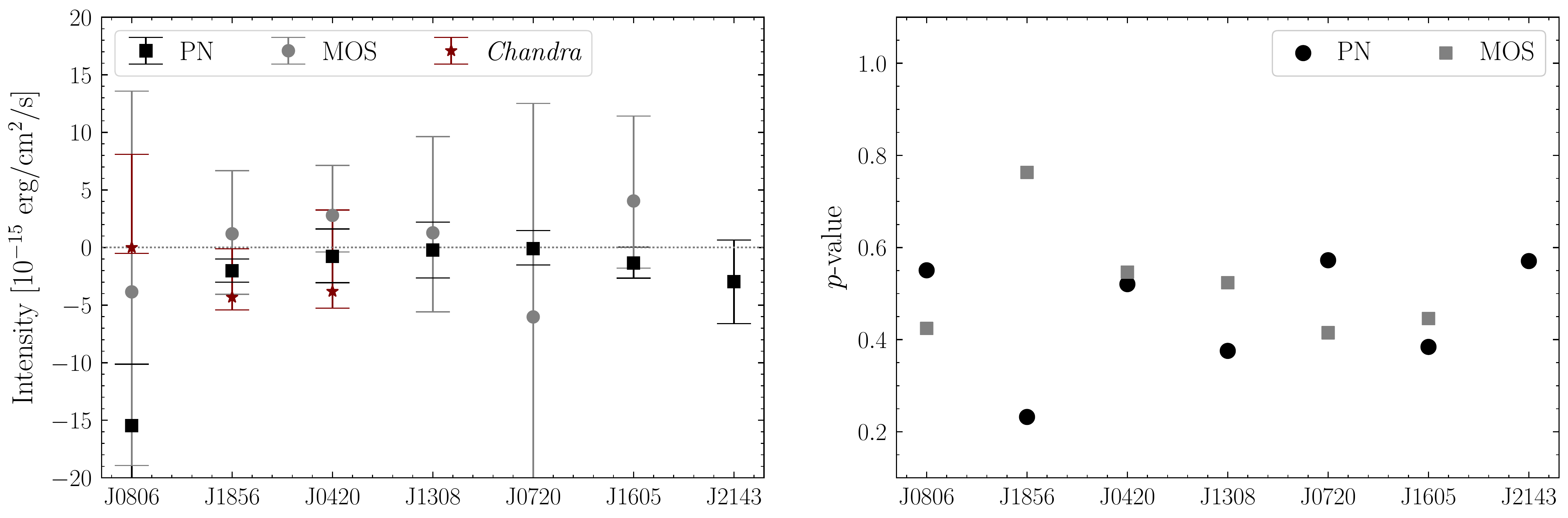}
\end{center}
\caption{(\textit{Left}) The 68\% confidence intervals for the reconstructed intensities in the 8-10 keV bin only in each instrument and for each NS. (\textit{Right}) The $p$-values for observing a pixel-by-pixel background with a likelihood less than the one observed in the data assuming the fitted background rate as its true rate, indicating the goodness of fit of the background model to the data.  In this figure, we restrict to counts at energies between 8 and 10 keV.  The $p$-value for PN data from RX J1856.6-3754 is quite poor, while the rest of the $p$-values are above 0.1.} 
\label{fig:Fourth_Bin}
\end{figure*}

For completeness, we include the best-fit intensities in the 8-10 keV bin for each NS along with the $p$-value for its goodness-of-fit in the background region under the null model in Fig.~\ref{fig:Fourth_Bin}. Even discounting the systematic errors discussed above, the statistical uncertainties on the intensity tend to be quite weak in the 8-10 bin as compared to those uncertainties for energies between 2 and 8 keV.  The data also appear to demonstrate more frequent underfluctuations, which could be the result of systematic biases. 

\end{document}